\documentclass[twocolumn,trackchanges]{aastex7}

\usepackage{CJK}
\usepackage{amsmath}

\newcommand{\petit}{\texttt{petitRADTRANS}}

\newcommand{\teff}{T$_{\rm eff}$}

\newcommand{\kms}{$\rm km\,s^{-1}$}
\newcommand{\kpvsys}{$K_{\rm p} - \Delta v_{\rm sys}$}
\newcommand{\kp}{$K_{\rm p}$}
\newcommand{\vsys}{$\Delta v_{\rm sys}$}
\newcommand{\ktb}{KELT-20~b}
\newcommand{\wtb}{WASP-33~b}
\newcommand{\mob}{MASCARA-1~b}
\newcommand{\knb}{KELT-9~b}
\newcommand{\wob}{WASP-189~b}
\newcommand{\tob}{TOI-1518~b}
\newcommand{\cpo}{$\rm [(C+O)/H]$}

\newcommand{\caltech}{Department of Astronomy, California Institute of Technology, Pasadena, CA 91125, USA}
\newcommand{\gps}{Division of Geological \& Planetary Sciences, California Institute of Technology, Pasadena, CA 91125, USA}
\newcommand{\ucsc}{Department of Astronomy \& Astrophysics, University of California, Santa Cruz, CA95064, USA}
\newcommand{\keck}{W. M. Keck Observatory, 65-1120 Mamalahoa Hwy, Kamuela, HI 96743, USA}
\newcommand{\ucla}{Department of Physics \& Astronomy, 430 Portola Plaza, University of California, Los Angeles, CA 90095, USA}
\newcommand{\jpl}{Jet Propulsion Laboratory, California Institute of Technology, 4800 Oak Grove Dr.,Pasadena, CA 91109, USA}
\newcommand{\ucsd}{Center for Astrophysics and Space Sciences, University of California, San Diego, La Jolla, CA 92093}

\newcommand{\northwestern}{Center for Interdisciplinary Exploration and Research in Astrophysics (CIERA) and Department of Physics and Astronomy,
Northwestern University, Evanston, IL 60208, USA}
\newcommand{\arizona}{James C. Wyant College of Optical Sciences, University of Arizona,
Meinel Building 1630 E. University Blvd., Tucson, AZ 85721, USA}
\newcommand{\steward}{Steward Observatory, University of Arizona, 933 N Cherry Ave, Tucson, AZ, USA 85719}

\received{\today}

\shorttitle{KPIC UHJ survey}
\shortauthors{Finnerty et al.}
\graphicspath{{./}{}}

\begin{document}
\begin{CJK*}{UTF8}{gbsn}

\title{Atmospheric characterization of six ultra-hot Jupiters from $K$-band high-resolution spectroscopy}

\correspondingauthor{Luke Finnerty}
\email[show]{lfinnert@umich.edu}

\author[0000-0002-1392-0768]{Luke Finnerty}
\affiliation{\ucla}
\email{lfinnert@umich.edu}

\author[0000-0002-0176-8973]{Michael P. Fitzgerald}
\affiliation{\ucla}
\email{mpfitz@ucle.edu}

\author[0000-0002-6171-9081]{Yinzi Xin}
\affiliation{\caltech}
\email{yxin@caltech.edu}

\author[0000-0002-6618-1137]{Jerry W. Xuan}
\altaffiliation{51 Pegasi b Fellow}
\affiliation{Department of Earth, Planetary, and Space Sciences, University of California, Los Angeles, CA 90095, USA}
\email{jerryxuan@g.ucla.edu}

\author{Julie Inglis}
\affiliation{\caltech}
\email{jinglis@caltech.edu}

\author[0000-0003-2429-5811]{Shubh Agrawal}
\affiliation{Department of Physics and Astronomy, University of Pennsylvania, Philadelphia, PA 19104, USA}
\email{shubh@sas.upenn.edu}

\author[0000-0002-6525-7013]{Ashley Baker}
\affiliation{\caltech}
\email{abaker@caltech.edu}

\author{Randall Bartos}
\affiliation{\jpl}
\email{randall.d.bartos@jpl.nasa.gov}

\author{Geoffrey A. Blake}
\affiliation{\gps}
\email{gab@gps.caltech.edu}

\author[0000-0003-4737-5486]{Benjamin Calvin}
\affiliation{\caltech}
\affiliation{\ucla}
\email{bcalvin@astro.ucla.edu}

\author{Sylvain Cetre}
\affiliation{\keck}
\email{scetre@keck.hawaii.edu}

\author[0000-0001-8953-1008]{Jacques-Robert Delorme}
\affiliation{\keck}
\affiliation{\caltech}
\email{jdelorme@keck.hawaii.edu}

\author{Greg Doppmann}
\affiliation{\keck}
\email{gdoppmann@keck.hawaii.edu}

\author[0000-0002-1583-2040]{Daniel Echeverri}
\affiliation{\caltech}
\email{dechever@caltech.edu}

\author{Katelyn Horstman}
\affiliation{\caltech}
\email{khorstma@astro.caltech.edu}

\author[0000-0002-5370-7494]{Chih-Chun Hsu}
\affiliation{\northwestern}
\email{chsu@northwestern.edu}

\author[0000-0001-5213-6207]{Nemanja Jovanovic}
\affiliation{\caltech}
\email{nem@caltech.edu}

\author[0000-0002-4934-3042]{Joshua Liberman}
\affiliation{\caltech}
\affiliation{\arizona}
\affiliation{\steward}
\email{jliberman@arizona.edu}

\author[0000-0002-2019-4995]{Ronald A. L\'opez}
\affiliation{\ucla}
\email{rlopez@astro.ucla.edu}

\author{Dimitri Mawet}
\affiliation{\caltech}
\affiliation{\jpl}
\email{dmawet@astro.caltech.edu}

\author{Evan Morris}
\affiliation{\ucsc}
\email{ecmorris@ucsc.edu}

\author{Jacklyn Pezzato-Rovner}
\affiliation{\caltech}
\email{jpezzatorovner@gmail.com}

\author[0000-0003-2233-4821]{Jean-Baptiste Ruffio}
\affiliation{\ucsd}
\email{jruffio@ucsd.edu}

\author[0000-0003-1399-3593]{Ben Sappey}
\affiliation{\ucsd}
\email{bsappey@ucsd.edu}

\author{Tobias Schofield}
\affiliation{\caltech}
\email{toby.s117@gmail.com}

\author{Andrew Skemer}
\affiliation{\ucsc}
\email{askemer@ucsc.edu}

\author[0000-0001-5299-6899]{J. Kent Wallace}
\affiliation{\jpl}
\email{james.k.wallace@jpl.nasa.gov}

\author[0000-0003-0354-0187]{Nicole L. Wallack}
\affiliation{Earth and Planets Laboratory, Carnegie Institution for Science, Washington, DC 20015, USA}
\email{nwallack@carnegiescience.edu}

\author[0000-0003-0774-6502]{Jason J. Wang (王劲飞)}
\affiliation{\northwestern}
\email{jason.wang@northwestern.edu}

\author[0000-0002-4361-8885]{Ji Wang (王吉)}
\affiliation{Department of Astronomy, The Ohio State University, 100 W 18th Ave, Columbus, OH 43210 USA}
\email{wj198414@gmail.com}

\begin{abstract}

We present new Keck/KPIC high-resolution spectroscopic detections of three ultra-hot Jupiters (UHJs) in the $K$ band: \wob\ ($\rm SNR = 7.2$), \mob\ ($\rm SNR = 8.6$), and \tob\ ($\rm SNR = 7.1$), as well as a tentative detection of \knb\ ($\rm SNR = 5.0$). We perform a uniform set of atmospheric retrieval analysis on these objects, as well as previously reported KPIC observations of \wtb\ ($\rm SNR = 11.2$) and \ktb\ ($\rm SNR = 10.5$), We perform atmospheric retrievals for the pressure-temperature ($P-T$) profile, orbital velocity parameters, $v\sin i$, and abundances of CO, H$_2$O, OH, and Fe, with parameterized mixing profiles to account for the expected vertical abundance variations of H$_2$O and OH. We also perform a set of retrievals assuming chemical equilibrium, which are generally in good agreement with the free retrievals. Except for \knb, the retrieved spectra are dominated by CO emission features, with additional weak H$_2$O or OH features consistent with thermal dissociation of H$_2$O. \knb, which is significantly hotter, appears to have very weak molecular features. Dissociation limits our ability to reliably constrain H$_2$O or OH abundances from $K$ band data alone, resulting in poor constraints on the C/O ratio. For all objects, the atmospheric abundances from detected carbon and oxygen species are $1-10\times$ solar. These results highlight the importance of wide spectral coverage for high-resolution retrievals. Additional observations to expand phase and wavelength coverage are needed to better constrain oxygen species and possible spatial inhomogeneities from dissociation. 

\end{abstract}


\section{Introduction} \label{sec:intro}

 \begin{deluxetable*}{ccccccc}
    \tabletypesize{\footnotesize}
    \tablecaption{Stellar and planetary properties}
    
    \tablehead{ & \colhead{ WASP-33~A\tablenotemark{e}} & \colhead{KELT-9~A\tablenotemark{f}} & \colhead{MASCARA-1~A\tablenotemark{g}} & \colhead{WASP-189~A\tablenotemark{h}} & \colhead{KELT-20~A\tablenotemark{i}} & \colhead{TOI-1518~A\tablenotemark{j}} }
    \startdata
        RA\tablenotemark{a}                     & 02:26:51   & 20:31:27  & 21:10:13  & 15:02:45  & 19:38:39  & 23:29:04  \\
        Dec\tablenotemark{a}                    & +37:33:02  & +39:56:13 & +10:44:20 & -03:01:53 & +31:13:09 & +67:02:05 \\
        $K_{\rm mag}$\tablenotemark{b}          & 7.47 & 7.48 & 7.74 & 6.06 & 7.42 & 8.33  \\
        Mass [$\rm M_\odot$]\tablenotemark{c}   & $1.495\pm0.031$ & $2.52^{+0.25}_{-0.20}$ & $1.72\pm0.07$& $2.03\pm0.07$ & $1.76^{+0.1}_{-0.2}$ & $1.8\pm0.3$\tablenotemark{j} \\
        Radius [$\rm R_\odot$]\tablenotemark{c} & $1.444\pm0.034$ & $2.36^{+0.08}_{-0.06}$ & $2.1\pm0.2$  & $2.36\pm0.03$ & $1.57\pm0.06$        & $1.95\pm0.05$\tablenotemark{j}          \\
        \teff [K]\tablenotemark{c}              & $7430\pm100$    & $10170\pm450$          & $7550\pm150$ & $8000\pm80$   & $8720^{+250}_{-260}$ & $7300^{+100}_{-100}$\tablenotemark{j}  \\
        $\log g$\tablenotemark{c}               & $4.25\pm0.1$    & $4.09\pm0.014$           & $4.1\pm0.2$  & $3.9\pm0.2$   & $4.29\pm0.02$        & $4.1\pm0.2$\tablenotemark{j}   \\
        $v\sin i$ [\kms]                        & $90\pm10$       & $111.4\pm1.3$          & $109\pm4$    & $93.1\pm1.7$  & $117\pm3$          & $85.1\pm6.3$           \\
        $v_{\rm rad}$ [\kms]                    & $-0.3^{+5.3}_{-5.6}$\tablenotemark{k}          & $-20.6\pm0.1$          & $8.52\pm0.02$      & $-24.45\pm0.01$\tablenotemark{l}    & $-23.3\pm0.3$        & $-13.9\pm0.2$                \\
        \smallskip \\
        \hline
         &  \wtb & \knb & \mob & \wob & \ktb & \tob \\
        \hline
        Period [d]\tablenotemark{d}               & $1.219867\pm1\times10^{-6}$ & $1.4810903\pm4\times10^{-7}$ & $2.14879\pm2\times10^{-5}$ & $2.724031\pm3\times10^{-6}$ & $3.4741012\pm2\times10^{-7}$    & $1.902606\pm7\times10^{-6}$\\
        $\rm t_{\rm transit}$ [JD\tablenotemark{d}& $2454163.2237\pm0.0003$     & $2459883.1989\pm0.0002$      & $2459798.2859\pm0.0002$    & $2456706.457\pm0.002$       & $2460337.98638\pm5\times10^{-5}$& $2459854.4146\pm1\times10^{-4}$\\
        $a$ [AU]                                  & $0.0256\pm0.0002$           & $0.0346\pm0.001$             & $0.0404\pm0.0005$\tablenotemark{m} & $0.0501\pm0.001$    & $0.057^{+0.001}_{-0.002}$       & $0.039\pm0.001$  \\
        $i$ [deg]                                 & $87.7\pm1.8$                & $86.8\pm0.3$                 & $87^{+2}_{-3}$             & $84.03\pm0.14$              & $86.1\pm0.3$                    & $77.8\pm0.2$ \\
        $\lambda$ [deg]                           & $251\pm1$                   & $84.8\pm1.4$                 & $69.5\pm3$                 & $86.4^{+2.9}_{-4.4}$        & $3.4\pm2.1$                     & $240\pm1$ \\
        Mass [$\rm M_J]$                          & 2.8                         & $2.88\pm0.84$                & $3.7\pm0.9$                & $2.0^{+0.2}_{-0.1}$         & $<3.37$                         & $<2.3$ \\
        Radius [$\rm R_J$]\tablenotemark{d}       & $1.5\pm0.05$                & $1.89^{+0.06}_{-0.05}$       & $1.5\pm0.3$                & $1.62\pm0.02$               & $1.90\pm0.05$                   & $1.75\pm0.13$ \\
        $\rm T_{\rm eq}$ [K]                      & $2700\pm40$                 & $4060\pm180$                 & $2630\pm50$                & $2650\pm30$                 & $2210\pm60$                     & $2500\pm30$ \\
        $K_{\rm p}$ [\kms]                        & $230.9^{+6.9}_{-7.4}$\tablenotemark{k} & $254\pm7$         & $204\pm3$                  & $200\pm4$                   & $176.7\pm0.6$\tablenotemark{n}  & $198^*$\\
    \enddata
    \tablerefs{\tablenotemark{a}\citet{gaiaedr3}, \tablenotemark{b}\citet{cutri2003}, \tablenotemark{c}\citet{tic8},\tablenotemark{d}\citet{guerrero2021}, \tablenotemark{e}\citet{collier2010}, \tablenotemark{f}\citet{gaudi2017}, \tablenotemark{g}\citet{talens2018m1}, \tablenotemark{h}\citep{lendl2020},  \tablenotemark{i}\citet{lund2017}, \tablenotemark{j}\citep{cabot2021}, \tablenotemark{k}\citep{nugroho2021}, \tablenotemark{l}\citep{anderson2018}, \tablenotemark{m}\citep{hooton2022}, \tablenotemark{n}\citep{yan2022k20} }
    \tablecomments{$\rm T_{\rm eq}$ and $K_{\rm p}$ were calculated based on the listed parameters unless a reference is provided. For all other parameters, the corresponding reference is given primarily by row, secondarily by column, if the value does not have a reference specifically indicated. *-- the reported \kp\ for \tob\ range from $207\pm4.5$ \kms\ \citep{simonnin2024} to $158^{+68}_{-44}$ \kms\ \citep{cabot2021}. The adopted value of 198 \kms\ is intended to split the difference, and was close to the best-fit \kp\ found in initial cross-correlation tests.}
    
    \label{tab:props}
\end{deluxetable*}

The hottest exoplanets, known as ultra-hot Jupiters (UHJs), present an ongoing puzzle in our understanding of planet formation and a formidable challenge to our models of atmospheric physics and chemistry. Observationally, there is evidence for supersonic winds in UHJs \citep[e.g.][]{ehrenreich2020, lesjak2023}, which vary with altitude \citep{kesseli2022, kesseli2024}, and potentially also in time \citep{pai2022}. However, definitively measuring circulation patterns remains challenging, because large temperature contrasts between the day and night sides of UHJs can lead to condensation \citep{ehrenreich2020} and cold trapping \citep{pelletier2023}. These changes in atmospheric composition with longitude can produce net Doppler shifts in the observed spectrum similar to those resulting from winds and rotation. Disentangling the effects of winds, chemistry, temperature variations, and nightside/terminator clouds is a persistent challenge in interpreting spectroscopy of UHJs \citep[e.g.][]{savel2022}. Constraints on very stable molecules such as CO, which has strong spectral features but minimal 3D abundance variations, even under UHJ conditions, may provide a stable reference point to begin isolating these effects from each other \citep{savel2023}.

Even a simplified 1D approach to UHJ atmospheres can lead to significant biases in inferred abundances due to molecular dissociation. While the CO bond is strong enough that the atmospheric abundance is constant with altitude near the infrared photosphere for all but the hottest UHJs, H$_2$O is strongly impacted by thermal dissociation, leading to weaker-than-expected H$_2$O spectral features for a given deep-atmosphere abundance \citep{parmentier2018}. Observationally, this biases free retrievals of UHJ spectra using vertically fixed abundances towards high C/O ratio and metallicity relative to solar \citep[e.g.][]{brogi2023, finnerty2023}, while retrieval analysis based on equilibrium chemistry models prefer lower C/O ratios and metallicities more compatible with solar values \citep[e.g.][]{brogi2023, ramkumar2023}. Recent analyses have also accounted for dissociation by using an approximate analytic model for the H$_2$O vertical mixing profile from \citet{parmentier2018}, which avoids this bias without requiring a full equilibrium chemistry calculation \citep{gandhi2024, pelletier2025} that often entails assumptions about abundance ratios other than C/O. Alternatively, \citet{finnerty2025k20} took an entirely numerical approach to dissociation, fitting for a pressure below which the H$_2$O abundance decays following a power law. 

Despite these challenges, UHJs are the most promising exoplanet population for efforts aiming to measure bulk atmospheric composition. The high temperature of UHJs, while challenging, also leads to nearly all species, including refractory elements, being present in the gas phase and accessible to spectroscopy. For example, \citet{pelletier2023} reported detection of 15 metal species in the atmosphere of the UHJ WASP-76~b, including simultaneous detection of the neutral and first ionized states of both calcium and iron. This enables precise measurements of the refractory/volatile ratio in UHJs \citep[e.g.][]{pelletier2025}, which can break degeneracies in formation history which arise from considering only C/O ratio and overall metallicity \citep{lothringer2021, chachan2023}.

The formation history of UHJs is a particularly interesting challenge. While many UHJs show significant spin-orbit misalignment \citep[e.g.][]{collier2010, gaudi2017, talens2018m1, cabot2021}, this is not universal \citep{talens2018}. This suggests that either some UHJs are formed through disk migration processes, preserving low inclinations, or that the high-eccentricity migration process believed to be responsible for the misaligned UHJs \citep{naoz2011} does not always result in inclination excitation.

High-resolution cross-correlation spectroscopy (HRCCS) is a powerful tool for characterizing UHJ atmospheres. UHJs are generally found around hot, rapidly-rotating primaries, leading to few and very broad stellar spectral lines. In contrast, UHJs have strong, narrow emission features from a wide range of atoms and molecules, leading to a very large spectroscopic contrast between the planet and host star. The very high orbital velocity of UHJs and short orbital period makes scheduling HRCCS observations comparatively easy and presents unique opportunities for phase-resolved spectroscopy. While an increasing number of UHJs have been studied using HRCCS \citep[e.g.][]{pelletier2023, brogi2023, ramkumar2023, pelletier2025, finnerty2025k20}, multi-object comparison studies have been comparatively rare \citep{gandhi2023}. Understanding UHJs as a population will require more uniform analyses suitable for cross-comparisons between many objects, including at infrared wavelengths.

As part of the broader KPIC hot Jupiter survey from 2022--2025, we obtained $K$ band observations of six UHJs. \wtb\ was observed as an initial demonstration target, and the first analysis was reported in \citet{finnerty2023}. Observations of \ktb\ were reported in \citet{finnerty2025k20}, where we described a free-retrieval approach to the vertical mixing profiles of OH and H$_2$O in order to account for dissociation. Here we report the KPIC observations of \knb, \wob, \mob, and \tob, and perform a re-analysis of the \wtb\ and \ktb\ observations with the latest version of our atmospheric retrieval code. This uniform analysis of data obtained with a consistent spectral resolution and wavelength coverage provides a starting point for comparisons between UHJs to explore population-level diversity and trends. 

Section \ref{sec:obs} describes our observations, data reduction, and retrieval setup. Section \ref{sec:res} discusses the retrieval results for each object individually. Section \ref{sec:disc} discusses comparisons between the six objects and broader implications for the atmospheric dynamics and formation/evolutionary history of the ultra hot Jupiter population. Section \ref{sec:conc} concludes.

\section{Observations and Data Reduction}\label{sec:obs}

\begin{deluxetable*}{ccccccccc}
    \tablecaption{Summary of observations}
    \tabletypesize{\tiny}
     \tablehead{ & \colhead{UT Date} & \colhead{Wall time [h:mm]} & \colhead{t$_{exp}$ [s]}  & \colhead{N$_{spec}$} & \colhead{Science fibers} & \colhead{Orbital Phase} &  \colhead{Airmass} & \colhead{SNR/pixel}}
     \startdata
        \wtb & 2021 Nov. 21 & 3:12 & 90 & 46,48 & 1, 2 & 0.52--0.63 & 1.07--1.22 & 58 \\
        \knb & 2022 July 22 & 3:30 & 90 &  35, 35, 35 & 2, 3, 4 & 0.36--0.46 & 1.06--1.54 & 76 \\
        \wob & 2023 May 07 & 3:18 & 60 &  170 & 4 & 0.43--0.48 & 1.69--1.09 & 120 \\
        \mob & 2023 July 02 & 4:30 & 180  & 80 & 2 & 0.36--0.45 & 3.05--1.02 & 116 \\
        \ktb & 2023 July 30 & 6:29 & 300  & 71 & 4 & 0.54--0.61 & 1.72--1.25 & 133 \\
        \tob & 2023 July 30 & 3:00 & 180 & 53 & 4 & 0.52--0.59 & 1.50--1.60 & 63
    \enddata
    \tablecomments{N$_{spec}$ is the number of frames in each fiber's time series after coadding. Integration times and number of coadded frames were chosen to keep the planet velocity shift within the coadded integration time below 4 km/s 
    while staying well below the non-linear threshold for the NIRSPEC detector. Airmass is given for the start and end of each observation sequence, during which several targets transited zenith. SNR/pixel is given after extraction and coadding, and is not directly comparable across targets due to differences in the integration time. }
    \label{tab:obs}
\end{deluxetable*}

\begin{figure*}
    \centering
    \includegraphics[width=0.95\linewidth]{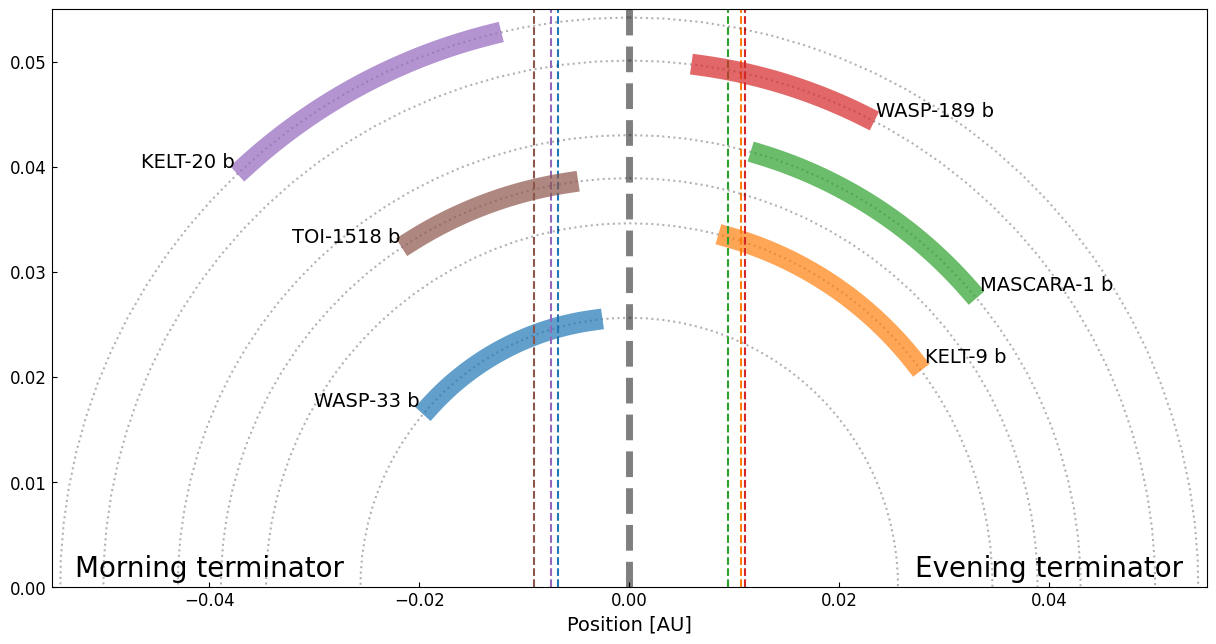}
    \caption{Orbital phase coverage of our observations, with semi-major axes drawn to-scale. The stellar radii are shown as dashed vertical lines in the colors corresponding to each planet to illustrate the start/end of secondary eclipse. We omit portions of the time series for \tob, \wtb, \wob, and \knb\ falling within secondary eclipse. \wtb, \knb, and \mob\ were observed to significant phase angles.}
    \label{fig:phasecoverage}
\end{figure*}

\subsection{Observations}

Table \ref{tab:obs} summarizes our observations with Keck II/ KPIC \citep{nirspec, nirspecupgrade, nirspecupgrade2, kpic, echeverri2022, kpicII}. KPIC was a series of upgrades to both Keck II/NIRSPEC and Keck II Adaptive Optics (AO) to provide diffraction-limited high-resolution spectroscopy in the $H$, $K$, and $L$ bands. The use of AO greatly reduced the thermal background relative to slit-fed spectrographs, and the use of single mode fibers caused changes in pointing or AO performance to lead to a loss of flux, but with no change in the wavelength solution or blaze function. After the 2022 phase II upgrade \citep{kpicII}, KPIC could achieve top-of-atmosphere throughputs up to 5\% under excellent conditions (estimated $H$-band Strehl $\sim0.4$). More typical conditions after the phase II upgrade corresponded to an estimated $H$-band Strehl $\sim0.2$ and throughput $\sim2.5-3\%$, while the record throughput in KPIC phase I was $\sim3\%$. 

The observations of \wtb\ were previously described in \citet{finnerty2023} and those of \ktb\ in \citet{finnerty2025k20}.  We refer the reader to those papers for further details. The following subsections summarize the previously unreported observations along with changes to the previously-reported reduction process as described in \citet{finnerty2025k20, finnerty2025hd143}. The orbital phase coverage of our observations is illustrated in Figure \ref{fig:phasecoverage}.
\subsubsection{\wtb}

These observations were previously described in \citet{finnerty2023}. The science frames were obtained in an ABBA nodding sequence using science fibers 1 and 2 and 90 second exposures. While \citet{finnerty2023} coadded the science data from the two fibers to create a single time series, for this reanalysis we instead keep the data from each science fiber separate and sum the resulting log-likelihoods. Coadding data from different fibers results in frame-to-frame variations in the line-spread function (LSF) due changes in the measured flux impacting the effective weight of each fiber in the coadded spectrum. Handling each fiber separately eliminates this source of LSF variation. The first 7 exposures in science fiber 1 and the first 8 exposures in science fiber 2 were obtained during the secondary eclipse and are omitted from the log-likelihood calculation. The nominal observed planet radial velocities out of eclipse (in the stellar reference frame) ranged from $-57$ to $-171$ \kms, with a maximum change within a single exposure of 2 \kms. 

Conditions were good throughout the observations, with typical top-of-atmosphere throughput around 2.5\%, close to the maximum value achieved for KPIC phase I. HIP 95771 was observed prior to \wtb\ for wavelength calibration.

\subsubsection{\knb}

The science frames were obtained in an ABC nodding pattern between science fibers 2, 3, and 4 with 90 second exposures. As with \wtb, we treat the data from each science fiber as a separate time series and sum the resulting log-likelihoods. The last five exposures in each fiber were taken in secondary eclipse and are omitted from the calculation of the log-likelihood. The nominal out-of-eclipse planet velocities ranged from 200 \kms\ to 90 \kms, with a maximum change of 2 \kms\ within a single exposure. 

Conditions were excellent throughout the observations, with top-of-atmosphere throughput up to 5\%. The KPIC phase II upgrade \citep{echeverri2022, kpicII} is responsible for the significant sensitivity improvement compared with the KPIC phase I observations of \wtb.  HIP 95771 was observed prior to \knb\ for wavelength calibration. 

\subsubsection{\wob}

Observations were obtained staring on science fiber 4 with 60 second exposures and no nodding. Afternoon dark frames were used for background subtraction. The last fifty exposures were taken after the start of secondary eclipse and are omitted from the log-likelihood calculation. The nominal star-frame planet velocity ranged from 89 \kms\ at the start of observations to 46 \kms\ at the start of secondary eclipse, with a maximum change of $\sim0.5$ \kms\ within a single frame.  

Conditions were very good throughout the observations, with typical throughputs $\sim4\%$. Brief patchy cirrus caused some extinction towards the start of the observation sequence. HIP 62944 was observed prior to \wob\ for wavelength calibration.

\subsubsection{\mob}

Observations were obtained staring on science fiber 2 with 180 second exposures. The observations ended just before the start of secondary eclipse, and we use the entire observation sequence for calculating the log-likelihood. The nominal planet velocity changes from 154 \kms\ to 59 \kms\ in the stellar reference frame over the course of the observations, with a maximum single-frame shift of 1.5 \kms. 

Conditions were good throughout the observations, but the large airmass at the beginning of the observation sequence led to relatively poor throughput early on. The top-of-atmosphere throughput was $\sim2$\% at airmass 2.2, but improved to $\sim4$\% by the end of the observations when \mob\ reached its highest elevation. HIP 81497 was observed at the start of the night for wavelength calibration. 

\subsubsection{\ktb}

These observations were previously described in \citet{finnerty2025k20}. Observations were obtained staring on science fiber 4 with 300 second exposures. Observations began after the end of secondary eclipse, and covered nominal planet velocities from $-43$ \kms\ to $-119$ \kms\ relative to the host star. The maximum single-frame velocity shift was $\sim1.2$ \kms. 

Conditions were good throughout the observations, with throughputs ranging from 3-4\%. HIP 95771 was observed after \ktb\ for wavelength calibration. 

\subsubsection{\tob}

These observations commenced immediately after the end of the \ktb\ observations, again staring on science fiber 4. The exposure time was reduced to 180 seconds in order to minimize the intra-frame velocity smearing of the planet lines. Observations began as secondary eclipse was ending, and we therefore omit the first three frames from the log-likelihood calculation. The planetary radial velocity changed from $-32$ \kms\ to $-107$ \kms\ relative to the host star reference frame over the observations, with a maximum frame-to-frame change of 1.6 \kms. 

Conditions were similar observations of \ktb\ earlier that night, but the top-of-atmosphere throughput on \tob\ was somewhat lower, 2.7-3.5\%. \tob\ was observed at a significantly higher airmass than \ktb\ and the primary star is $\sim1$ mag fainter, making the AO correction for \tob\ significantly more challenging than it was for \ktb. We use the same observations of HIP 95771 as were used for \ktb\ for wavelength calibration.

\subsection{Data Reduction}

We use the modified KPIC DRP\footnote{\href{https://github.com/kpicteam/kpic_pipeline/}{https://github.com/kpicteam/kpic\_pipeline/}} described in \citet{finnerty2025k20, finnerty2025hd143} which accounts for the variable line spread function (LSF) of NIRSPEC. Background subtraction was performed using daytime calibration frames obtained with the same exposure time as the science observations. For the spectral extraction, we fit a Gaussian-Hermite model to the trace profile, which is used for optimum extraction, as described in \citet{finnerty2025k20}. The resulting trace profile is then broadened by a factor of 1.14 based on measurements of OH lines described in \citet{Finnerty2022} and used as the instrumental LSF in the forward model. As discussed in \citet{finnerty2025k20, finnerty2025hd143}, this scaling factor is somewhat uncertain, leading to a systematic uncertainty in retrieved $v\sin i$ values. For targets which were observed with multiple fibers, a separate spectral time series was created for each fiber in order to eliminate frame-to-frame variations in the LSF due to variable relative flux between fibers. The resulting log-likelihoods from each separate time series were summed during sampling. 

Wavelength calibration was performed using the KPIC DRP, which compares observations of a late-type giant star to a stellar $\times$ telluric model. Using late-type giants significantly increases the number of spectral features available for fitting, particularly compared to the early-type UHJ host stars, which have only a few, very broad lines in the $K$-band. In previous analyses, the bluest three orders (37--39, 1.94--2.09$\mu$m) were omitted due to strong telluric CO$_2$ features, leaving six orders (31--36, 2.1-2-2.49 $\mu$m) for the cross-correlation and retrieval analysis. In \citet{finnerty2023} and \citet{finnerty2024}, we also omitted orders 35 and 36 (2.1-2.2 $\mu$m) due to poor wavelength calibration, as these orders have relatively few spectral features even when observing late-type stars. As additional observations were obtained, the wavelength solutions for these orders have significantly improved, allowing their inclusion in retrieval analysis \citep{finnerty2025hd143, finnerty2025k20, finnerty2025hd209}. For this work, we re-reduced all observations and obtained reliable wavelength solutions for these orders.  We verified the wavelength solutions by visual inspection of the wavelength calibration observations with the stellar$\times$telluric model, and by comparing the coefficients of a 4th-order Legendre polynomial fit to the wavelength solution for each order. We also include order 37--39 in the analysis, despite the telluric contamination. Inspection of the detrended time series showed that the combination of PCA and telluric masking is similarly effective in these orders as it is in the orders less impacted by tellurics. This region of the spectrum is potentially useful for UHJ retrievals due to the presence of strong OH and H$_2$O spectral features, motivating its inclusion. 

The data reduction process produces a flux array ($N_{frame}, N_{channel}$) for each order, as well as arrays of the mid-exposure times, barycentric velocities, wavelength solution for each order, and LSF for each order and channel. These data products are the inputs for our retrieval analysis, and are available online via \dataset[Zenodo]{10.5281/zenodo.20350230}. Reduction and retrieval codes will be made available upon reasonable request to the corresponding author.

\subsection{Atmospheric Retrieval}

We use the atmospheric retrieval framework described in \citet{finnerty2023, finnerty2024, finnerty2025hd143, finnerty2025k20}. Changes compared with \citet{finnerty2025k20} are minor, and primarily focused on improved masking of bad pixels or regions of high telluric absorption. The most significant change is a switch from the \citet{brogi2019} log-likelihood to the \citep{gibson2020} approach, and the associated error fitting, which is described in detail below. The full list of parameters and priors is included in Table \ref{tab:priors}. The same set of priors and masking parameters was used for all observations. 

\subsubsection{Data processing}

The time series for each order resulting from the data reduction procedure includes frame-to-frame flux variations and the instrumental blaze function, which must be removed before computing the log-likelihood. For each order, we first scale each spectrum to a consistent continuum level to remove the frame-to-frame flux variations. We then mask the 3\% of wavelength channels with the highest variance along the time axis, which masks channels strongly impacted by bad pixels or time-varying tellurics. The entire time series is then divided by the median spectrum taken along the time axis. In the limit of large planet velocity shift, the median spectrum is simply the stellar spectrum multiplied by the average telluric spectrum, $F_s \times \bar{T}$, and this division converts the spectral time series to $F_p/F_s \times T(t)$, where $T(t)$ is the time-dependent change in the telluric absorption spectrum. For smaller planet velocity shifts, the median spectrum may contain some contribution from the planet spectrum, leading to a self-division effect which can reduce the strength of planet features. We address this by similarly dividing the forward model $F_p+F_s$ time series by its own median to replicate any self-division. 

After the median division, we mask the first and last 50 wavelength channels, which have very low transmission due to the instrumental blaze function. We then mask any points deviating by more than 6 times the Median Absolute Deviation (MAD) from unity, in order to mask any remaining bad pixels. We finally mask the 3\% of the remaining wavelength channels with the highest temporal variance. 

At this stage, temporal variations in the telluric spectrum remain, and are strong enough to hide any planetary emission signals. Consistent with standard practice in HRCCS analysis \citep[e.g.][]{line2021, finnerty2023, finnerty2024, finnerty2025hd143, finnerty2025k20}, we remove this signal through Principal Component Analysis (PCA). For each order, the time series is mean-subtracted and masked data is interpolated over. We then perform the PCA and re-apply the original masking. Finally, we mask any points deviating by more than 4$\times$ the MAD. Following \citet{line2021}, we save the dropped principal components in order to add them to the forward model and project out the same number of components, replicating any distortion of the underlying planet signal resulting from PCA in our forward model. The re-injection is done as an addition, rather than a multiplication, in order to be consistent with the previous subtraction (as opposed to division) of the PCA components. We also mask any wavelength channels in the forward model with telluric transmission $<70\%$, in order to ensure that the higher-noise regions of the spectrum are not included in the likelihood calculation. We then compare the forward model and the data, computing the log-likelihood as described by \citet{gibson2020}. 

We performed retrievals omitting 4, 6, or 8 principal components for all targets. While the optimum number of components is likely different from target-to-target and order-to-order, previous work has found that overly-optimizing the number of omitted PCs can lead to false detections \citep{cheverall2023}. We therefore choose a single number of components to omit for all orders. The posteriors are generally consistent as the number of omitted components changes, but in some cases (particularly the last order for \wob), the four component case left obvious residuals upon visual inspection of the post-PCA time series. Thus, we present the six component retrievals as our fiducial results.  

\subsubsection{Error estimation}

Following \citet{gibson2020}, we take an empirical approach to estimating the uncertainties in the data. This avoids issues with error propagation during detrending by fitting a noise model to the detrended time series for each order. The noise is assumed to be composed of a photon term and a constant read/background term:

\begin{equation}
\sigma = \frac{\sqrt{aF + b}}{c\bar{F}}
\end{equation}

Where $a$ and $b$ are free parameters to be fit, $F$ is the observed flux prior to any detrending, $\bar{F}$ is the time-series median frame, and $c$ is the per-frame scaling constant applied when matching the continuum level of each frame. The $c\bar{F}$ term is required to match the treatment of the data prior to PCA.

After PCA, we are left with a median-divided residual array $R$. We use this to fit the $a$ and $b$ terms by maximizing the log-likelihood:

\begin{equation}
   \log L =  -0.5\sum \frac{R_i^2}{\sigma_i^2} - \sum\log\sigma_i^2
\end{equation}

As described in \citet{gibson2020}. As a final step, we take the best-fit $\sigma_i$ and replace it with a PCA reconstruction using the same number of principal components as omitted during the data detrending. This step serves to reduce the noise level in the uncertainty estimates themselves \citep{gibson2020}. Visual inspection of the final $\sigma_i$ time series is consistent with expectations. Values are higher in regions of low telluric or blaze transmission, and lower in the portions of the spectrum with greater fluxes. 

\subsubsection{Stellar models}

For each target, we use a PHOENIX \citep{phoenix} model for the host star, chosen to have the closest \teff\, $\log g$, and metallicity in the model grid to the assumed stellar parameters. Each model was rotationally broadened using the fast wavelength-dependent technique described in \citet{carvalho2023} to the respective $v\sin i$ values listed in Table \ref{tab:props}. For WASP-33~A, we used $T_{\rm eff} = 7400$ K, $\log g = 4.5$, $\rm Z = 0$, and $v\sin i = 86.9$ \kms. For KELT-9~A, we used $T_{\rm eff} = 9600$ K, $\log g = 4.0$, $\rm Z = 0$, and $v\sin i = 111$ \kms. For WASP-189~A, we used $T_{\rm eff} = 8000$ K, $\log g = 4.0$, $\rm Z = +0.5$, and $v\sin i = 93.1$ \kms. For MASCARA-1~A, we used $T_{\rm eff} = 7600$ K, $\log g = 4.0$, $\rm Z = 0$, and $v\sin i = 109$ \kms. For KELT-20~A, we used $T_{\rm eff} = 9000$ K, $\log g = 4.5$, $\rm Z = 0$, and $v\sin i = 114$ \kms. Finally, for TOI-1518~A, we used $T_{\rm eff} = 7400$ K, $\log g = 4.0$, $\rm Z = 0$, and $v\sin i = 85.1$ \kms. 

The nearest-neighbor approach to stellar model selection leads to discrepancies with some of the stellar properties listed in Table \ref{tab:props}. However, we note that many of the stellar properties listed in Table \ref{tab:props} differ significantly between different sources. For the effective temperature of TOI-1518~A, for example, \citet{exofop3} gives $9900^{+500}_{-2300}$ K, \citep{cabot2021} and \citet{simonnin2024} both prefer values of $7300\pm100$ K. For WASP-33~A, reported radii include $1.444\pm0.034\rm\ R_\odot$ \citep{collier2010}, $1.55\pm0.05\rm\ R_\odot$ \citep{stassun2017}, and $1.60^{+0.06}_{-0.05}\rm\ R_\odot$ \citep{tic8} These discrepancies suggest the stellar properties of the rapidly rotating early-type host stars are poorly constrained by existing observations.

Fortunately, these discrepancies appear to have minimal impact on the analysis presented here. As a test case, we compared the log-likelihood as a function of \vsys\ for the TOI-1518 observations with stellar models having $\rm T_{eff} = 7400$ and $\rm T_{eff} = 9600$. The change in stellar model led to a global shift of the log-likelihood values, but no significant change to the relative log-likelihoods for a given stellar model or to the overall detection significance. The rapidly rotating early-type stellar spectra are effectively divided out during the model processing, and therefore the only effect of a significant mismatch in stellar properties is a global shift in $F_p/F_s$. Our retrieval results demonstrate that the data presented here have very poor sensitivity to such shifts, and are therefore unlikely to be significantly impacted but such errors in the stellar models. 

Despite the rapid rotation rates of the host stars, we do not include limb or gravity darkening, which may lead to minor offsets in the absolute $F_p/F_s$ values. HRCCS techniques are only weakly sensitive to the absolute $F_p/F_s$, as discussed in Section \ref{sec:disc}, and we therefore do not expect this omission will significantly impact our interpretation of the retrievals. 

\subsubsection{Planet models and pressure-temperature profiles}

We compute 1D model atmospheres for the planets using \petit 3\ \citep{prt:2019, prt:2020, Nasedkin2024}. We use 80 log-uniform spaced pressure layers from $10^2 - 10^{-6}$ bar, following \citep{finnerty2025k20}. We adopt the 4-parameter model for the $P-T$ profile from \citet{guillot2010}, fitting for the log infrared opacity ($\log \kappa$), log optical/infrared opacity ratio ($\log \gamma$), intrinsic temperature, and equilibrium temperature. We also performed test retrievals using a more flexible parameterization of the $P-T$ profile, treating the temperatures at six evenly spaced pressures as free parameters with a prior on the second derivative of the profile, similar to the approach taken in \citet{pelletier2021}. We found that the $P-T$ profiles were very poorly constrained using this approach, but that the molecular abundances were consistent with the abundances obtained from retrievals using the \citet{guillot2010} profile. Tests of the free $P-T$ parameterization over a wider wavelength range (e.g. $H+K$ or $K+L$ data sets) were able to meaningfully constrain the temperature profile, suggesting that the apparent lack of constraints in the $K$ band data is a result of the limited wavelength coverage and resulting poor sensitivity to continuum. Based on these findings, we present the results of the retrievals using the \citet{guillot2010} parameterization. 

Despite the uncertain masses, and therefore surface gravities, of several targets, we do not include $\log g$ as a free parameter. As described in \citet{finnerty2026hd806}, the implementation of the \citet{guillot2010} $P-T$ profile in \petit\ takes a pressure as an input, which is then converted to an optical depth as $\tau = \kappa P/g$. This creates a direct correlation between $\log \kappa$ and $\log g$, which prevents constraints on $\log g$ if $\log \kappa$ is included in the retrieval. 

Following \citet{finnerty2025k20}, we use a truncated log-normal prior for the scale factor applied to the planet model with a standard deviation of 0.1. Values above $\pm0.75$ are rejected for numerical stability. As discussed in \citet{finnerty2025k20}, the use of this prior somewhat, but not completely, breaks the temperature/scale factor degeneracy which impacts HRCCS observations with limited wavelength coverage.

\subsubsection{Parameterized vertical mixing profiles}

Dissociation of H$_2$O is a major process in the upper atmospheres of UHJs, which significantly impacts interpretation of the $K$ band spectra \citep{brogi2023, ramkumar2023, finnerty2025k20}. If not incorporated into the forward model, either through free retrieval of the H$_2$O and OH mixing profiles or by assuming equilibrium chemistry, dissociation will bias retrieval results towards high C/O ratios, as the only remaining way to match the relative strengths of CO and H$_2$O features is to reduce the H$_2$O abundance and/or increase the CO abundance. In the case of \knb, this may be further complicated by the dissociation of CO at even lower pressures, which we ignore in favor of constant-with-altitude abundances for both CO isotopologues and Fe.

We initially addressed this effect by incorporating dissociation and formation pressures into our free retrieval framework, as previously described in \citet{finnerty2025k20}. For H$_2$O, OH, and CO, we fit a break pressure, and set the corresponding abundance to decay as $(P/P_{\rm dissoc})^4$ at lower pressures, with the power law index chosen to roughly match equilibrium chemistry predictions from \texttt{easyCHEM} \citep{molliere2017, lei2024}. For OH, we also fit a formation pressure, with the abundance decreasing as the pressure increases beyond the formation pressure. See \citet{finnerty2025k20} for a further discussion of this setup. While this approach minimizes assumptions about the vertical mixing profiles, and was used successfully to constrain H$_2$O dissociation in \ktb\ \citep{finnerty2025k20}, our attempts to apply this technique to the other five targets considered in this work resulted in unconstrained dissociation pressures in all cases except for H$_2$O in \wtb. This suggests that this approach will work in the high-SNR limit, but that the data quality for most of our targets does not support free retrieval of even the H$_2$O dissociation pressure.

Instead, we adopted the parameterized H$_2$O mixing profile described in \citet{parmentier2018}. The water volume mixing ratio is given by:

\begin{equation}
    {A_{\rm H_2 O}} = \left(\frac{1}{A_{\rm H_2 O,0}^{0.5} } + \frac{1}{A_{\rm H_2 O,d}^{0.5} } \right)^{-2}
\end{equation}

Where $A_{\rm H_2 O,0}$ is the deep-atmosphere H$_2$O abundance, the logarithm of which is fit as a free parameter, and $A_{\rm H_2 O,d}$ is given by:

\begin{equation}
    A_{\rm H_2 O,d} = P^2 10^{4.83\times10^{4}/T -15.9}
\end{equation}

Where $P$ is the pressure of each atmospheric layer in log bars and $T$ is the corresponding temperature of each pressure level. 

\citet{parmentier2018} does not provide a parameterization for the OH mixing profile. OH is formed in UHJ atmospheres primarily from H$_2$O dissociation, and so we set the OH abundance to decay as a power law with an index of two at pressures greater than the pressure where $A_{\rm H_2 O,0}= A_{\rm H_2 O,d}$, which is the point at which the H$_2$O dissociation becomes significant. At lower pressures, OH will itself dissociate, and so we set the OH VMR to decay as a power law, again with an index of 2, at pressures below $1/3$ of the $A_{\rm H_2 O,0}= A_{\rm H_2 O,d}$ pressure, motivated by the vertical mixing profiles from \texttt{easyCHEM} \citep{lei2024} for chemical equilibrium. The maximum OH abundance is a free parameter in the retrieval. This decouples the OH and H$_2$O features in the retrieval, so that the retrieval can return a spectrum with features from only one species or the other. In the event where the OH and H$_2$O abundances are linked but only one species is preferred by the data, this could potentially bias the abundance of the preferred species to lower values, e.g. if H$_2$O is preferred but OH is not the retrieval may prefer a somewhat lower H$_2$O abundance in order to reduce the model mismatch arising from the associated OH features. In practice, the presented observations have only minimal sensitivity to OH, and any such abundance biases would therefore be minimal. However, by taking this approach we can assess whether the presence of OH is independently supported by the observations via the retrieved posterior.

For H$_2$, we apply a similar parameterization to H$_2$O using the coefficients from \citet{parmentier2018}, and set the H abundance to conserve the total quantity of H+$\rm H_2$. He is then used to fill the remainder of the atmosphere. Finally, we apply a penalty to any model in which the mean molecular weight exceeds 2.8 AMU in any pressure layer based on a Gaussian prior as follows:

\begin{equation}
p = 
\left\{
\begin{array}{cc}
     &  0,\ {\rm MMW_{max}} < 2.8\\
     & \frac{(\rm MMW_{max} - 2.8)^2}{2\times0.3^2},\ {\rm MMW_{max}}>2.8
\end{array}
\right\}
\end{equation}

This was constructed to apply no penalty to atmospheres with maximum mean molecular weights $<2.8$ AMU, corresponding to $\sim10\times$ solar metallicity, while higher metallicities fall off as a Gaussian, with atmospheres at $\sim100\times$ solar metallicity penalized at approximately $3\sigma$. This quantity is then subtracted from the computed log-likelihood. We find that without this penalty, the retrieved mean molecular weights can reach as high as 10 AMU as the retrieval attempts to fit the strengths of spectral lines by increasing abundances instead of via changes to the $P-T$ profile. Including this penalty results in more plausible metallicities and avoids violations of the H$_2$-dominated atmosphere assumption, while still permitting very metal-enriched atmospheres if required to fit the observations.  

\subsubsection{Chemical equilibrium profiles}

We also ran a set of retrievals with vertical mixing profiles computed assuming chemical equilibrium. Rather than fitting individual molecular abundances, we fit for $\rm [C/H]$, $\rm [C/H]$, and $\rm [M/H]$, where the latter parameter scales the abundances of all atoms except for hydrogen, helium, carbon, and oxygen. These atomic abundances were then put in to \texttt{easyCHEM} \citep{lei2024} for each temperature and pressure in the $P-T$ profile to obtain the vertical mixing profiles for H$_2$O, CO, OH, and Fe under local chemical equilibrium. $^{13}\rm CO$ opacity was omitted for the equilibrium retrievals. The penalty for high mean molecular weights described in the previous subsection was similarly applied in the chemical equilibrium case.

\subsubsection{Kinematic parameters}

We assume the UHJ orbits are circular, which is consistent with the prior ephemerides for these targets. The planet velocity at each orbital phase is then given by: 

\begin{equation}
    v_{pl} = K_p\sin\phi + v_{rad} - v_{bary} + \Delta v_{sys}
\end{equation}

\noindent
where $\phi$ is the orbital phase, computed from the assumed ephemeris, $v_{rad}$ is the systemic radial velocity listed in Table \ref{tab:props}, and the planetary radial velocity semi-amplitude, \kp, and systemic radial velocity offset, \vsys, are free parameters in the retrieval. We also fit for the width of the rotational broadening kernel applied to the planet. 

We additionally assume that the stellar reflex motion is negligible, constant stellar barycentric velocity, and no stellar offset from the assumed radial velocity, so that the stellar velocity is simply given by $v_{st} = v_{rad} - \bar{v}_{bary}$. These assumptions are reasonable for the rapidly rotating host stars of the observed UHJs, and reduce the computational complexity of the retrievals by minimizing the number of Doppler shifts applied to the stellar template.

\subsubsection{Opacity sources}

The opacity tables used for OH, $^{12}$CO, and $^{13}$CO were previously described in \citet{finnerty2023}. For H$_2$O we used the default tables included in \petit\ based on the \citet{polyansky2018} POKAZATEL linelist. For both CO isotopologues we used the HITEMP 2019 lists \citep{hitemp2020} and the \citet{li2015} partition function. For OH we used the partition function from \citet{YOUSEFI2018} and HITEMP 2020 linelist \citep{hitemp2020}. Opacities for CO and OH were generated via ExoCross \citep{exocross2018} following instructions in the \petit~documentation. For Fe and H$_2$, we use the default \petit\ opacities. The Fe opacity is based on the Kurucz atomic line lists, and H$_2$ opacity is from \citet{rothman2013}. We also include CIA opacities for $\rm H_2 - H_2$ \citep{borysow2001, borysow2002} and $\rm H_2-He$ \citep{borysow1988, borysow1989a, borysow1989b}, and used the scattering mode for \petit. 

While UHJs are in the temperature regime where H$^-$ may be a significant source of continuum opacity, and H$^-$ features have been detected in the atmosphere of \knb\ \citep{jacobs2022}, we do not include H$^-$ opacity in our analysis, following \citet{finnerty2023, finnerty2025k20}. \citet{finnerty2023} found that in the $K$ band, where the H$^-$ opacity is free-free, including H$^-$ results in a {\em uniform} reduction in the strength of planet lines relative to the continuum. HRCCS analysis has poor sensitivity to the absolute flux level of the planet, and as a result the H$^-$ abundance parameters are unconstrained when included in the retrieval. \citet{jacobs2022} found that H$^-$ opacity results in a turnoff at 1.4 $\mu$m in the spectrum of \knb, corresponding to the transition from free-free opacity at longer wavelengths to bound-free opacity at shorter wavelengths. This result suggests that combining $K$ band observations with additional data at shorter wavelengths would result in sensitivity to H$^-$ abundances via the change in continuum opacity at the bound-free/free-free transition. 

\subsubsection{Sampling and log-likelihood}

We use \texttt{MultiNEST} \citep{feroz2008, feroz2009, buchner2014, feroz2019} to estimate posteriors via nested sampling \citep{skilling2004}. We employ 1024 live points and a convergence criteria $\Delta \log z < 0.01$, similar to our previous HRCCS retrievals \citep[e.g.][]{finnerty2023, finnerty2026hd806}. The log-likelihood is computed using the \citet{gibson2020} mapping for each fiber separately and summed, avoiding issues associated with combining data with slightly different line-spread functions or interpolating wavelength solutions. 

We considered retrievals with three different log-likelihood functions. From \citet{brogi2019}:
\begin{equation}
\begin{split}
    \log L = -\frac{N}{2} \log \left(\frac{\sum f_i^2}{N} + \frac{\sum g_i^2}{N} - 2\frac{\sum f_ig_i}{N} \right) = \\
    -\frac{N}{2} \log \left(\frac{1}{N}\sum(f_i -g_i)^2 \right)
\end{split}
\end{equation}

From \citet{gibson2020}:
\begin{equation}
    \log L = -\frac{N}{2}\log\left(\frac{1}{N} \sum \frac{(f_i-g_i)^2}{\sigma_i^2}\right)
\end{equation}

And finally the traditional $\chi^2$:
\begin{equation}
    \log L = -\frac{N}{2} \sum \frac{(f_i-g_i)^2}{\sigma_i^2})
\end{equation}

Where in all three equations, $f$ is the observed spectrum, $g$ is the forward model, $N$ is the number of spectral points, and $\sigma$ is the uncertainty fit to the detrended time series as described above. The log-likelihood is calculated for each order and frame and then summed across all orders, frames, and fibers, the latter only if observations were taken in multiple science fibers.

For the best detected targets, \ktb\ and \wtb, we find that retrievals with the \citet{brogi2019}, \citet{gibson2020}, or $\chi^2$ log-likelihoods all return consistent posteriors, which is the expected result in the low-noise limit. For the remaining targets, $\chi^2$ retrievals have a tendency to prefer solutions significantly offset from the expected reference frame of the planet, while retrievals using the \citet{brogi2019} tend to return posteriors which are broader than, but consistent with, the posteriors obtained using the \citet{gibson2020} log-likelihood. Retrievals with the \citet{gibson2020} log-likelihood also tended to converge faster. 

Comparing the retrievals using the \citet{brogi2019} and \citet{gibson2020} likelihoods suggests that accounting for time- and wavelength-dependent errors in these data improves retrievals compared to the constant-error assumption made by \citet{brogi2019}, even with detrending procedures that aim to mask high-noise portions of the observed spectra. At the same time, the discrepancies between the \citet{gibson2020} retrievals and retrievals using $\chi^2$ suggests that there are significant errors in the estimated flux uncertainties which can bias results in low or moderate signal-to-noise cases. The \citet{gibson2020} log-likelihood depends on the logarithm of the uncertainties, which would significantly lessen the impact of inaccurate uncertainties on the final log-likelihood compared with $\chi^2$, which depends directly on the uncertainties. 

We emphasize that the above discussion is largely supposition to explain the results seen in these particular, rather noisy, data sets, and should not be taken as a general statement regarding the effects of different log-likelihood functions on HRCCS analyses. Additional work is necessary to understand whether and how errors in estimated uncertainties are impacting retrievals, how to improve the robustness of uncertainty estimates in HRCCS data sets, and how the quality of the data interacts with the error estimation and choice of log-likelihood function. This would require a significant number of dedicated simulations, which are beyond the scope of this paper. 

Based on these results, we choose to present results obtained using the \citet{gibson2020} log-likelihood function, which has been frequently used in the existing literature.  

\section{Results}\label{sec:res}

\begin{figure*}
    \centering
    \includegraphics[width=0.3\linewidth]{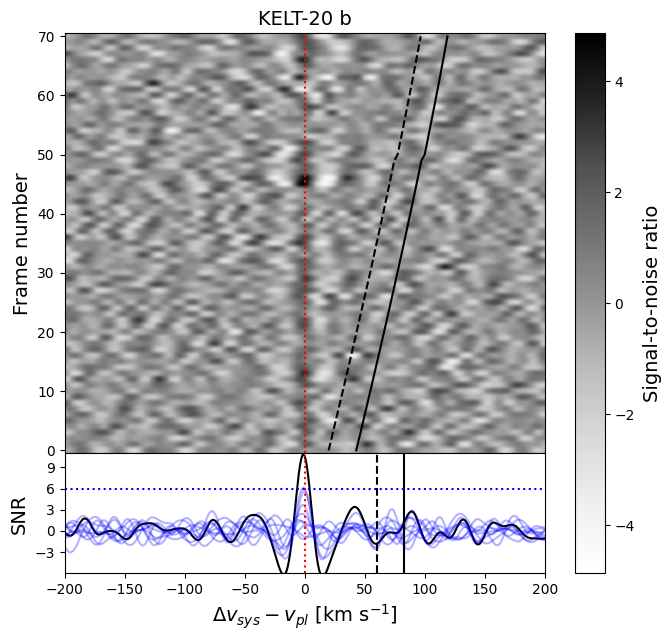}
    \includegraphics[width=0.3\linewidth]{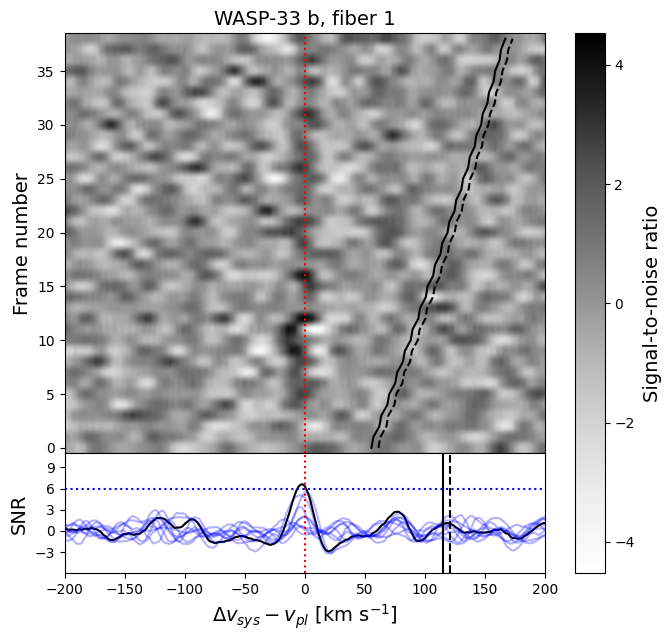}
    \includegraphics[width=0.3\linewidth]{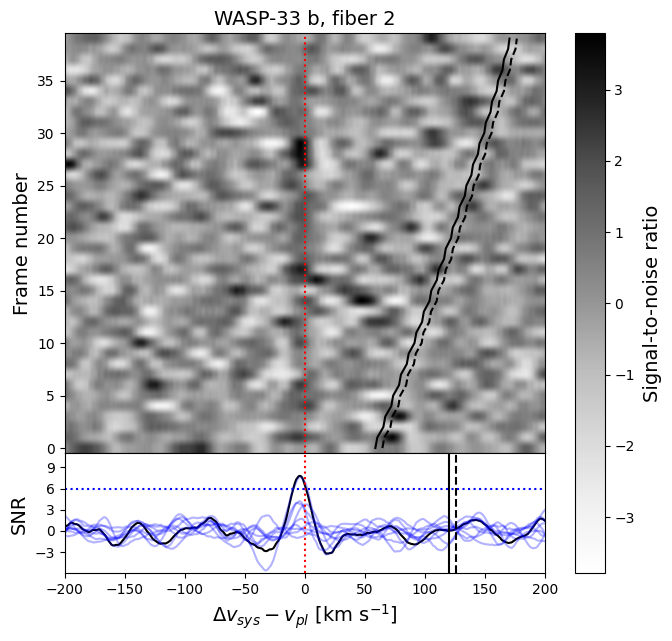}
    \includegraphics[width=0.3\linewidth]{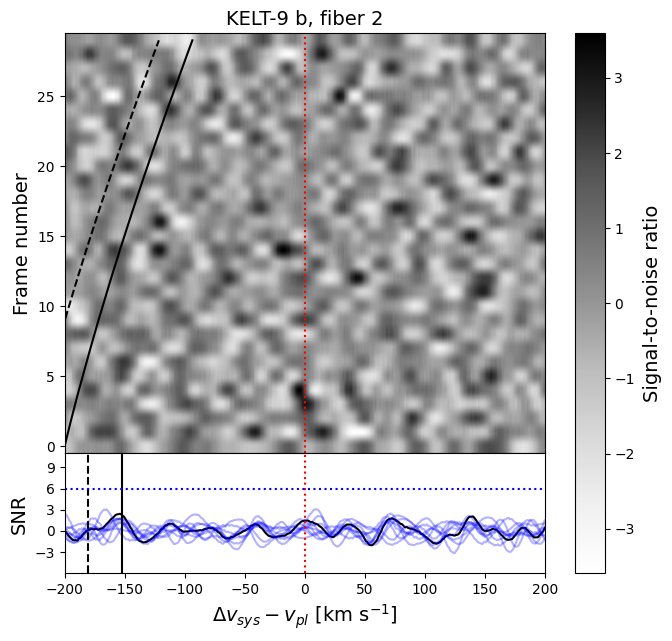}
    \includegraphics[width=0.3\linewidth]{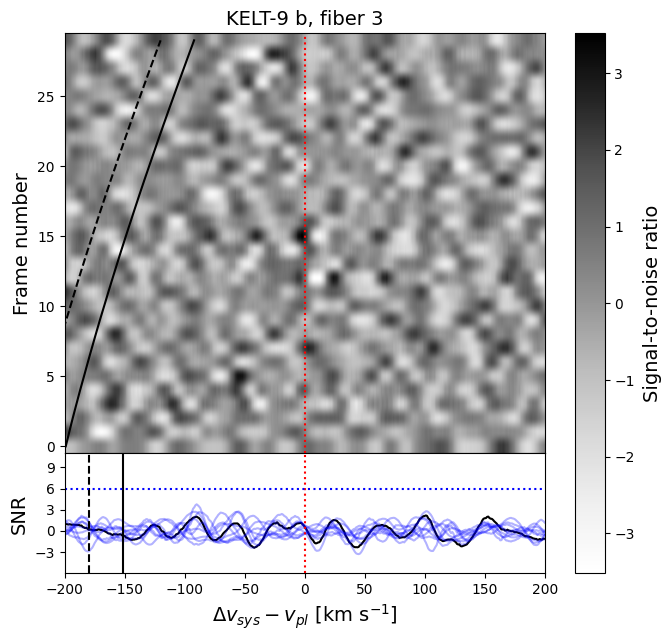}
    \includegraphics[width=0.3\linewidth]{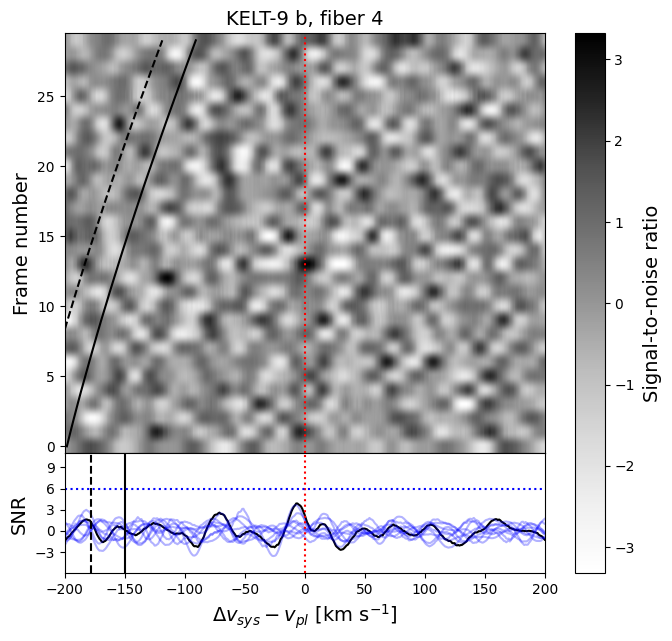}
    \includegraphics[width=0.3\linewidth]{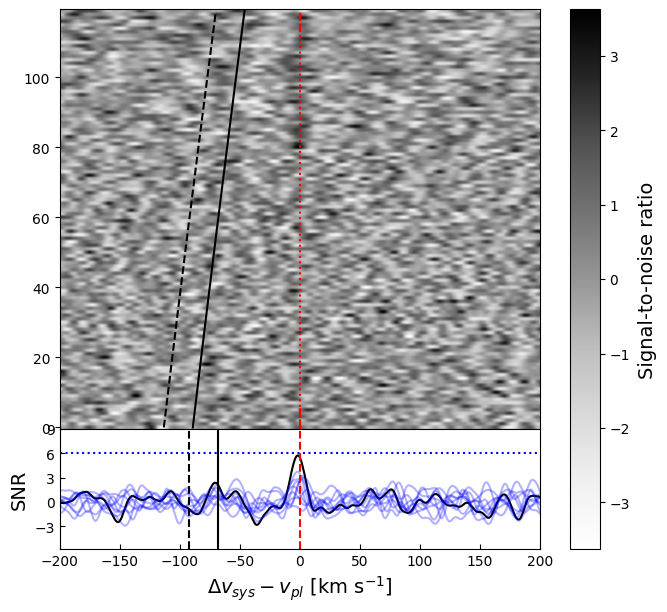}
    \includegraphics[width=0.3\linewidth]{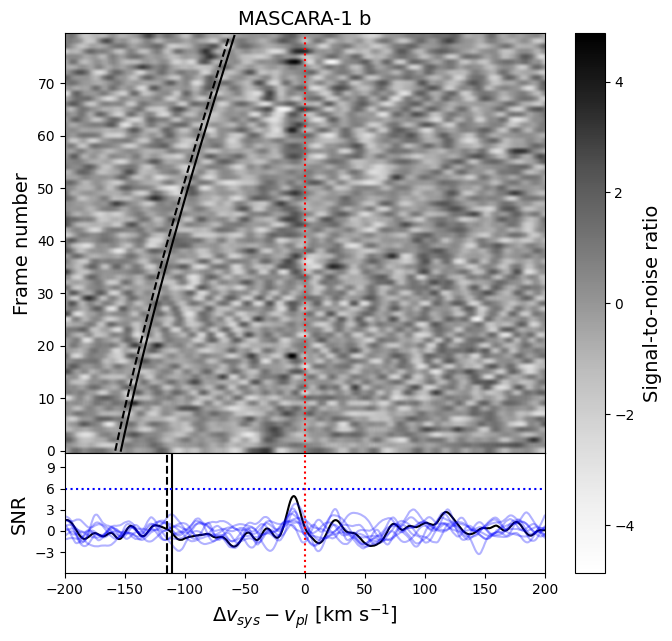}
    \includegraphics[width=0.3\linewidth]{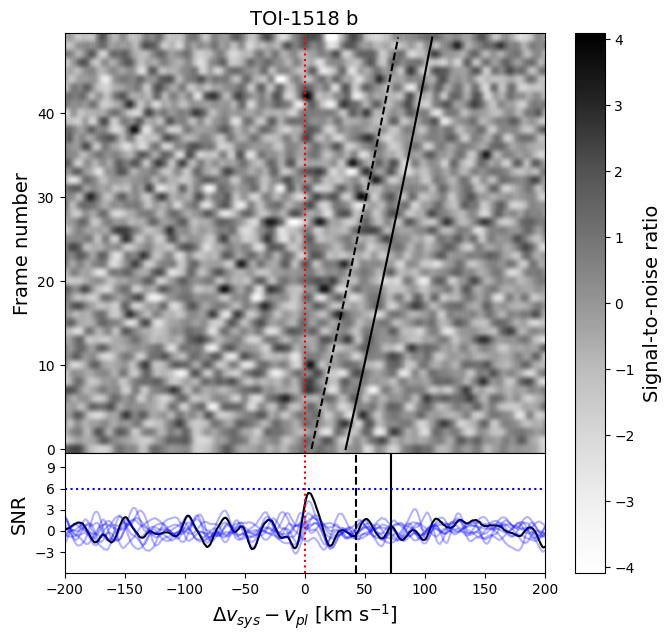}
    \caption{Time series log-likelihoods in the nominal planet rest frame computed for each frame (top) and summed in the planet reference frame (bottom) for each target and time series. The nominal planet reference frame is indicated in dotted red, the stellar reference frame in solid black, and the telluric reference frame in dashed black. In the lower panel, individual orders are shown in blue, while the sum of all orders is shown in black. For the top panel, the log-likelihoods for each frame are median-subtracted and divided by the standard deviation of the $\Delta v_{sys}-v_{pl} > 50$ \kms\ region to reduce frame-to-frame variations for clarity and convert to a noise map. The the bottom panel, we convert to signal-to-noise by dividing by the standard deviation of the $\Delta v_{sys}-v_{pl} > 50$ \kms\ region after summing over the entire time series. As a result of the different noise region definitions, the signal-to-noise ratios differ from the \kpvsys\ plots in Figure \ref{fig:kpvsysall}. \ktb\ and \wtb\ are obvious even in the 2D plots, while \wob, \mob, and \tob\ show $\rm SNR\sim5$ detections after summing over the time series. The three individual time series for \knb\ do not show a significant detection, but coadding all three time series results in a weak detection in the \kpvsys\ maps.}
    \label{fig:vtracks}
\end{figure*}

\begin{figure*}
    \centering
    \includegraphics[width=0.3\linewidth]{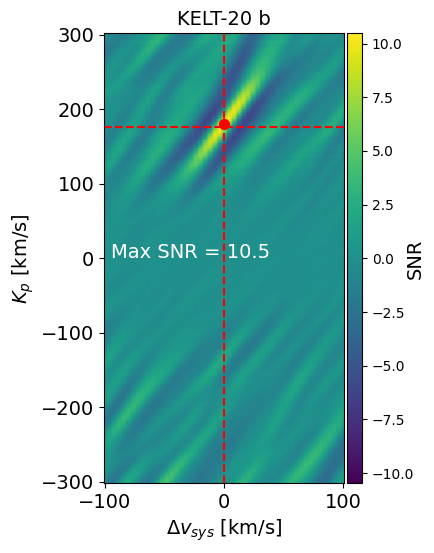}
    \includegraphics[width=0.3\linewidth]{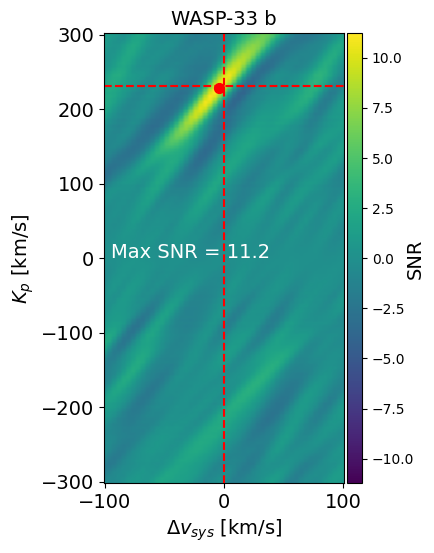}
    \includegraphics[width=0.3\linewidth]{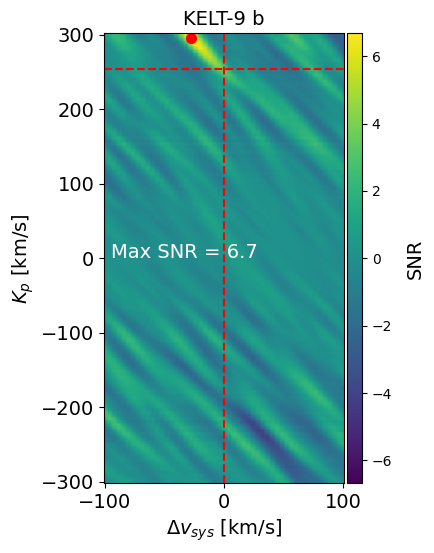}
    \includegraphics[width=0.3\linewidth]{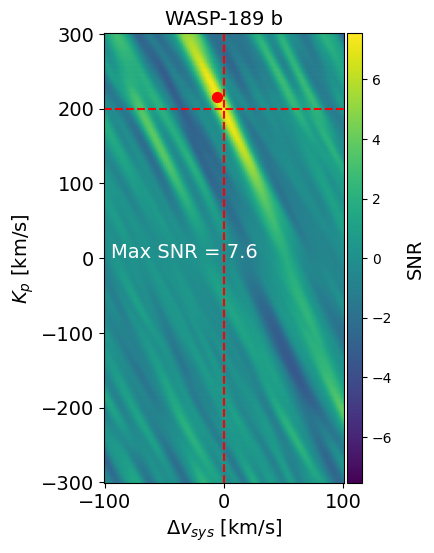}
    \includegraphics[width=0.3\linewidth]{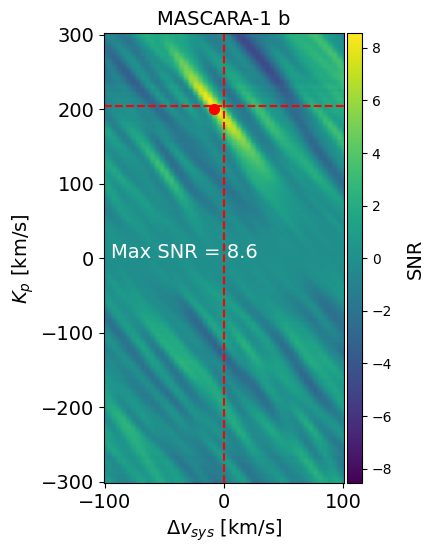}
    \includegraphics[width=0.3\linewidth]{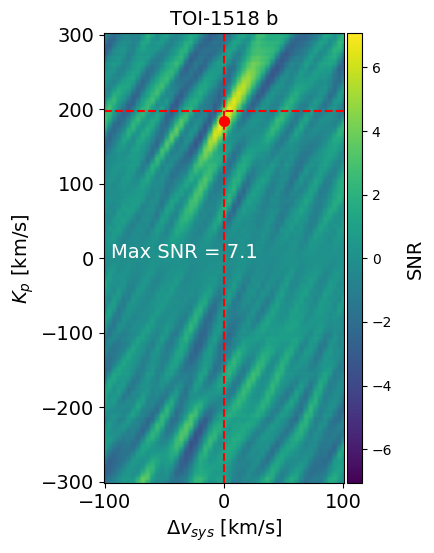}
    \caption{\kpvsys\ diagrams for the free retrieval maximum-likelihood model for each target, omitting six principal components. The maps are computed with the log-likelihood function, and converted to signal-to-noise by first median-subtracting each row of constant \kp\ and then dividing by the standard deviation of the \kp$<0$ region. The nominal planetary \kp\ and \vsys\ are indicated in dashed red, and the red dot indicates the location of maximum signal-to-noise in the \kpvsys\ map. All targets are detected at SNR$>$6.  \ktb\ and \wtb\ are very strongly detected at $\rm SNR>10$, while \wob, \mob, and \tob\ are detected at $\rm SNR>7$. \knb\ is the most weakly detected at $\rm SNR = 6.7$, despite a total observation signal-to-noise substantially greater than was obtained for \wtb. This discrepancy could be explained by dissociation of nearly all molecules in \knb, which is the hottest target by a margin of more than 1000 K. For all targets, the maximum in the \kpvsys\ diagram is consistent with the expected \kp\ and \vsys, and dominates over other features in the map, providing confidence that the features being fit in the retrieval are associated with the planet and not with uncorrected residuals. }
    \label{fig:kpvsysall}
\end{figure*}

\begin{figure*}
    \centering
    \includegraphics[width=0.3\linewidth]{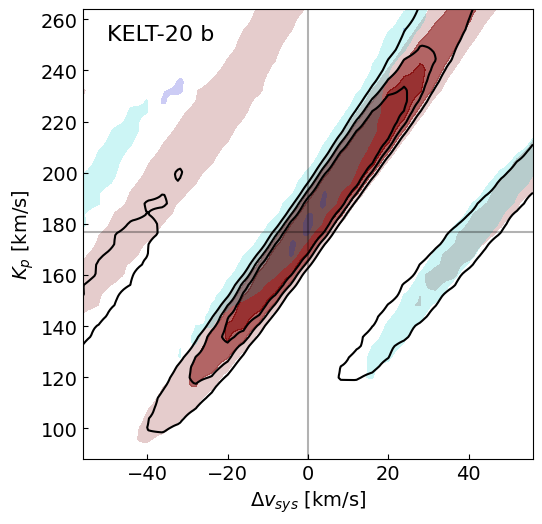}
    \includegraphics[width=0.3\linewidth]{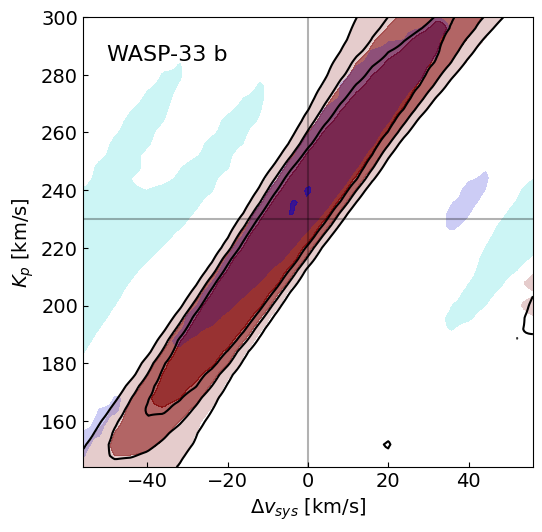}
    \includegraphics[width=0.3\linewidth]{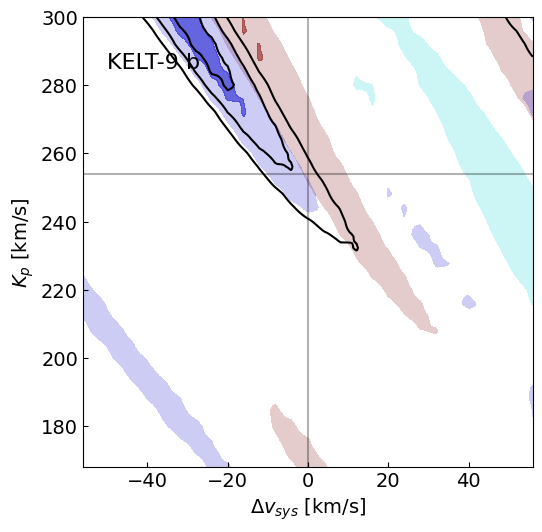}
    \includegraphics[width=0.3\linewidth]{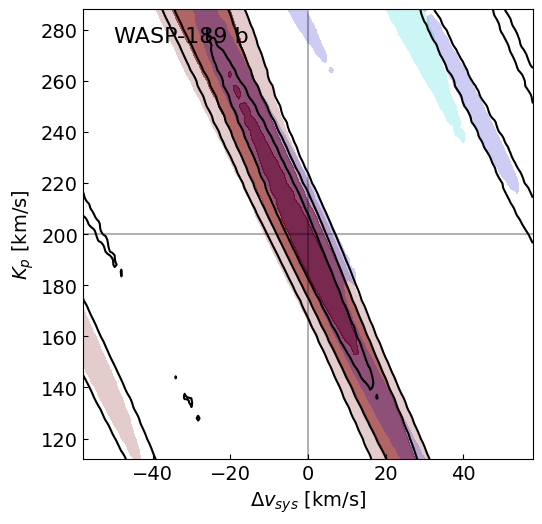}
    \includegraphics[width=0.3\linewidth]{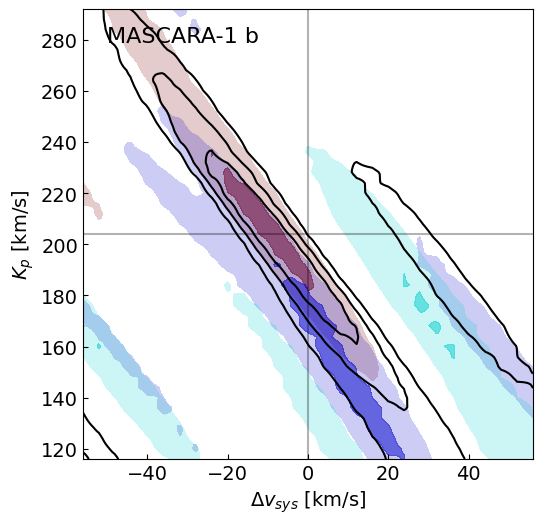}
    \includegraphics[width=0.3\linewidth]{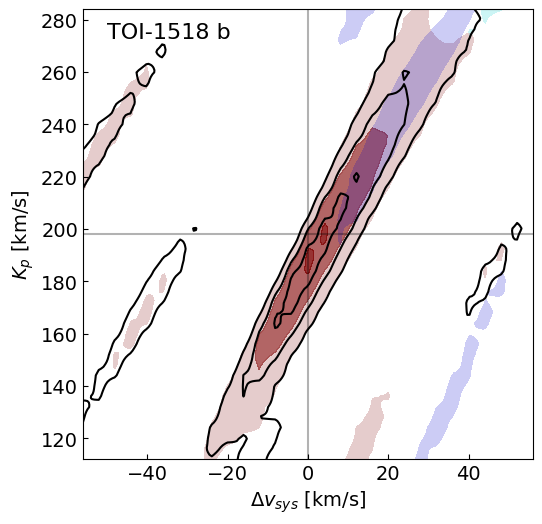}
    \caption{\kpvsys\ contours for the free retrieval maximum-likelihood model (black) as well as CO-only (filled dark red), H$_2$O-only (filled blue) and OH-only (filled cyan) models. Each model was generated using the retrieved maximum-likelihood parameters, setting the abundance of the included species to $10^{-3}$ and the abundances of all other species to $10^{-15}$. The contours are drawn at $\rm SNR = 2, 4,6$. The contours enable us to asses both which species are present and whether there are velocity offsets between different species.}
    \label{fig:kpvsys_contours}
\end{figure*}

Parameters, priors, and retrieved median/max-likelihood values for each target from the retrieval omitting six principal components are presented in Table \ref{tab:priors}. Figure \ref{fig:vtracks} shows the normalized 2D and 1D cross-correlation functions in the planet reference frames for the maximum-likelihood models from the free retrieval for each time series. The corresponding \kpvsys\ diagrams are shown in Figure \ref{fig:kpvsysall}. We assess the detection strengths of CO, H$_2$O, OH, and Fe through a series of single-molecule tests, and present the resulting \kpvsys\ contours in Figure \ref{fig:kpvsys_contours}. The following subsections present the results from each object individually.

\begin{figure*}
    \centering
    \includegraphics[width=0.3\linewidth]{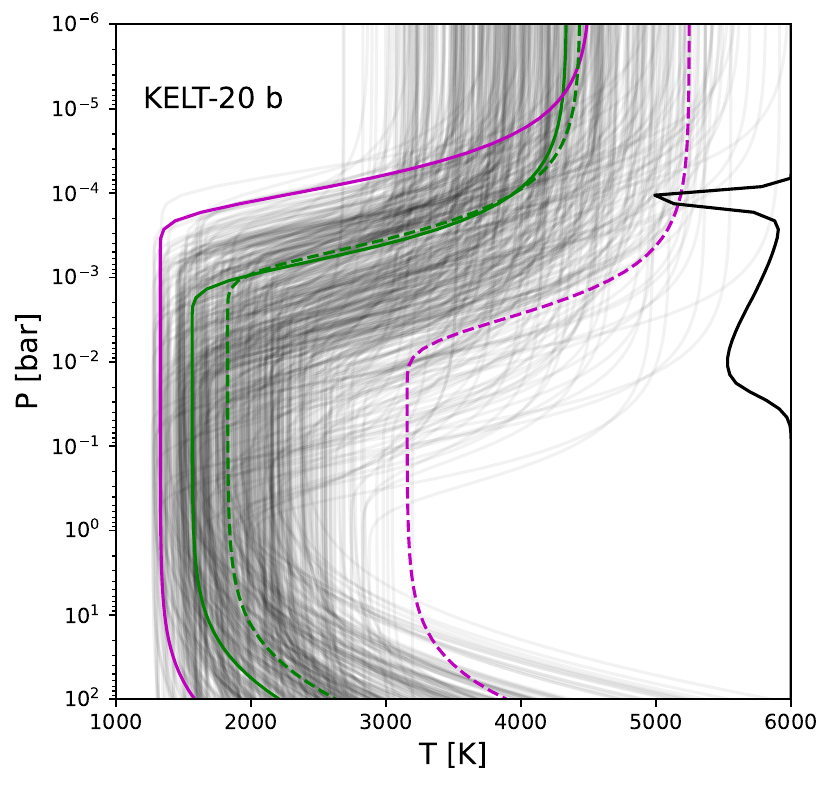}
    \includegraphics[width=0.3\linewidth]{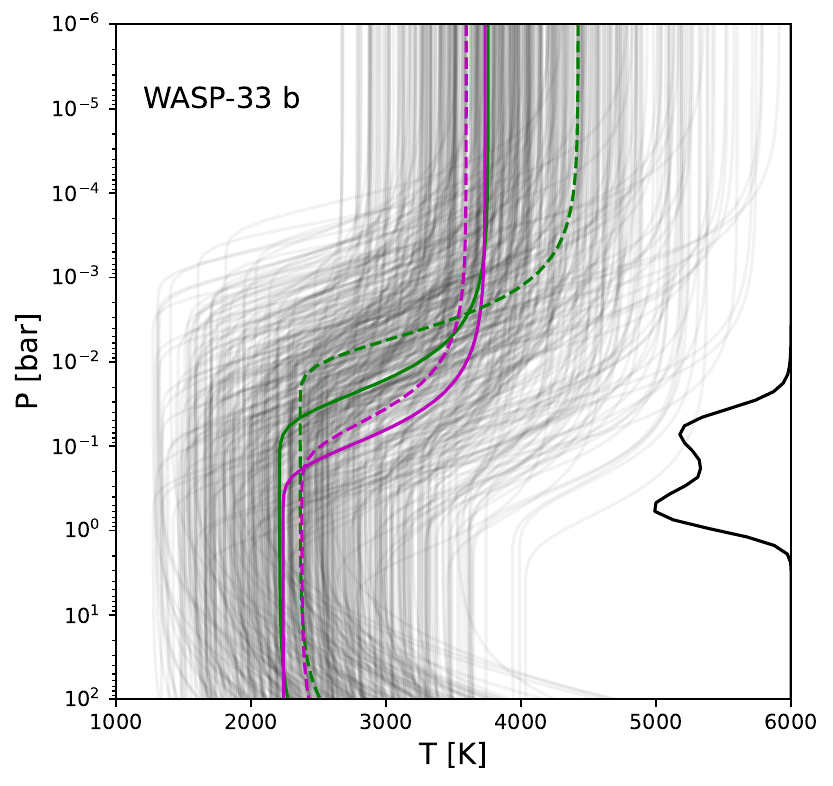}
    \includegraphics[width=0.3\linewidth]{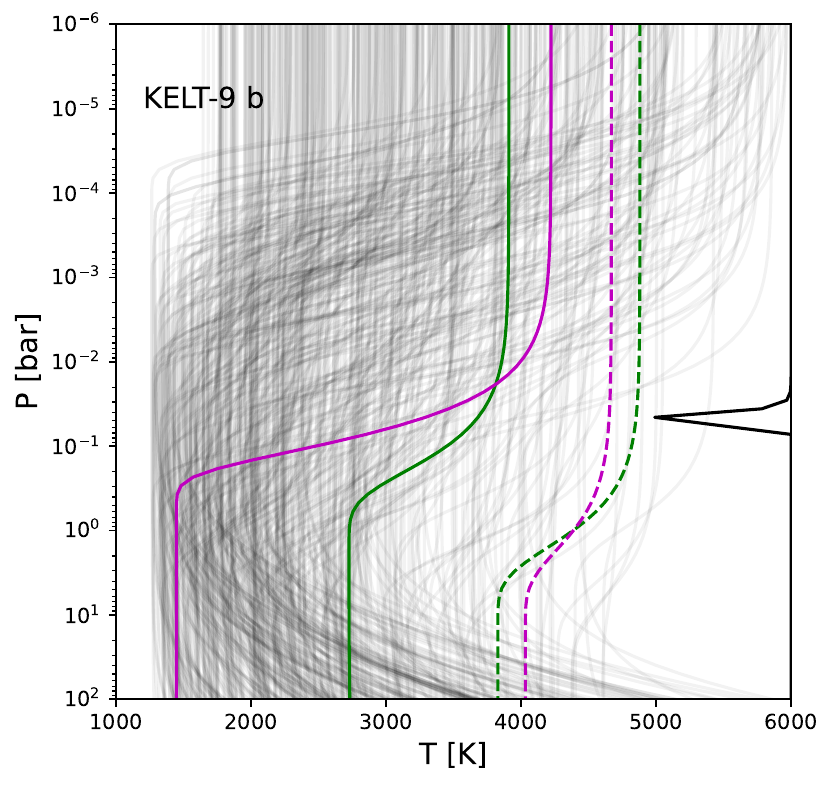}
    \includegraphics[width=0.3\linewidth]{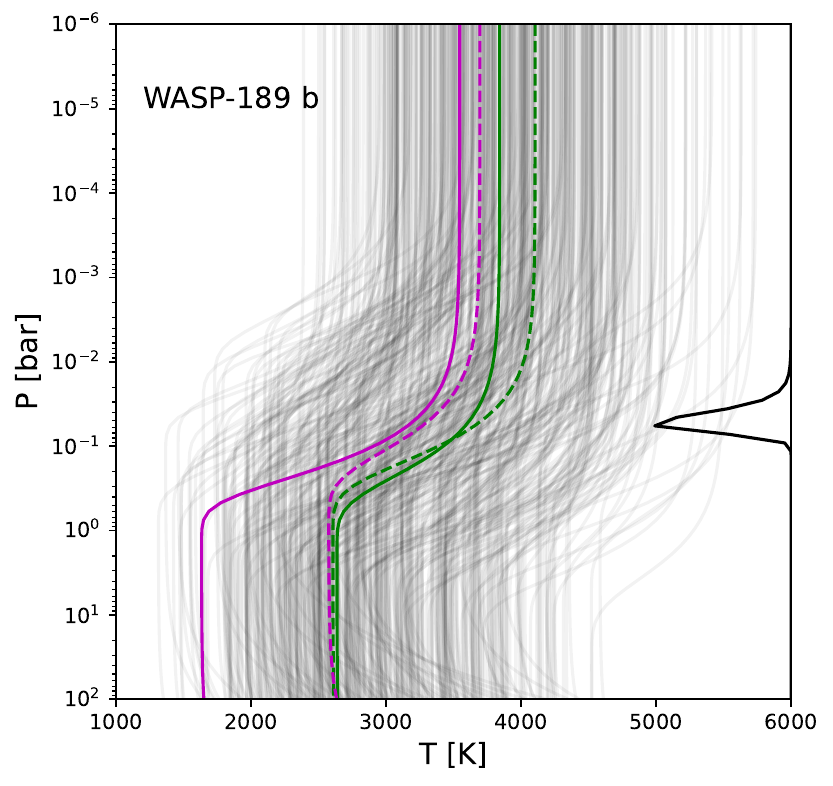}
    \includegraphics[width=0.3\linewidth]{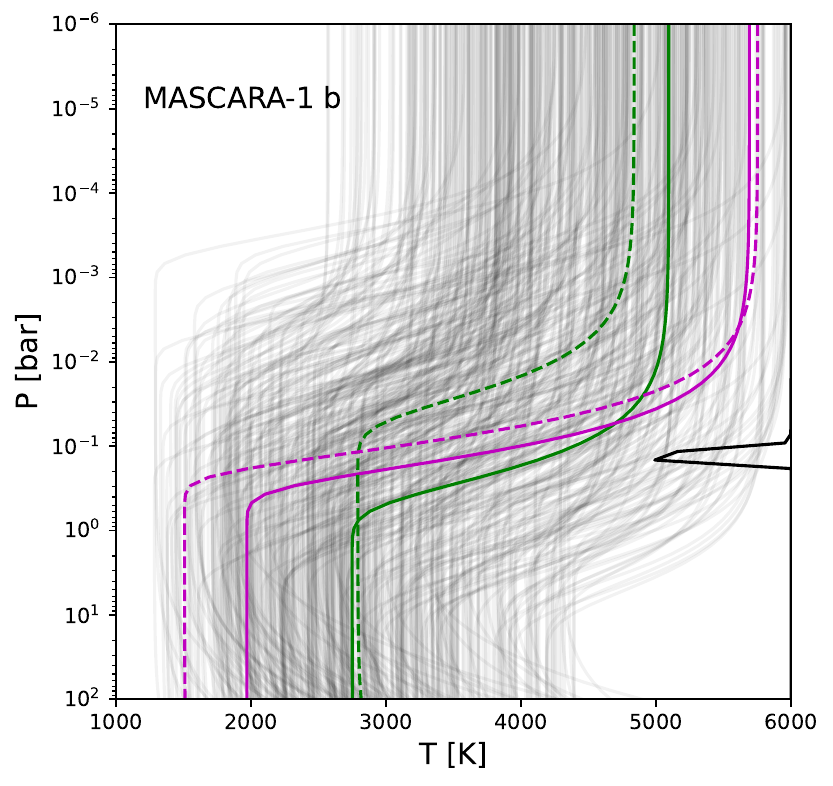}
    \includegraphics[width=0.3\linewidth]{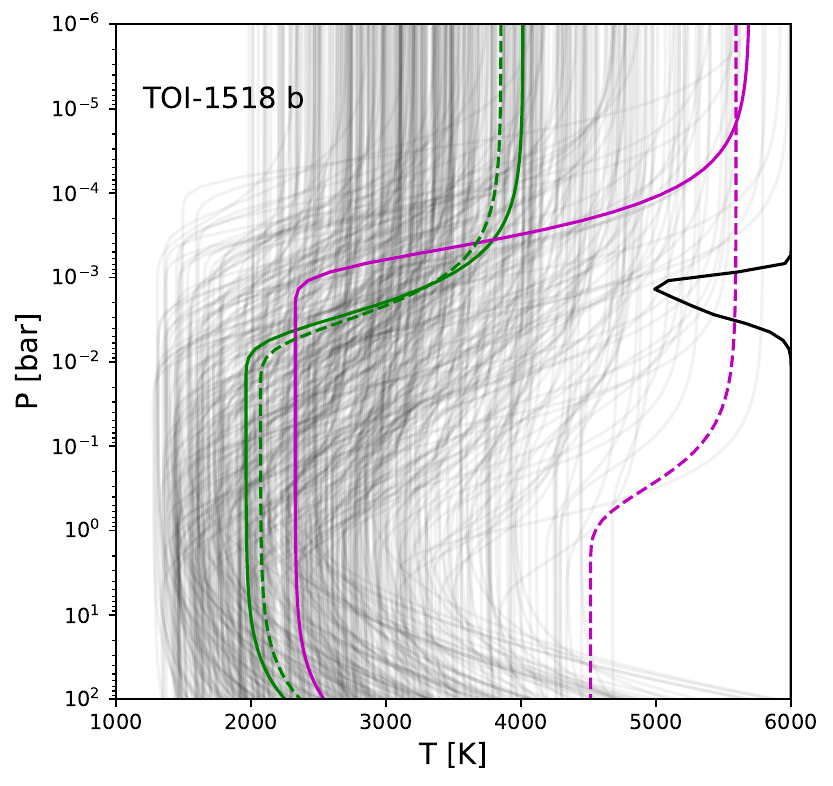}
    \caption{Retrieved $P-T$ profiles for each target. The median profiles are shown in green and maximum-likelihood are shown in purple, with the free retrievals shown as solid lines and the equilibrium as dashed lines. The black lines show 500 draws from the posterior. The black line to the right shows the wavelength-median emission contribution function for the maximum-likelihood profile from the free retrieval. Strong thermal inversions are preferred in all targets. However, the absolute temperatures of the $P-T$ profiles are poorly constrained, and the pressure level of the inversion shows a large scatter across targets. HRCCS is more sensitive to the thermal contrast of the $P-T$ profile,  which sets the strength of spectral features, rather than the absolute temperature, which sets the continuum level. Over short bandpasses, the data detrending procedure effectively removes the planetary continuum, and retrievals therefore provide only limited constraints on the absolute temperature and inversion pressure. The emission contribution functions generally peak where the temperature is rapidly changing with altitude, but can also be impacted by dissociation, which can be seen in \ktb. See \citet{finnerty2025k20} for further discussion.}
    \label{fig:pts}
\end{figure*}

\begin{figure*}
    \centering
    \includegraphics[width=0.925\linewidth]{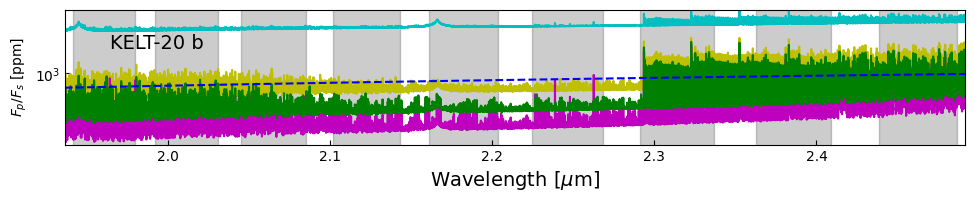}
    \includegraphics[width=0.925\linewidth]{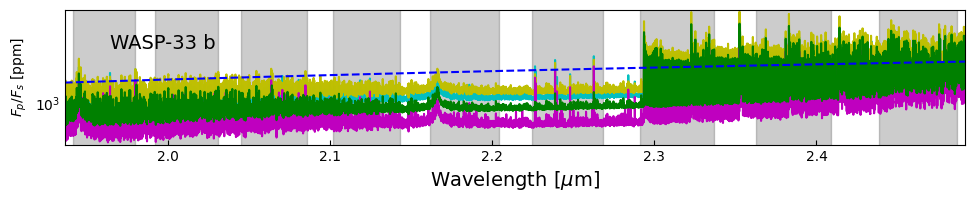}
    \includegraphics[width=0.925\linewidth]{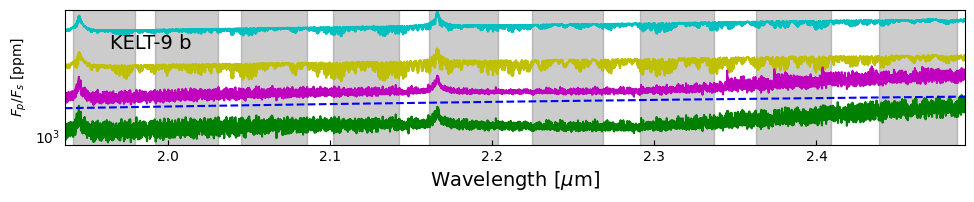}
    \includegraphics[width=0.925\linewidth]{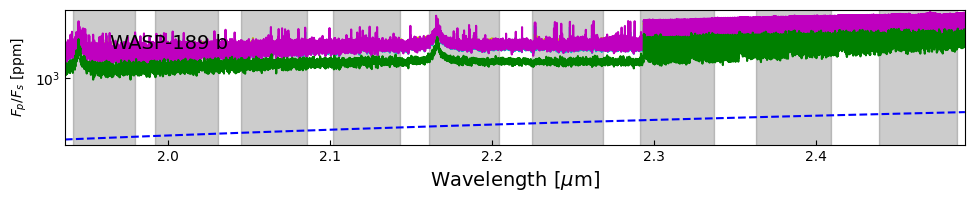}
    \includegraphics[width=0.925\linewidth]{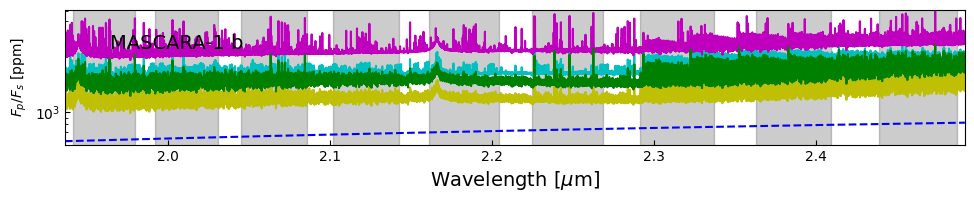}
    \includegraphics[width=0.925\linewidth]{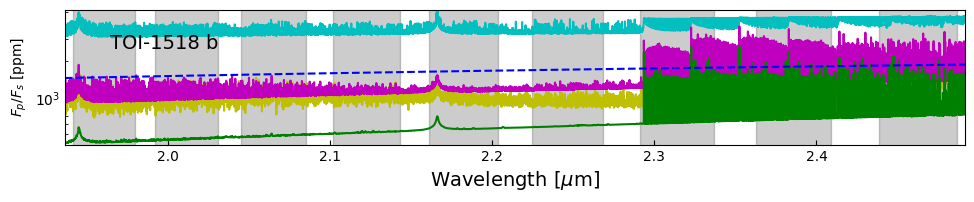}
    \caption{Planet/star flux ratios for the free retrieval maximum-likelihood (purple) free retrieval median (green), equilibrium max-likelihood (cyan), and equilibrium median (yellow) planet models, compared with a blackbody (dashed blue). Observed NIRSPEC orders are shaded. All retrieved spectra except \knb\ are dominated by emission from the $^{12}\rm CO$ $\nu = 2-0$ band beyond 2.3 $\mu$m, with features from mostly H$_2$O or OH in the three bluest orders. In contrast, the retrieved \knb\ spectra are dominated by H$_2$O emission, with contributions from the $^{13}\rm CO$ at longer wavelengths.}
    \label{fig:specs}
\end{figure*}

\begin{figure*}
    \centering
    \includegraphics[width=0.3\linewidth]{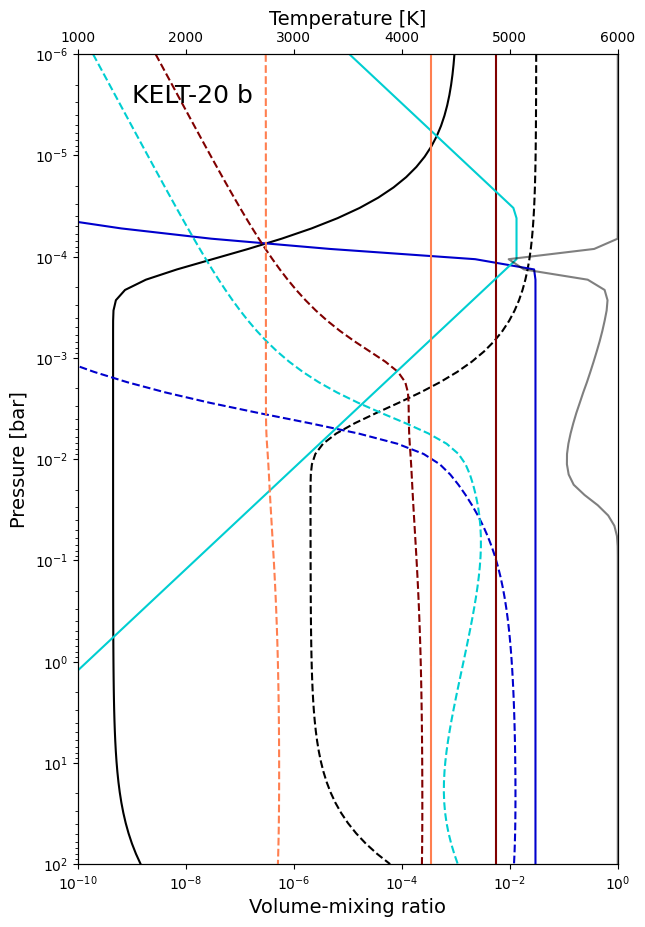}
    \includegraphics[width=0.3\linewidth]{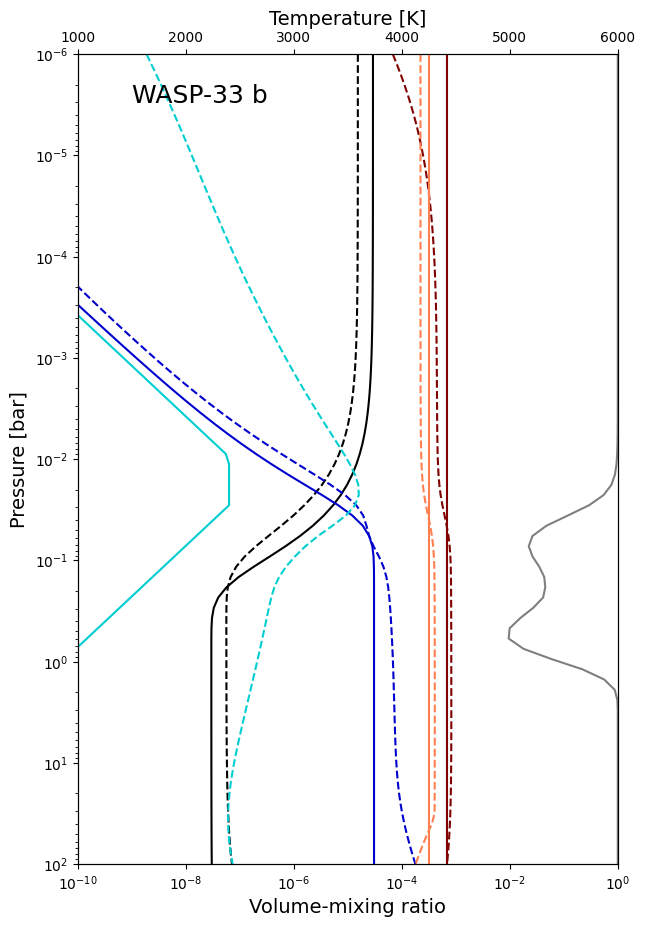}
    \includegraphics[width=0.3\linewidth]{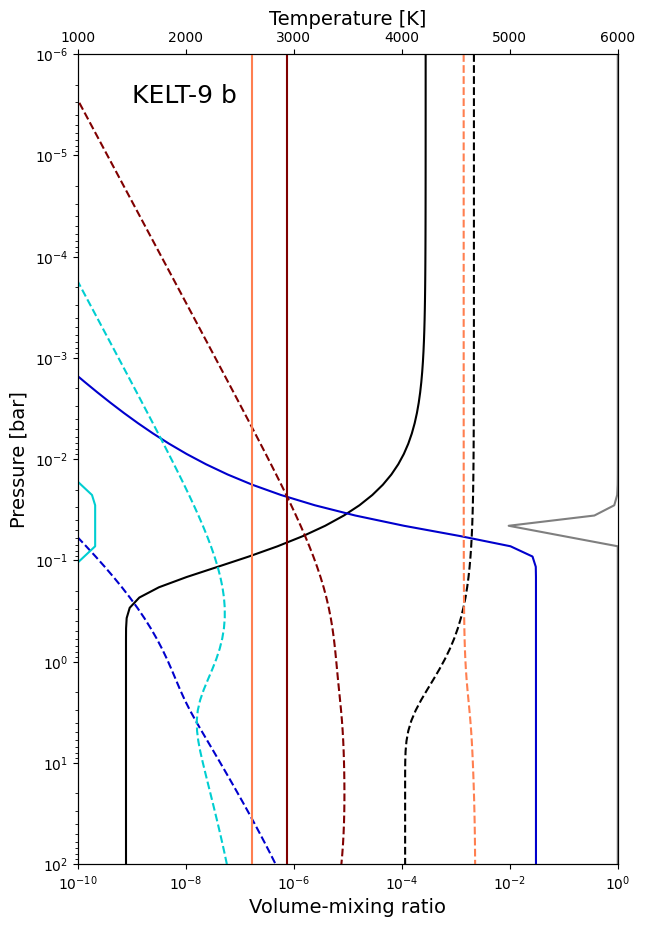}
    \includegraphics[width=0.3\linewidth]{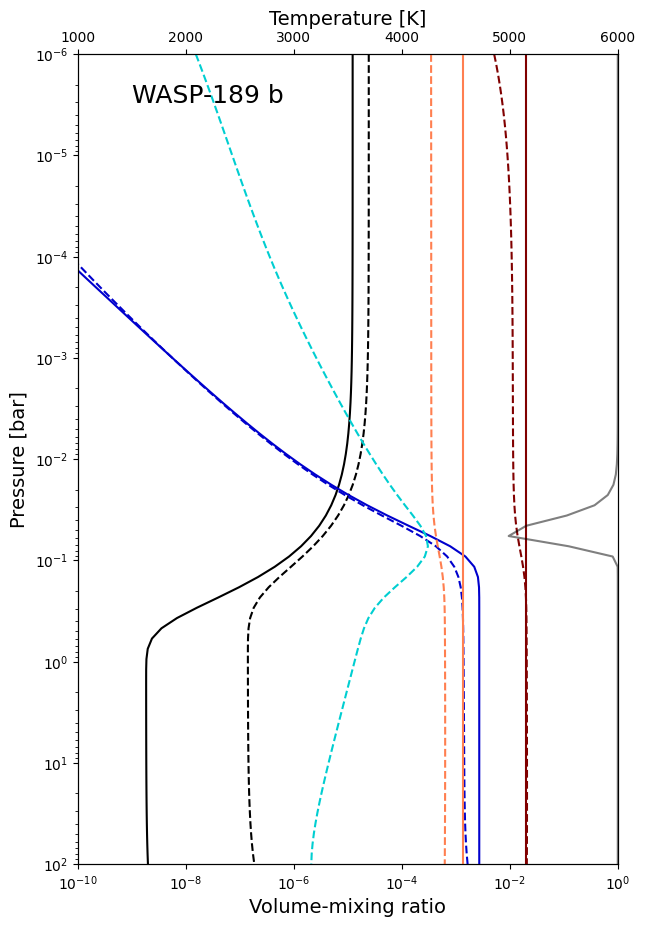}
    \includegraphics[width=0.3\linewidth]{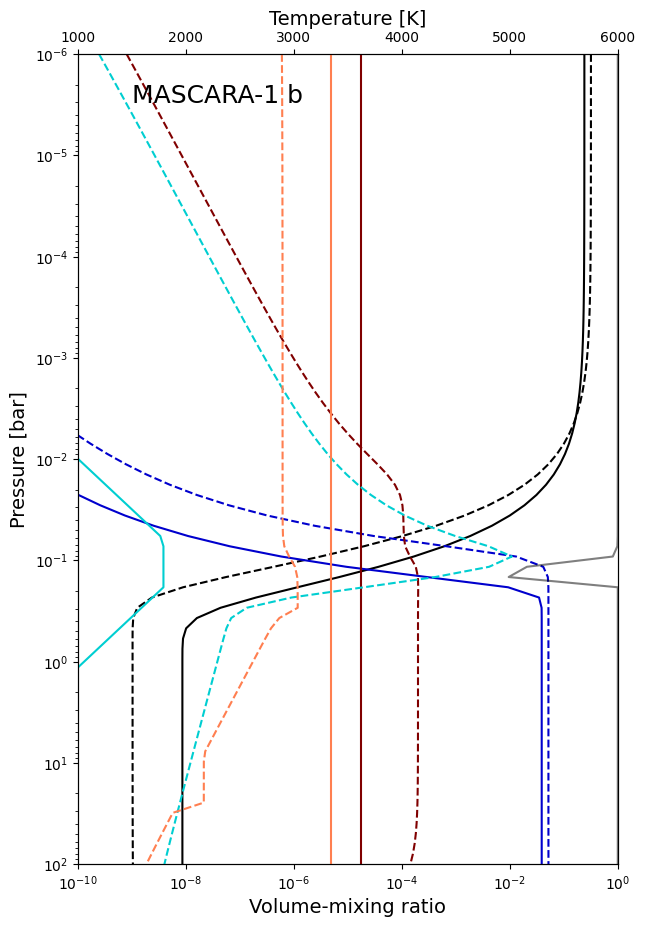}
    \includegraphics[width=0.3\linewidth]{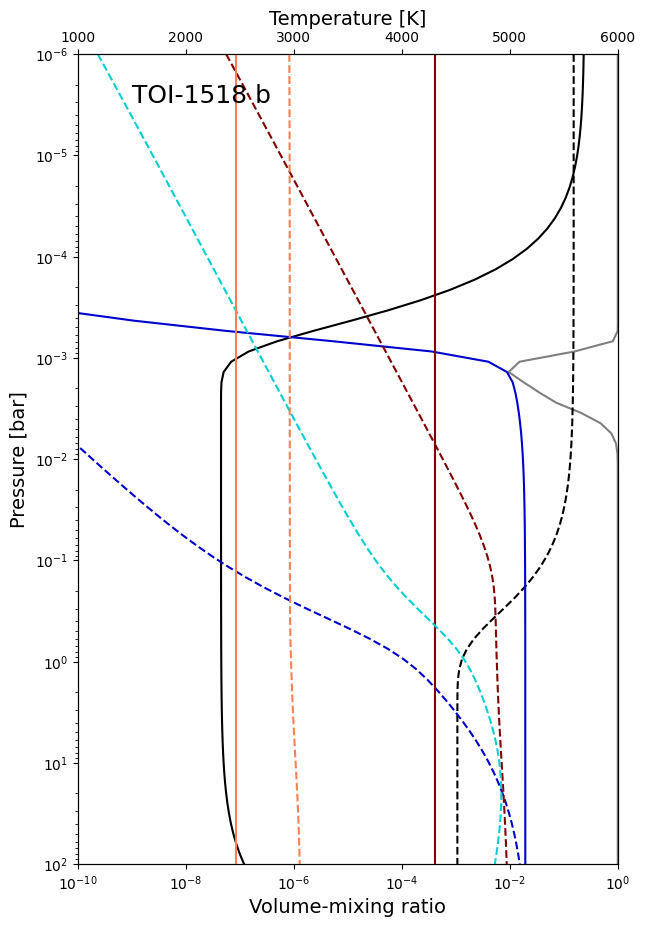}
    \caption{Retrieved max-likelihood $P-T$ profiles (black) and volume mixing ratios for  H$_2$O (blue), CO (dark red), OH (cyan), and Fe (light orange) for all targets. The free retrievals are shown as solid lines while the equilibrium retrievals are shown as dashed lines. The gray lines at the right show the wavelength-median emission contribution function. \knb\ shows the largest discrepancy between the retrievals, with the equilibrium case preferring an atmosphere dominated by Fe features, and the free retrieval preferring strong H$_2$O, despite the high temperature of \knb. For \wtb and \wob, the equilibrium and free retrievals are in good agreement, though the free retrieval does not support significant OH. For \ktb, the equilibrium case prefers a higher temperature and smaller $P-T$ contrast than the free retrieval, leading to significantly different mixing profiles, but both cases support the presence of H$_2$O, CO, and OH, in similar relative quantities. In \mob, the free retrieval appears to be under-estimating the deep CO abundance due to neglecting CO dissociation, but the H$_2$O profiles are in good agreement. For \tob, the retrievals differ significantly, likely because the atmospheric detection is dominated by CO alone and a wide range of $P-T$ profiles and abundances can produce a CO-dominated spectrum. }
    \label{fig:vmrs}
\end{figure*}

\begin{figure*}
    \centering
    \includegraphics[width=0.3\linewidth]{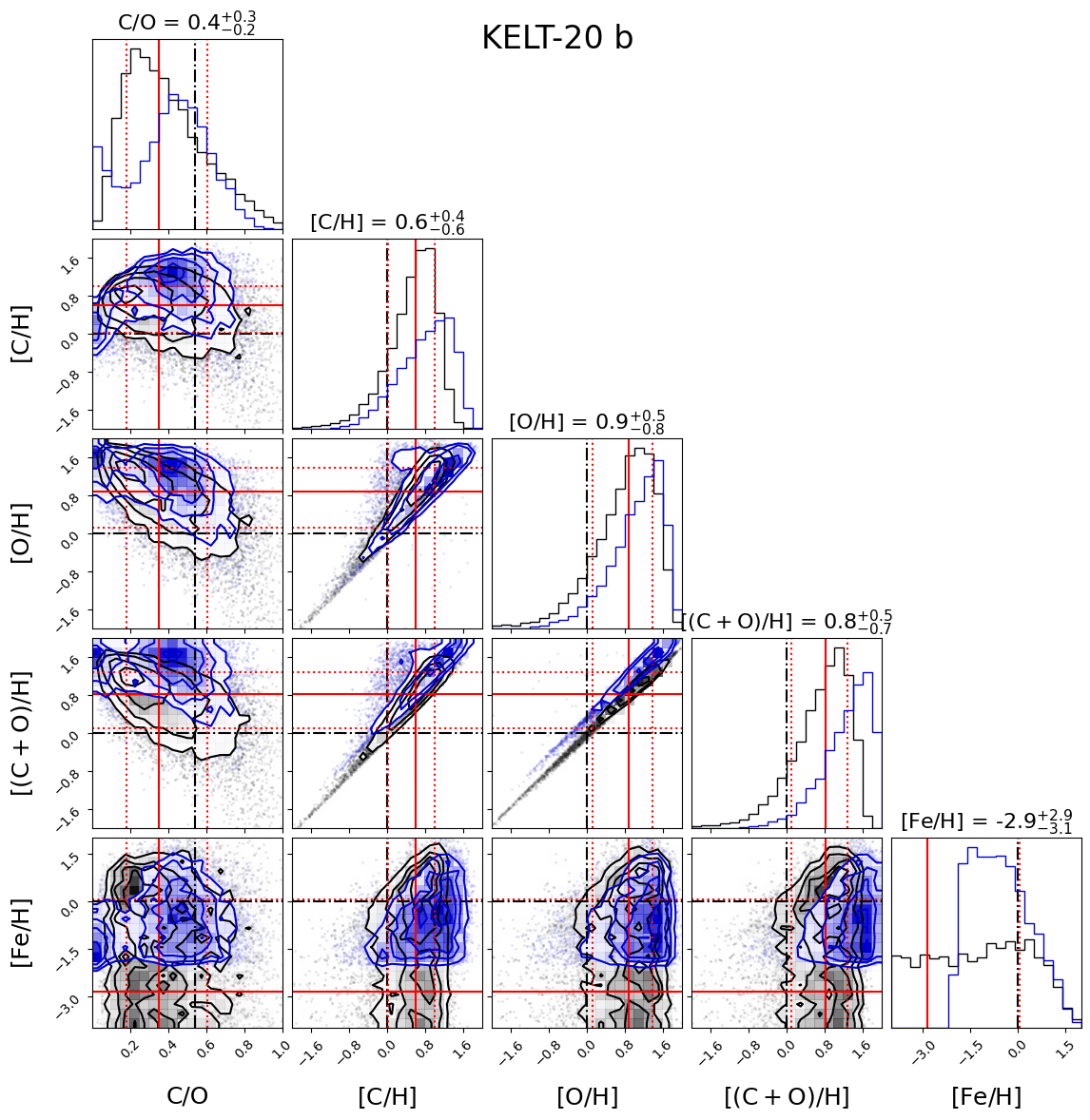}
    \includegraphics[width=0.3\linewidth]{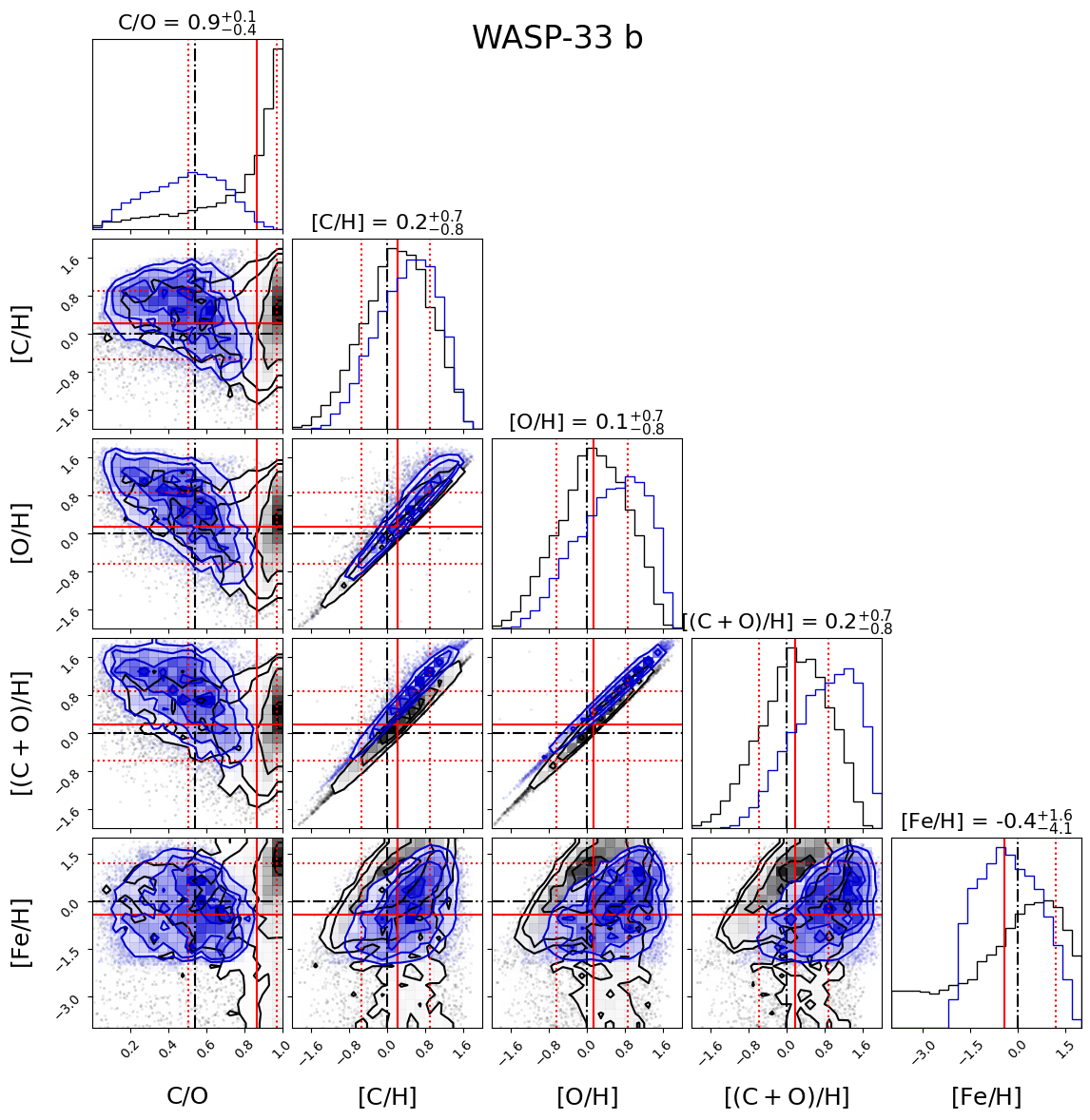}
    \includegraphics[width=0.3\linewidth]{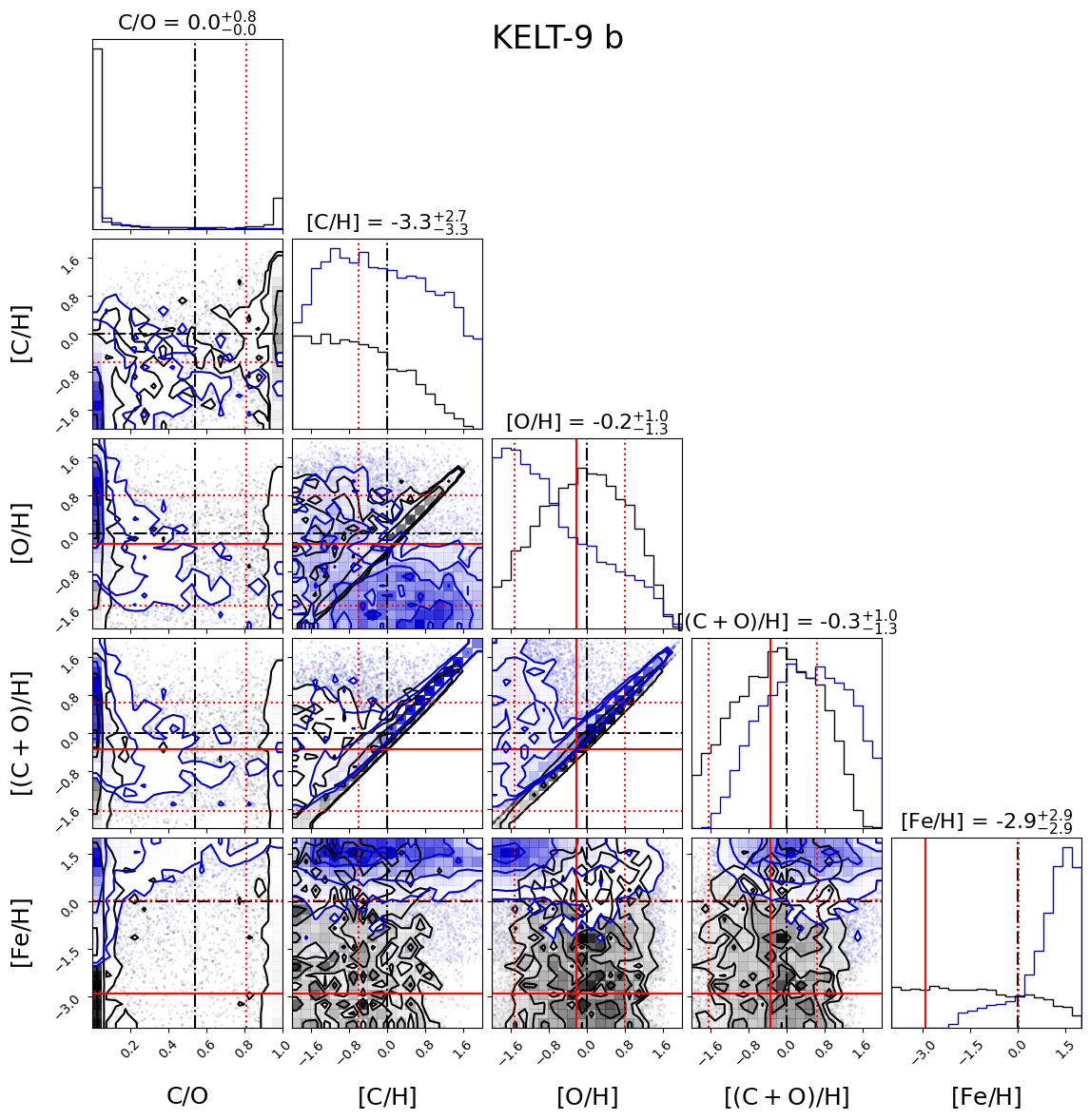}
    \includegraphics[width=0.3\linewidth]{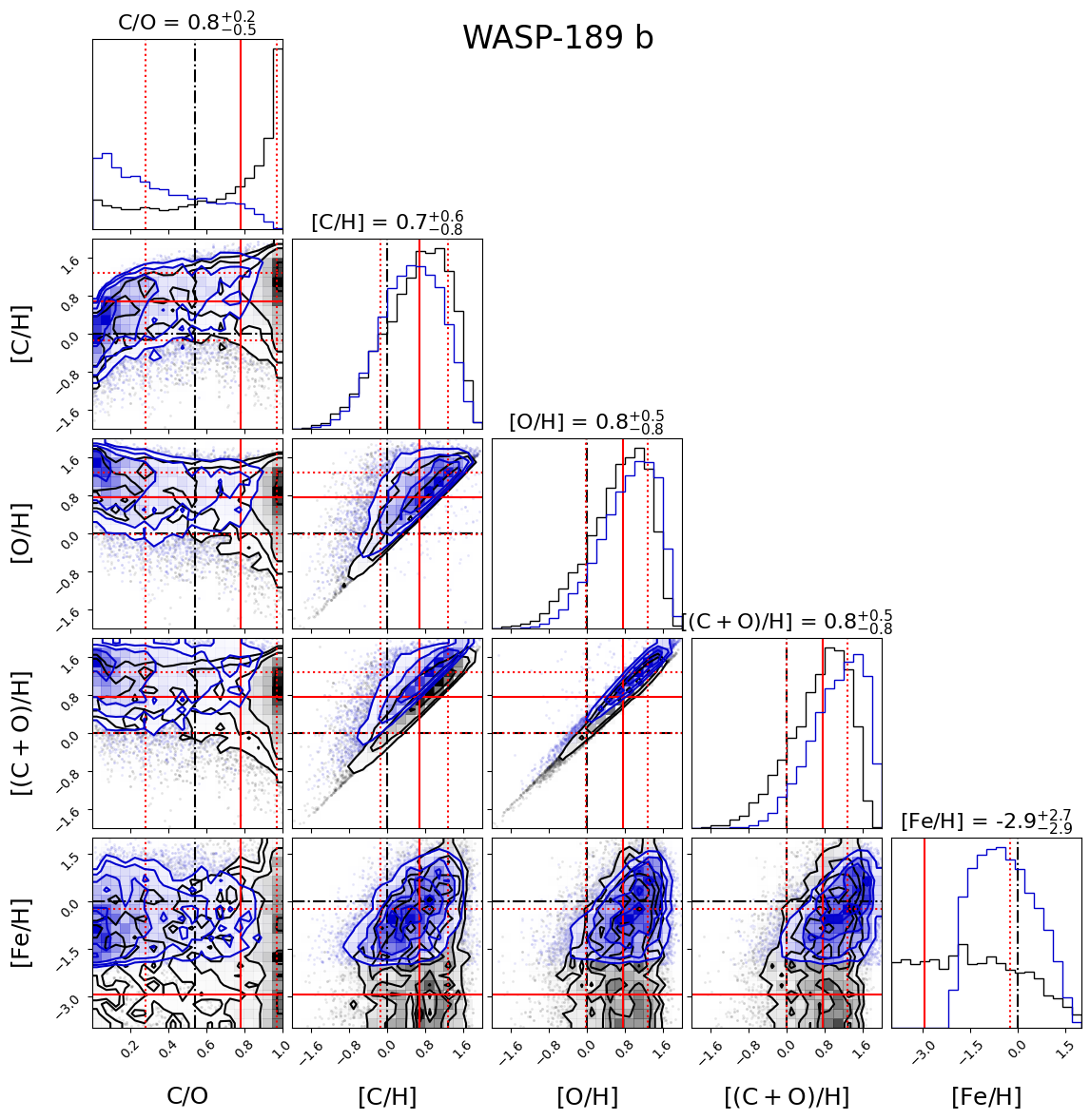}
    \includegraphics[width=0.3\linewidth]{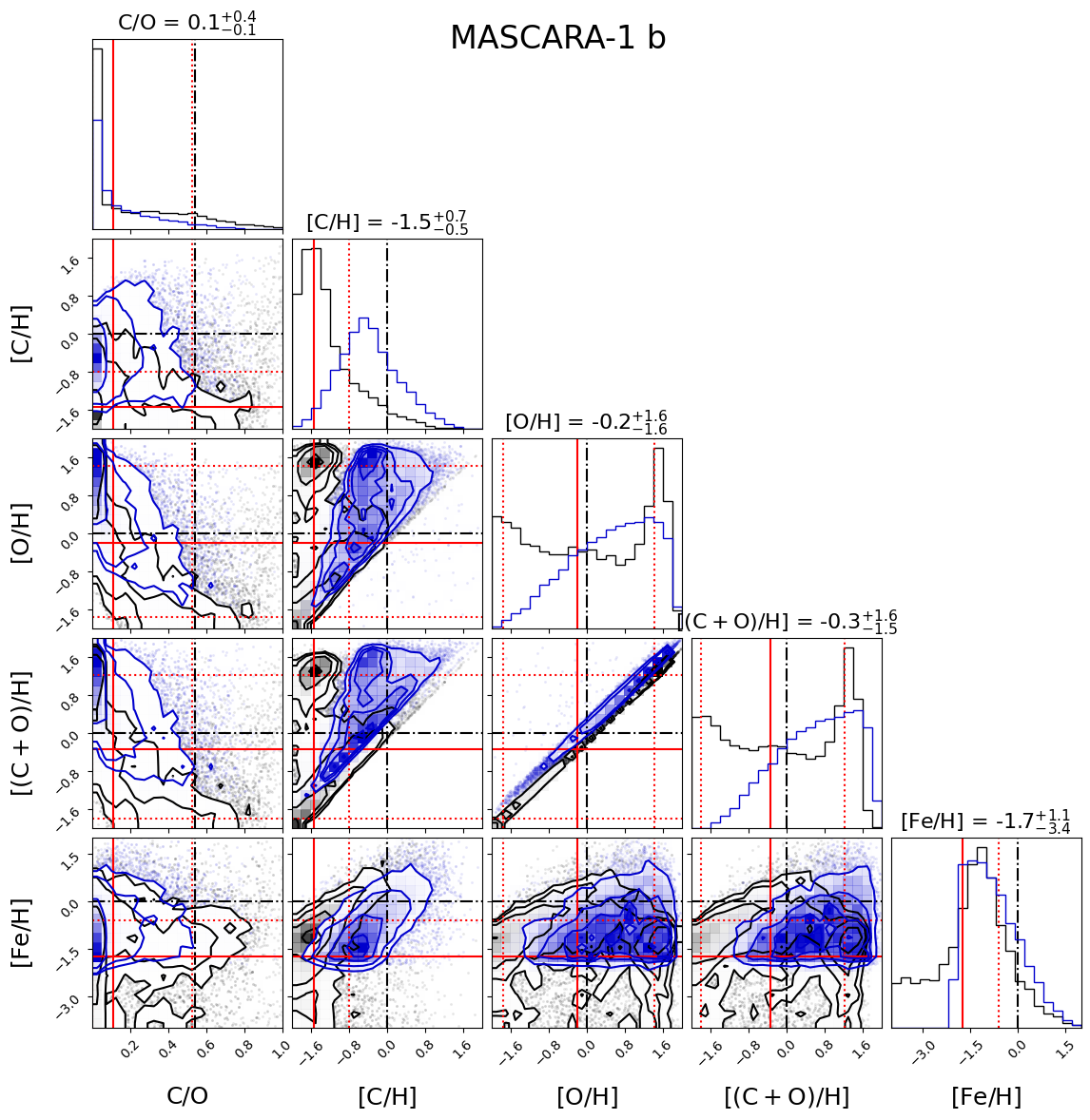}
    \includegraphics[width=0.3\linewidth]{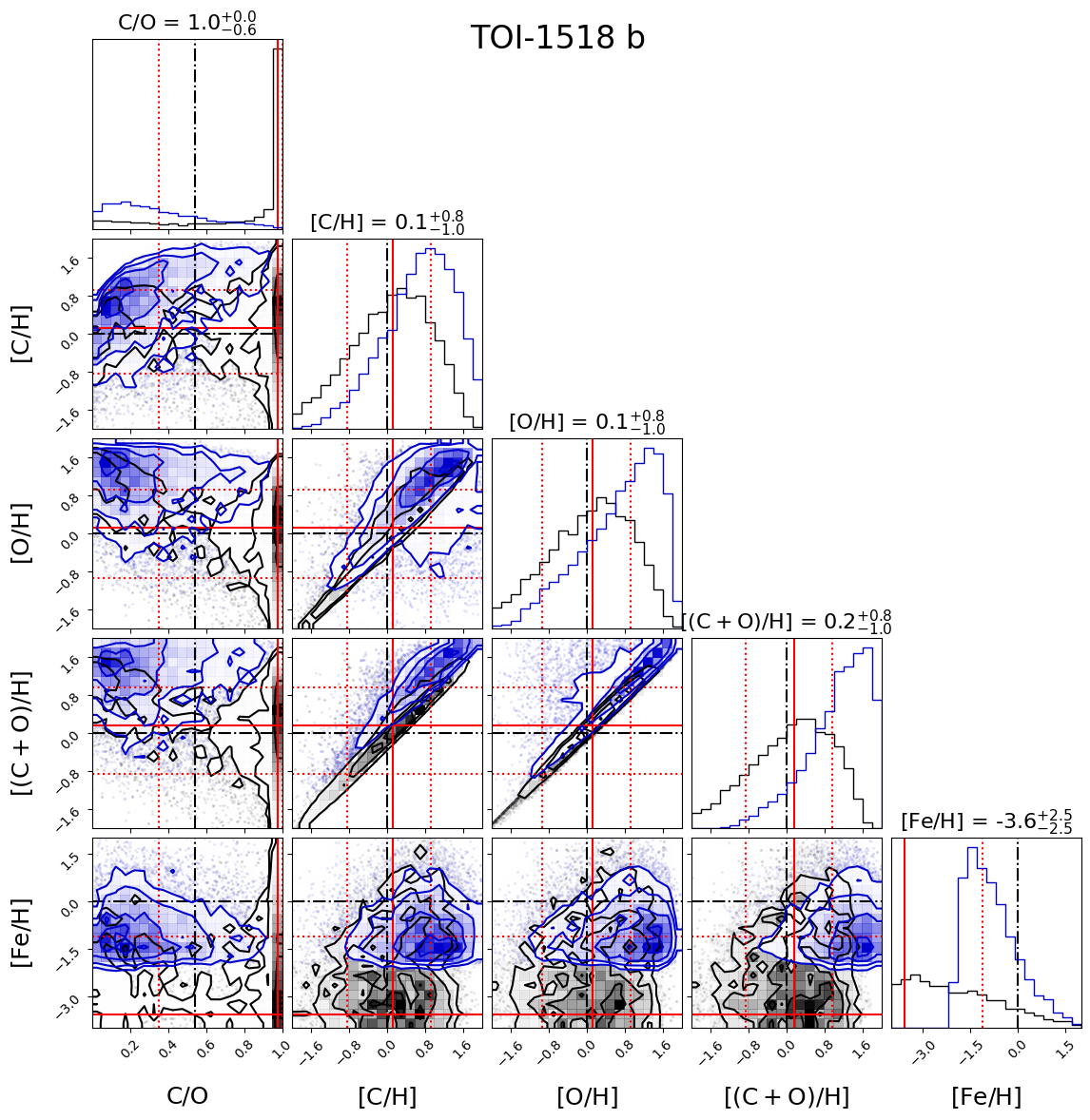}
    \caption{Derived posteriors for the C/O ratio, carbon and oxygen abundances, total volatile abundance, and iron abundance. Median and $\pm34$\% confidence intervals are indicated in solid and dotted red, respectively, and the solar values are shown in dash-dot black. The black distribution is computed from the free retrieval, and the blue distribution from the equilibrium retrieval. The retrieved carbon and oxygen abundances are generally consistent with solar-to-supersolar volatile abundances, but have very large uncertainties that extend from sub-solar values to the upper end of the mean molecular weight prior. The Fe abundances are at best weak upper limits. The retrieved C/O ratios are bimodal, preferring either 0 or 1 depending on whether the retrieval prefers low H$_2$O abundances dissociating at high altitudes, resulting in a high C/O, or high H$_2$O abundances and deep dissociation, resulting in a low C/O. The actual retrieved spectra are similar despite the range of C/O ratios, indicating that our retrievals are not reliably measuring the bulk composition of these UHJ atmospheres.  }
    \label{fig:abunds}
\end{figure*}

The retrieved $P-T$ profiles for each object are shown in Figure \ref{fig:pts}. The retrieved median and maximum-likelihood spectra are plotted in Figure \ref{fig:specs}, along with a comparison to the expected $F_p/F_s$ assuming both star and planet are blackbodies at their respective effective and equilibrium temperatures. Figure \ref{fig:vmrs} compares the retrieved mixing ratio profiles to the corresponding profiles from the \texttt{easyCHEM} \citep{lei2024} chemical equilibrium retrievals. Figure \ref{fig:abunds} shows the corner plots for the derived abundance parameters, C/O, [C/H], [O/H], [(C+O)/H], and [Fe/H]. The full corner plots for the retrievals are presented in Appendix \ref{app:corner}, including the four and eight component retrievals and the equilibrium retrievals.

For all targets, the retrieved posteriors are generally consistent as the number of principal components changes. As mentioned in \S \ref{sec:obs}, in a few cases, particularly the last order of \wob, omitting only four components leaves obvious residuals in the spectral time series, most likely arising from the fringing effects in KPIC data described in \citet{Finnerty2022, Xuan2024, horstman2025}. We therefore adopt the six component retrieval as our fiducial results.

\begin{deluxetable*}{ccccccccc}
    \tabletypesize{\tiny}
    \tablecaption{List of parameters, priors, and results for atmospheric retrievals}
    \tablehead{\colhead{Name} & Symbol  & \colhead{Prior} & \ktb & \wtb & \knb & \wob & \mob & \tob }
    \startdata
          & & Free retrievals & \\
        \hline
        log infrared opacity [$\rm cm^{2} g^{-1}$] & $\log \kappa$ &  Uniform(-4, 0)   & $-1.2^{+0.4}_{-0.6} (-0.7)$  & $-1.9^{+0.6}_{-0.6} (-2.4)$ & $-2.7^{+1.4}_{-0.8} (-3.5)$      & $-2.9^{+0.7}_{-0.7}\ (-3.4)$     & $-3.0^{+0.9}_{-0.6} (-3.4)$ & $-1.6^{+0.8}_{-1.0} (-1.0)$ \\
        log infrared/optical opacity & $\log \gamma$  &  Uniform(0, 3)                 & $1.8^{+0.3}_{-0.4} (2.2)$    & $1.0^{+0.4}_{-0.3} (0.9)$     & $0.7^{+0.4}_{-0.4} (1.9)$      & $0.7^{+0.5}_{-0.3}\ (1.4)$       & $1.1^{+0.6}_{-0.4} (1.9)$ & $1.3^{+0.6}_{-0.6} (1.6)$ \\ 
        Intrinsic temperature [K] & $\rm T_{int}$ & Uniform(100,500)                   &  $300^{+100}_{-100} (100)$   & $300^{+100}_{-100} (200)$     & $300^{+100}_{-100} (100)$      & $290^{+130}_{-120}\ (380)$       & $300^{+100}_{-100} (200)$ & $300^{+100}_{-100} (200)$ \\
        Equilibrium temperature [K] & $\rm T_{equ}$ & Uniform(1500, 5000)              &  $1900^{+400}_{-200} (1600)$ & $2600^{+500}_{-500} (2600)$   & $3100^{+900}_{-800} (1700)$    & $3010^{+550}_{-610}\ (1920)$     & $3200^{+800}_{-700} (2300)$ & $2300^{+800}_{-500} (2800)$ \\
        $K_{\rm p}$ offset [\kms] & $\Delta K_{\rm p}$  & Uniform(-40, 40)             &  $-4.7^{+4.0}_{-4.2} (-1.5)$ & $-13.6^{+4.4}_{-4.4} (-12.5)$ & $15.1^{+9.6}_{-24.1} (24.7)$   & $-23.0^{+18.0}_{-11.0}\ (-30.0)$ & $1.1^{+5.9}_{-7.6} (0.0)$ & $-5.8^{+19.5}_{-12.7} (-16.7)$ \\ 
        $v_{\rm sys}$ offset [\kms] & $\Delta v_{\rm sys}$ & Uniform(-30, 30)          &  $-3.2^{+1.9}_{-2.0} (-2.1)$ & $-10.2^{+2.3}_{-2.3} (-9.6)$  & $-11.2^{+17.0}_{-6.8} (-18.6)$ & $5.0^{+3.4}_{-5.4}\ (7.3)$       & $-9.9^{+3.7}_{-3.0} (-9.1)$ & $3.4^{+7.1}_{-4.5} (-1.9)$ \\
        Rotational velocity [\kms] & $v_{\rm rot}$ & Uniform(0, 15)                    &  $8.1^{+1.0}_{-1.1} (8.4)$   & $9.6^{+1.2}_{-1.2} (9.0)$     & $7.4^{+3.7}_{-3.9} (8.5)$      & $5.3^{+2.4}_{-2.7}\ (0.0)$       & $4.3^{+2.8}_{-2.5} (4.3)$ & $7.7^{+3.7}_{-4.0} (2.4)$ \\
        log H$_2$O volume mixing ratio & log H$_2$O  &  Uniform(-12, -0.3)             & $-2.5^{+0.6}_{-0.9} (-1.5)$  & $-4.0^{+1.1}_{-0.9} (-4.5)$   & $-4.1^{+1.5}_{-4.2} (-1.5)$    & $-3.2^{+0.9}_{-1.2}\ (-2.6)$     & $-3.5^{+1.8}_{-1.7} (-1.4)$ & $-5.6^{+2.6}_{-4.1} (-1.7)$ \\
        log $\rm ^{12}CO$ volume mixing ratio & $\log\rm ^{12}CO$ &  Uniform(-12, -0.3)&  $-2.8^{+0.4}_{-0.6} (-2.3)$ & $-3.2^{+0.7}_{-0.8} (-3.2)$   & $-6.7^{+2.7}_{-3.3} (-6.1)$    & $-2.7^{+0.6}_{-0.8}\ (-1.7)$     & $-4.9^{+0.7}_{-0.5} (-4.8)$ & $-3.3^{+0.8}_{-1.0} (-3.4)$ \\
        log OH volume mixing ratio & log OH &  Uniform(-12, -0.5)                      &  $-2.7^{+0.6}_{-1.2} (-1.9)$ & $-8.0^{+2.7}_{-2.6} (-7.2)$   & $-7.8^{+2.7}_{-2.6} (-9.7)$    & $-7.7^{+2.8}_{-2.7}\ (-11.1)$     & $-8.5^{+2.3}_{-2.2} (-8.4)$ & $-7.5^{+2.8}_{-2.7} (-12.0)$ \\
        log Fe volume mixing ratio & log Fe &  Uniform(-12, -1)                        &  $-7.3^{+2.9}_{-3.1} (-3.5)$ & $-4.8^{+1.6}_{-4.1} (-3.5)$   & $-7.3^{+2.9}_{-2.9} (-6.8)$    & $-7.3^{+2.7}_{-2.9}\ (-2.9)$      & $-6.2^{+1.1}_{-3.4} (-5.3)$ & $-8.0^{+2.5}_{-2.5} (-7.1)$ \\
        log $\rm ^{13}CO$ volume mixing ratio & $\log\rm ^{13}CO$ &  Uniform(-12, -1)  &  $-8.2^{+2.5}_{-2.5} (-4.9)$ & $-8.4^{+2.3}_{-2.3} (-11.7)$  & $-5.5^{+1.6}_{-3.6} (-4.4)$    & $-8.8^{+2.1}_{-2.0}\ (-7.8)$      & $-8.6^{+2.2}_{-2.1} (-5.4)$ & $-8.2^{+2.4}_{-2.3} (-7.2)$ \\
        log H$_2$ volume mixing ratio & $\log \rm H_2$ &  Uniform(-0.4, -0.05)         &  $-0.2^{+0.1}_{-0.1} (-0.1)$ & $-0.2^{+0.1}_{-0.1} (-0.2)$   & $-0.2^{+0.1}_{-0.1} (-0.1)$    & $-0.2^{+0.1}_{-0.1}\ (-0.1)$      & $-0.2^{+0.1}_{-0.1} (-0.1)$ & $-0.2^{+0.1}_{-0.1} (-0.1)$ \\
        Scale factor & scale & LogNormal(0, 0.1)                                       & $0.0^{+0.1}_{-0.1} (0.1)$    & $-0.0^{+0.1}_{-0.1} (-0.2)$   & $0.0^{+0.1}_{-0.1} (0.2)$      & $0.02^{+0.09}_{-0.09}\ (0.16)$    & $0.0^{+0.1}_{-0.1} (0.1)$ & $0.0^{+0.1}_{-0.1} (0.1)$ \\ 
        \smallskip \\
        \hline
          & & Chemical equilibrium retrievals & \\
        \hline
        log infrared opacity [$\rm cm^{2} g^{-1}$] & $\log \kappa$ &  Uniform(-4, 0) & $-0.9^{+0.4}_{-0.5}\ (-1.2)$ & $-1.3^{+0.6}_{-0.7}\ (-2.1)$   & $-3.4^{+1.3}_{-0.4}\ (-3.3)$   & $-2.8^{+0.8}_{-0.7}\ (-2.5)$     & $-1.9^{+0.6}_{-0.8}\ (-3.6)$ & $-1.5^{+0.8}_{-1.3}\ (-2.9)$  \\
        log infrared/optical opacity & $\log \gamma$  &  Uniform(0, 3)               & $1.6^{+0.5}_{-0.6}\ (0.9)$   & $1.1^{+0.5}_{-0.4}\ (0.8)$     & $0.5^{+0.2}_{-0.2}\ (0.3)$     & $0.8^{+0.6}_{-0.3}\ (0.7)$       & $1.0^{+0.4}_{-0.3}\ (2.4)$ & $1.1^{+0.6}_{-0.6}\ (0.4)$  \\
        Intrinsic temperature [K] & $\rm T_{int}$ & Uniform(100,500)                 & $290^{+130}_{-120}\ (470)$   & $300^{+130}_{-130}\ (340)$     & $300^{+120}_{-130}\ (190)$     & $300^{+130}_{-130}\ (440)$       & $300^{+130}_{-130}\ (310)$ & $290^{+130}_{-120}\ (130)$  \\
        Equilibrium temperature [K] & $\rm T_{equ}$ & Uniform(1500, 5000)            & $2160^{+870}_{-470}\ (3660)$ & $2770^{+430}_{-530}\ (2730)$   & $4260^{+440}_{-800}\ (4360)$   & $3010^{+570}_{-740}\ (2930)$     & $3250^{+510}_{-580}\ (1790)$ & $2420^{+1310}_{-590}\ (4980)$  \\
        $K_{\rm p}$ offset [\kms] & $\Delta K_{\rm p}$  & Uniform(-40, 40)           & $-4.7^{+4.4}_{-4.3}\ (-4.8)$ & $-13.5^{+4.4}_{-4.2}\ (-15.3)$ & $-2.0^{+19.0}_{-14.0}\ (-7.0)$ & $-18.0^{+23.0}_{-14.0}\ (-37.0)$ & $-5.8^{+7.2}_{-9.1}\ (-1.9)$ & $4.0^{+20.0}_{-21.0}\ (-23.0)$  \\
        $v_{\rm sys}$ offset [\kms] & $\Delta v_{\rm sys}$ & Uniform(-30, 30)        & $-3.0^{+2.0}_{-2.0}\ (-2.9)$ & $-10.1^{+2.3}_{-2.2}\ (-10.8)$ & $1.0^{+9.0}_{-11.0}\ (2.0)$    & $4.7^{+4.1}_{-6.5}\ (10.0)$      & $-6.7^{+4.4}_{-3.7}\ (-8.7)$ & $7.6^{+6.0}_{-6.1}\ (3.0)$  \\
        Rotational velocity [\kms] & $v_{\rm rot}$ & Uniform(0, 15)                  & $7.6^{+1.1}_{-1.1}\ (6.9)$   & $9.8^{+1.1}_{-1.2}\ (8.8)$     & $6.6^{+4.6}_{-4.1}\ (0.6)$     & $5.0^{+2.3}_{-2.4}\ (3.6)$       & $4.4^{+3.2}_{-2.6}\ (2.7)$ & $9.6^{+3.0}_{-4.3}\ (3.4)$  \\
        Carbon abundance & $\rm [C/H]$ & Uniform(-2,2)                               & $0.9^{+0.5}_{-0.7}\ (-0.3)$  & $0.5^{+0.6}_{-0.7}\ (0.3)$     & $-0.2^{+1.3}_{-1.2}\ (-0.4)$   & $0.5^{+0.6}_{-0.7}\ (1.7)$       & $-0.4^{+0.7}_{-0.6}\ (-0.4)$ & $0.7^{+0.7}_{-0.9}\ (1.4)$  \\
        Oxygen abundance & $\rm [O/H]$ & Uniform(-2,2)                               & $1.1^{+0.4}_{-0.7}\ (1.2)$   & $0.6^{+0.7}_{-0.9}\ (0.2)$     & $-0.9^{+1.3}_{-0.8}\ (-1.9)$   & $0.9^{+0.5}_{-0.7}\ (1.5)$       & $0.6^{+0.8}_{-1.1}\ (1.8)$ & $0.8^{+0.6}_{-1.1}\ (1.6)$  \\
        Metal abundance  & $\rm [M/H]$ & Uniform(-2,2)                               & $-0.6^{+1.1}_{-0.9}\ (-2.0)$ & $-0.3^{+1.1}_{-1.0}\ (0.9)$    & $1.2^{+0.5}_{-1.1}\ (1.7)$     & $-0.5^{+1.1}_{-1.0}\ (1.1)$      & $-0.9^{+1.0}_{-0.7}\ (-1.7)$ & $-1.0^{+1.0}_{-0.7}\ (-1.5)$  \\
        Scale factor & scale & LogNormal(0, 0.1)                                     & $0.0^{+0.1}_{-0.1}\ (0.2)$   & $-0.00^{+0.1}_{-0.1}\ (-0.1)$  & $0.0^{+0.1}_{-0.1}\ (0.3)$     & $0.02^{+0.09}_{-0.09}\ (0.12)$   & $-0.0^{+0.1}_{-0.1}\ (-0.1)$ & $-0.0^{+0.1}_{-0.1}\ (-0.2)$  \\
    \enddata 
    \tablecomments{Values with errors are the retrieved medians and 68$\% / 1\sigma$ confidence interval, and limits are given at 95\% confidence. Values in parentheses without errors are the values for the maximum-likelihood model. In addition to the listed priors, we required that the atmospheric temperature stay below 6000 K at all pressure levels, and penalize atmospheres with mean molecular weights greater than 10 AMU. The full corner plots are included included in Appendix \ref{app:corner}.}
    \label{tab:priors}
\end{deluxetable*}

\subsection{\ktb}

The the posterior from the free retrieval for \ktb\ is consistent with the previous analysis in \citet{finnerty2025k20}, to which we refer the reader for a detailed discussion. This consistency establishes that the changes in the priors, detrending approach, and log-likelihood function compared with \citet{finnerty2025k20} do not substantially impact the retrieved atmospheric composition for a well-detected target. The most significant compositional change compared to \citet{finnerty2025k20} is the clear preference for OH, which we attribute to the inclusion of the three bluest orders on the NIRSPEC detector, which cover significant H$_2$O and OH spectral features. Additionally, the retrieved $P-T$ profile prefers a stronger, lower pressure inversion compared with \citet{finnerty2025k20}. These parameters' respective impacts on the final strength of emission lines are inversely correlated, and this behavior is therefore not surprising given our poor overall sensitivity to the continuum level. 

Figure \ref{fig:vtracks} shows a clear detection of the maximum-likelihood model for \ktb, even in the 2D cross-correlation function. Figure \ref{fig:kpvsysall} shows \ktb\ is the second-best detected of the six targets, at SNR = 10.5. Figure \ref{fig:kpvsys_contours} demonstrates that the detection is CO-dominated. The H$_2$O and OH models both produce weak features consistent with the nominal \kp\ and \vsys. This is consistent with the presence of weak H$_2$O spectral features in the maximum-likelihood model plotted in Figure \ref{fig:specs}, and with the preference of the retrieval for a high deep-atmosphere H$_2$O abundance and OH features. 

The center of the OH-only contour in Figure \ref{fig:kpvsys_contours} is slightly blueshifted compared to the maximum-likelihood model. While this could be indicative of weak OH emission with an apparent kinematic offset from circulation effects, the presence of other features of comparable strength in the \kpvsys\ map suggests this could simply be noise. We defer further discussion of possible spatial inhomogeneities in \ktb\ to Finnerty {\it et al.} in prep., which also considers pre-eclipse observations from KPIC not reported here.  

As discussed in \citet{finnerty2025k20}, the dissociation results in a degeneracy between the CO abundance and the H$_2$O dissociation pressure, complicating abundance inferences. While the H$_2$O abundance is not included as an explicit parameter in this analysis, changing the pressure at the base of the thermal inversion effectively sets H$_2$O dissociation pressure, and Figure \ref{fig:pts} shows there is a substantial scatter of this pressure. \citet{finnerty2025k20} reported a significantly greater dissociation pressure when this pressure is included as a free parameter, combined with a higher pressure thermal inversion and reduced temperature contrast over the $P-T$ profile. For the C/O ratio, the result of these degeneracies is a preference for modestly low C/O, but with a long tail extending to approximately solar values (see Figure \ref{fig:abunds}). The retrieved carbon and oxygen abundances are both consistent with solar-to-supersolar values.

The retrieved \kp, \vsys, and $v_{rot}$ values are all quite similar to the values reported in \citet{finnerty2025k20}, to which we refer the reader for a more detailed discussion. In brief, the small retrieved offset from the nominal reference frame can be explained by a combination of ephemeris uncertainty and uncertainty in the stellar radial velocity, precluding using the retrieved offsets in \kp\ and \vsys\ to constrain circulation patterns. The retrieved $v_{rot}$ is higher than expected for 1:1 tidal synchronization, but our physical interpretation of this parameter is limited by systematic uncertainties in the LSF width. 

The results from the equilibrium retrieval are generally consistent with the free retrieval, given the degeneracies discussed above. Figure \ref{fig:pts} shows the median $P-T$ profiles are similar for the two retrievals, but the maximum-likelihood profile in the equilibrium case is hotter, with a smaller thermal contrast and deeper start to the inversion. The leads to a significantly higher continuum flux level, but weaker features relative to the planet continuum in the planet models plotted in Figure \ref{fig:specs}, consistent with HRCCS being only weakly sensitive to planet continuum over a narrow bandpass. The equilibrium retrieval is consistent with the free retrieval for C/O, [C/H], and [O/H], as shown in Figure \ref{fig:abunds}, indicating that the VMR parameterization in the free retrievals can capture the impact of dissociation on bulk composition similarly to chemical equilibrium models. 

\subsection{\wtb}

The previous analysis of these data, reported in \citet{finnerty2023}, used an older version of the same retrieval pipeline employed here. Significant changes were made to the detrending/PCA procedures since that analysis, which did not incorporate vertical abundance variations. In this analysis, the maximum-likelihood model from the free retrieval produces a clear trace in the 2D cross-correlation plots for both science fibers, and summing the two cross-correlation time series for the maximum-likelihood model results in a detection in the \kpvsys\ space at $\rm SNR = 11.2$, slightly better than that of \ktb\ (see Figure \ref{fig:kpvsysall}).

Figure \ref{fig:kpvsys_contours} indicates the overall detection is dominated by CO emission features, with a weak contribution from H$_2$O generally consistent in \kp\ and \vsys. The OH template does not produce any significant features near the nominal planet reference frame. The retrieved spectrum is consistent with these results and with the spectrum retrieved for \ktb, showing strong CO emission features beyond $2.3\ \mu\rm m$ and weak H$_2$O features at shorter wavelengths. In contrast to \ktb, however,  we retrieve a higher deep-atmosphere CO abundance compared to H$_2$O, shown in Figure \ref{fig:vmrs}, which leads to a preference for high C/O ratios shown in Figure \ref{fig:abunds}. The overall carbon and oxygen abundances are both roughly consistent with solar values, with uncertainties of $\pm0.8$ dex. 

Retrievals on these data have preferred a high C/O ratio with both the fixed and parameterized vertical mixing profile treatments. \citet{finnerty2023} used fixed-with-altitude abundances for all species, which is known to produce a bias towards high C/O ratios \citep{brogi2023}, but both the retrieval presented here and tests with freely retrieving the H$_2$O dissociation pressure similar to \citep{finnerty2025k20} have continued to prefer $\log \rm CO > \log \rm H_2O$ in the deep atmosphere, suggesting that the preference for high C/O ratio may not be an artifact of vertical mixing assumptions. An elevated C/O ratio has also been reported for the UHJ WASP-121~b \citep{smith2024}. These results suggest at least some portion of the UHJ population may be carbon-enriched, in contrast with the oxygen-rich composition apparently preferred for \ktb. 

While \citet{finnerty2023} retrieved a \kp\ and \vsys\ consistent with the assumed ephemeris, the retrieval presented here prefers significant offsets in both parameters. The retrieved lower value of \kp\ and negative value for \vsys\ partially cancel, reducing the offset compared to the nominal planetary reference frame, and the planetary feature in Figure \ref{fig:kpvsysall} still overlaps with the assumed \kp\ and \vsys. While \citet{finnerty2023} used $v_{rad} = -9.2$ \kms, as reported in \citet{gontcharov2006}, for this work we adopt $v_{rad} = -0.3$ \kms\ from \citet{collier2010}. We also computed a slightly larger value for the nominal \kp\ in this work, and used updated values for the orbital period and transit time. That such apparently minor changes in the assumed ephemeris producing a $>10$ \kms\ change in the retrieved \vsys\ values suggests that we cannot interpret \vsys\ physically, and should instead approach it as a nuisance parameter for uncertainties in the stellar radial velocity and orbital phase. Broader phase coverage would break the degeneracy between \kp\ and \vsys, enabling physical interpretation of these offsets which is not possible with the data presented here.

The equilibrium retrieval results are very similar to the free chemistry results. Figure \ref{fig:pts} shows the median/maximum-likelihood $P-T$ profiles for both cases are consistent with the general spread of different profiles drawn from the free chemistry posterior. The retrieved spectra in Figure \ref{fig:specs} are all very similar, with minor differences in the continuum level. Figure \ref{fig:vmrs} demonstrates that the maximum-likelihood mixing profiles from both retrievals are nearly identical, confirming that in the high-SNR case our parametrized approach reproduces chemical equilibrium results. The most significant difference in the mixing profiles is for OH, where the free chemistry approach prefers a significantly lower abundance, which is consistent with the $K$-band OH opacity simply being too weak for the retrieval to independently prefer the presence of OH features. The equilibrium case prefers a slightly higher deep H$_2$O abundance compared to the free chemistry approach, as well as minor CO dissociation well above the $K$-band photosphere, but Figure \ref{fig:abunds} shows the [C/H] and [O/H] are still generally consistent between the two. However, the [O/H] in the equilibrium case is slightly elevated compared to the free retrieval, and this results in the free retrieval preferring higher C/O ratios than the equilibrium case, demonstrating that the apparent preference for high C/O ratios may still be an artifact of the chemical mixing treatment, and that minor changes in the mixing profile treatment can substantially change the inferred C/O ratio. 

\subsection{\knb}

It is not clear whether the data presented here provide a detection of \knb. Figure \ref{fig:kpvsysall} shows the retrieved maximum-likelihood model from the free chemistry retrieval is detected at a cross-correlation $\rm SNR = 6.7$, which appears to constitute a detection. While the cross-correlation trail is not clearly visible in any of the individual fiber time series shown in Figure \ref{fig:vtracks}, a weak peak is evident in the science fiber 4 time series after summing over all frames.

The maximum-likelihood spectrum from the free retrieval, however, is dominated by H$_2$O emission features, which is inconsistent with our expectations for the atmosphere of \knb\ and with the results presented from other targets. As the hottest of the observed UHJs, we expected H$_2$O dissociation to be most complete for \knb, leading to an absence of H$_2$O features and a spectrum dominated by features from the more chemically stable CO. In Figure \ref{fig:kpvsys_contours}, the CO-only template, which is similar to what we had expected the retrieval to prefer, does produce a cross-correlation peak of comparable strength to the H$_2$O-only model preferred by the retrieval, but at a slight velcity offset. This suggests that the atmospheric composition we expected for \knb, one dominated by CO and transition metal features, is consistent with the data, but that significant H$_2$O features may also be present with a wind-induced velocity offset. 

Repeating the retrieval without the $1.9-2.1 \mu\rm m$ orders leads to a preference for a CO-dominated spectrum, with an upper limit on the H$_2$O abundance, possible contribution from Fe features, and an overall weak detection. This is much more consistent with chemical equilibrium expectations for \knb, which may be hot enough to significantly dissociate even CO, resulting in relatively weak infrared emission features. The omitted orders are heavily impacted by telluric absorption from CO$_2$ and H$_2$O, which may be responsible for the apparent H$_2$O detection when these orders are included. Figure \ref{fig:vtracks} shows that the nominal planet reference frame during our observations was never particularly close to the stellar or telluric reference frames, making such contamination surprising.

\begin{figure*}
    \centering
    \includegraphics[width=0.45\linewidth]{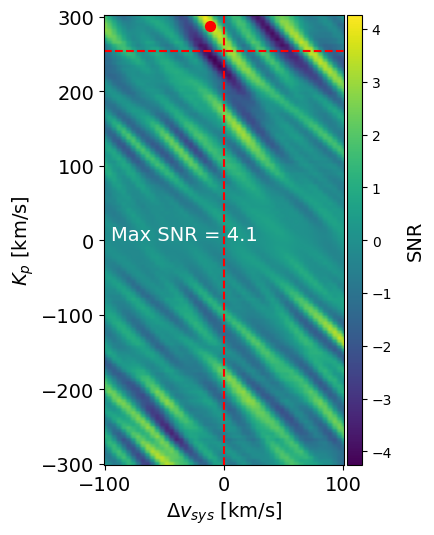}
    \includegraphics[width=0.45\linewidth]{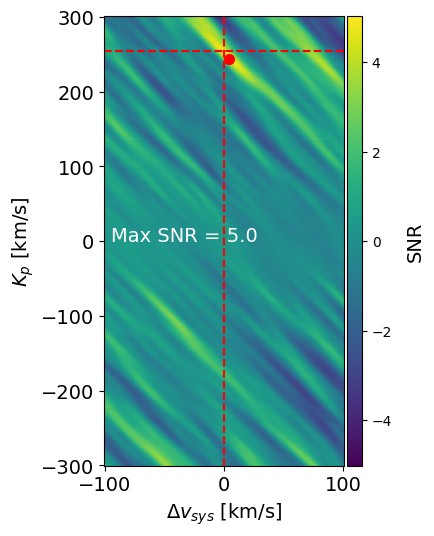}
    \caption{Left: \kpvsys\ map for \knb, similar to Figure \ref{fig:kpvsysall}, but using the $^{12}\rm CO$-only template used in Figure \ref{fig:kpvsys_contours}. The CO template is weakly detected at $\rm SNR\sim4$, consistent with expectations that the atmosphere of \knb\ is hot enough to dissociate nearly all molecules. Right: \kpvsys\ map for \knb, using the maximum-likelihood model from the chemical equilibrium retrieval, which is dominated by Fe emission features. This model is detected at $\rm SNR = 5$ and is consistent with the known ephemeris of \knb. An Fe-dominated atmosphere is consistent with results from optical emission spectroscopy \citep{zhang2026}.}
    \label{fig:k9co}
\end{figure*}

Figure \ref{fig:k9co} presents a \kpvsys\ map for the \knb\ observations similar to the maps presented in Figure \ref{fig:kpvsysall}, but using the planet model used for the \knb\ CO contours in Figure \ref{fig:kpvsys_contours} rather than the retrieved maximum-likelihood model. The CO-only model is weakly detected at $\rm SNR\sim4$, demonstrating that the weak CO feature-dominated spectrum is not clearly rejected.  The right panel of Figure \ref{fig:k9co} shows the \kpvsys\ map for the maximum-likelihood model for the equilibrium retrieval, which is recovered at $\rm SNR =5.0$. 

In contrast to the free chemistry results, the equilibrium retrieval is consistent with our prior expectations for the chemical composition of \knb, returning an atmosphere dominated by Fe and CO features. Figure \ref{fig:k9co} shows that the maximum-likelihood model from the equilibrium retrieval is detected at $\rm SNR =5$ and is much more consistent with the expected ephemeris of \knb\ than the free retrieval model presented in Figure \ref{fig:kpvsysall}. The retrieved equilibrium $P-T$ profiles shown in Figure \ref{fig:pts} are significantly hotter than the free chemistry profiles, with a thermal inversion deeper in the atmosphere and with smaller thermal contrast. The maximum-likelihood mixing profiles in the equilibrium case show an atmosphere dominated by Fe, with a relatively low CO abundance and H$_2$O strongly dissociated throughout the atmosphere. This composition is consistent with the strong Fe detection reported by \citet{zhang2026}. These results suggest the KPIC data provide a weak detection of Fe and CO in the atmosphere of \knb, but the all-order, free retrieval case is a clear demonstration of the potential failures of retrieval analysis in low-SNR detections.

\subsection{\wob}

The cross-correlation trail of the free retrieval maximum-likelihood model is weakly visible in Figure \ref{fig:vtracks}, where the planet detection appears to become significantly stronger as the planet approaches secondary eclipse. In the \kpvsys\ space, the maximum-likelihood \wob\ model is detected at $\rm SNR = 7.6$ (Figure \ref{fig:kpvsysall}), and Figure \ref{fig:kpvsys_contours} shows both the CO and H$_2$O templates produce significant features near the expected reference frame of \wob. Figure \ref{fig:vtracks} shows that the detection is dominated by the latter part of the time series as the planet approaches secondary eclipse. 

Figure \ref{fig:kpvsys_contours} suggests the detection is dominated by CO, with a weak H$_2$O contribution. This is consistent with the retrieved spectra plotted in Figure \ref{fig:specs}, which similar to \ktb\ and \wtb\ are dominated by CO features beyond 2.3 \micron\ with comparatively weak H$_2$O features at bluer wavelengths. The retrieved $P-T$ profile shows a thermal inversion beginning at $\sim1$ bar, and the retrieved mixing profiles plotted in Figure \ref{fig:vmrs} show significant H$_2$O dissociation at pressures lower than 100 mbar. Despite this, the retrieved deep-atmosphere H$_2$O abundances is still lower than the CO abundance, resulting in the preference for high C/O ratios seen in Figure \ref{fig:abunds}. Similar to \ktb, the retrieved [C/H] and [O/H] are $+0.7$ and $+0.8$, respectively, but the uncertainties are $\sim0.8$ dex and solar or even slightly subsolar compositions are not rejected. 

The retrieved velocity parameters for \wob\ are offset from the values based on the nominal ephemeris. The \wob\ observations have limited phase coverage, illustrated in Figure \ref{fig:phasecoverage}, and a significant portion of the data were obtained during secondary eclipse. The limited phase coverage leads to a particularly poor sensitivity to the exact \kp\ compared to the other targets, which can be seen in the extent of the planet feature in Figure \ref{fig:kpvsysall}. Figure \ref{fig:kpvsys_contours} shows that the maximum-likelihood, CO, and H$_2$O models are all consistent with each other and with the nominal reference frame of \wob, and we therefore do not believe the apparent offset is physically significant

The equilibrium retrieval returns similar $P-T$ profiles to the free retrieval, and the resulting spectra plotted in Figure \ref{fig:specs} are so similar that the equilibrium models cannot be easily distinguished from the corresponding free retrieval models. The retrieved mixing profiles in Figure \ref{fig:vmrs} are similarly consistent, with the exception of OH. As discussed for \wtb, the equilibrium model requires a significant OH abundance near where H$_2$O is dissociated, but the low OH opacity in the $K$-band is insufficient for the free retrieval to prefer a significant detection. Despite the consistence of the maximum-likelihood mixing profiles, the two chemical approaches differ in the retrieved C/O ratio, with the equilibrium retrieval preferring lower C/O than the free chemistry approach. This result is similar to the result for \wtb, and as in that case the retrieved C/O posteriors are wide enough as to be effectively unconstrained.

\subsection{\mob}

The 2D cross-correlation plot for \mob\ in Figure \ref{fig:vtracks} shows a weak, slightly blueshifted trace, which appears to be stronger closer to secondary eclipse. The maximum-likelihood model from the free retrieval is more clearly detected in the \kpvsys\ space, at $\rm SNR = 8.6$ (Figure \ref{fig:kpvsysall}). The CO-only template produces a slightly weaker figure near the reference frame of \mob\ in Figure \ref{fig:kpvsys_contours}, consistent with $\rm ^{12}CO$ being the dominant, but not sole, source of opacity. An overlapping but slightly offset feature from the H$_2$O template is evident in Figure \ref{fig:kpvsys_contours}, suggesting H$_2$O is opacity is significantly contributing to the detection of \mob. The OH contours in Figure \ref{fig:kpvsys_contours} are weak and significantly offset, and may not actually be associated with the planet.

The retrieved spectra show weak H$_2$O features at wavelengths shorter than 2.3 \micron, and strong CO features at redder wavelengths. Figure \ref{fig:vmrs} shows the the maximum-likelihood model prefers a very high H$_2$O abundance and a $P-T$ profile that leads to dissociation at $P\sim100\rm\ mbar$. These results are consistent with the retrieved posterior, which provides a weak constraint on the H$_2$O abundance. We note that omitting only 4 PCs for the analysis of \mob\ leads to a significantly different, strongly multi-model posterior, but that the 6 and 8 PC posteriors are consistent, suggesting that omitting only 4 PCs leaves significant residuals in the time series. 

The retrieved median and maximum-likelihood $P-T$ profiles plotted in Figure \ref{fig:pts} prefer a very strong inversion, with a thermal contrast of $\sim 3000\rm K$. Such a strong inversion over such a narrow pressure range may not be physical. Higher species abundances coupled with a decreased thermal gradient could preserve the strength of molecular features in the data, while also better matching the expected continuum level of \mob (see Figure \ref{fig:specs}). 

The retrieved velocity parameters prefer a \kp\ consistent with the assumed value and a slight blueshift in \vsys. Figure \ref{fig:kpvsys_contours} shows that the H$_2$O and CO templates differ somewhat in \kp\ and \vsys, but the overlap between the templates is significant, and the limited  phase coverage precludes interpreting this relatively minor offset in terms of global circulation. While a blueshift near secondary eclipse is surprising, once again the limited phase coverage and early spectral type of the host star combine to limit physical interpretation of this overall offset. 

The high H$_2$O abundance preferred by the free chemistry retrieval drives the C/O ratio to 0, and the carbon abundance to strongly sub-solar values, while the oxygen abundance is effectively unconstrained in Figure \ref{fig:abunds}. The resulting preference for extremely low C/O may be due at least in part to our treatment of vertical mixing. Figure \ref{fig:vmrs} shows that for the retrieved $P-T$ profile in chemical equilibrium CO is expected to begin dissociating at altitudes slightly above where H$_2$O begins to dissociate, but our models neglect this for a fixed-with-altitude CO VMR. If dissociation is indeed occurring, the retrieval will compensate for the reduced CO line strength in the data by reducing the forward model CO abundance, resulting in a bias to lower C/O ratios and low carbon abundances. 

The equilibrium retrieval is consistent with this scenario. The retrieved $P-T$ profile is roughly consistent with the free retrieval, as is the retrieved H$_2$O mixing profile. The CO mixing profile, in contrast, shows a preference for a higher abundance in the deep atmosphere but dissociation in the upper atmosphere, leading to a higher retrieved [C/H] in the equilibrium retrieval compared with the free retrieval, but the preference for low C/O persists in the equilibrium case. The preference of strong CO dissociation is surprising. \wtb\ and \wob\ are similar in equilibrium temperature to \mob, but those targets both prefer $P-T$ profiles that result in minimal CO dissociation. The retrieved planet models in Figure \ref{fig:specs} show a significantly greater continuum flux than the blackbody expectation, suggesting that the retrieved $P-T$ profile is over-estimating the planet temperature. A cooler $P-T$ profile would lead to minimal CO dissociation and weaker H$_2$O dissociation, which could then be compensated by a lower deep-atmosphere H$_2$O abundance compared to these retrievals in order to maintain the CO/H$_2$O feature strength ratio. This would lead to a weaker preference for low C/O ratios.

\subsection{\tob}

The retrieved \tob\ maximum-likelihood free chemistry model does not produce a clear trail in the 2D cross-correlation plot, but summing the time series reveals a weak, slightly redshifted peak (Figure \ref{fig:vtracks}). Figure \ref{fig:kpvsysall} shows the maximum-likelihood model is detected at $\rm SNR = 7.1$ in the \kpvsys\ space. Figure \ref{fig:kpvsys_contours} indicates that this detection is driven primarily by $^{12}\rm CO$, but the H$_2$O template does produce a weak, slightly offset feature in the \kpvsys\ contour map. While the retrieved median spectrum in Figure \ref{fig:specs} does not show H$_2$O features, these features are present in the maximum-likelihood model, suggesting these features may be preferred at marginal significance. 

Consistent with the \kpvsys\ maps, the posterior prefers a weak upper limit on the H$_2$O abundance. Larger H$_2$O abundances are more permitted at higher \kp, consistent with the offset contour in Figure \ref{fig:kpvsys_contours}. The lack of a constrained H$_2$O abundance leads to a preference for high C/O, shown in Figure \ref{fig:abunds}, and carbon and oxygen abundances compatible with solar with 1 dex uncertainties. The retrieved values of \kp\ and \vsys\ are consistent with the assumed ephemeris. While there is an offset between the CO and H$_2$O contours in Figure \ref{fig:kpvsys_contours}, similar to \mob, the weakness of the H$_2$O feature precludes further analysis. If this offset is a result of atmospheric circulation effects, this offset could significantly bias the inferred abundances.

Comparison of the free and equilibrium retrieval results shows similar discrepancies to those seen in \ktb. The equilibrium retrieval prefers a weaker inversion but overall higher temperature in the maximum-likelihood case, while the medians are similar. The \tob\ detection is comparatively weak, leading to a wider scatter in the retrieved posteriors compared to other targets. Figure \ref{fig:abunds} shows the equilibrium $P-T$ profile leads to significant H$_2$O dissociation at deep pressures and minor CO dissociation. Figure \ref{fig:abunds} shows that this leads to a preference for lower C/O in the equilibrium case, in contrast to the high C/O preferred by the free retrieval. Despite this disagreement, the retrieved spectra plotted in Figure \ref{fig:specs} are quite similar, dominated by CO features past 2.3 \micron\ with weak H$_2$O features at shorter wavelengths. This underscores that very similar finals spectra can result from very different C/O ratios depending on the $P-T$ and vertical mixing profiles of the atmosphere.

\section{Discussion}\label{sec:disc}

\subsection{Comparison to previous results}

\subsubsection{\ktb}

The results of this analysis are broadly consistent with the analysis of these data previously presented in \citet{finnerty2025k20}, to which we refer the reader for a complete discussion. Since that work, \citet{chachan2026} reported {\it HST} transmission observations which suggest a supersolar oxygen abundance and subsolar refractory abundance, consistent with the analysis presented here and in \citet{finnerty2025k20}. In contrast with \citet{chachan2026}, \citet{bonidie2026} report refractory abundances from high-resolution optical emission spectroscopy form LBT/PEPSI which are consistent with chemical equilibrium predictions for a $10-30\times$ solar refractory enrichment. However, \citet{bonidie2026} assumed fixed-with-altitude abundance profiles, while the equilibrium chemistry model predicts significant vertical abundance variations; this could account for the apparent discrepancy with other analyses of \ktb. 

\subsubsection{\wtb}

The results for \wtb\ are generally consistent with those reported in \citet{finnerty2023}, to which we refer the reader for additional details. Since \citet{finnerty2023}, both \citet{nugroho2021} and \citet{wright2023} demonstrated detection of individual vibrational bands of OH using data from Subaru/IRD in the $0.97-1.75\rm \mu m$ bandpass. \citet{choi2025} reported detection of CO, H$_2$O, and OH emission features from commissioning observations of the Gemini-N/IGRINS2 spectrograph, which covers the entirety of the $H$ and $K$ bands, and analysis of low-resolution secondary eclipse spectroscopy reported evidence for H$_2$O and CO emission features \citep{zou2026}. These results clearly demonstrate that significant OH features are present in the infrared emission spectrum of \wtb, and that our non-detection in the $K$-band data is most likely an artifact of the low OH opacity in this bandpass and the limited pressure range over which the OH abundance is significant. 

\subsubsection{\knb}

As one of the hottest known exoplanets, \knb\ has been a frequent observational target, especially in the visible. Optical high-resolution transmission \citep{hoeijmakers2018, yan2019, hoeijmakers2019, borsato2023, darpa2024} and emission \citep{pino2020, kasper2021, zhang2026} features have confirmed a wide range of atomic and ionic species in the atmosphere of \knb, while generally preferring a molecule-free atmosphere. Observations in the UV have found evidence for escaping Mg and Fe ions \citep{baldwin2026}.

In the infrared, observations with {\it HST} have provided evidence for strong H$^-$ absorption in the atmosphere of \knb\ \citep{jacobs2022}. Post-eclipse $K$-band emission observations with CFHT/SPIROU detected Fe and OH features in emission, and tentative evidence for CO emission \citep{yang2025}. The retrieved free chemistry spectrum presented in Table \ref{tab:priors} is inconsistent with these results, providing further reason to view the free retrieval results with skepticism. Free retrievals omitting the bluest three orders return a posterior consistent with the planet properties reported in the existing literature, and Figure \ref{fig:k9co} demonstrates a weak detection of CO emission, suggesting our data are consistent with a weak detection of \knb\ with the expected composition, and that the H$_2$O-dominated spectrum preferred by the retrieval may be an artifact of uncorrected systematics on top of weak planetary spectral features. This interpretation is strengthened by the equilibrium retrieval results, which suggest an atmosphere dominated by Fe features with weak CO, consistent with \citet{pino2020} and \citet{zhang2026}.The origin of the H$_2$O signal in the free retrieval is unclear, given that the planet radial velocity track is well-separated from both the stellar and telluric reference frames. This result is a reminder that retrieval results could be interpreted cautiously and with consideration of known physical and chemical processes. 

\knb\ poses an additional potential challenge to retrieval analysis in the form of extreme winds. \citet{pai2022}, for example, reported day-to-night wind speeds up to 10 \kms\ in high resolution optical transmission observations, varying with a range of $5-8$ \kms\ on week-to-year timescales. Analysis of optical transmission data focused on Fe lines found evidence for a strong day-to-night wind, and that including non-local thermodynamic equilibrium effects significantly improved model fits for \knb~\citep{Stangret2024}. \citet{zhang2026} found that optical emission from the dayside of \knb\ becomes detectable shortly after the end of the planetary transit, and that the combination of winds and rotation introduces a time-dependent Doppler shift of up to 15 \kms\ from the nominal reference frame, driven by a 11.6 \kms\ supersonic wind. These results suggest features in the \kpvsys\ map which appear to be significantly offset from the nominal reference frame of \knb\ may still be attributable to the planetary atmosphere, and that there may be detectable kinematic offsets between different atomic/molecular species in \knb, which our retrieval code does not account for. The offset between the H$_2$O and CO contours in Figure \ref{fig:kpvsys_contours} is consistent with the previously reported wind speeds. In the absence of a clear explanation for the origin of the H$_2$O signal, we cannot rule out that the H$_2$O signal is associated with \knb, but is separated from the CO reference frame due to global circulation.  

\subsubsection{\wob}

\wob\ is more than one magnitude brighter in $K$ than the next-brightest UHJs observed here, and is accessible to observatories in both the northern and southern hemispheres. This has made \wob\ an attractive target for high-resolution spectroscopy since its initial discovery by \citet{anderson2018}. In dayside emission, optical observations quickly confirmed the presence of Fe features in emission, indicating the presence of a thermal inversion \citep{yan2020}, and subsequent observations in the near infrared detected CO emission features \citep{yan2022}. Observations of dayside emission using VLT/CRIRES+ in the $K$-band resulted in detections of CO and Fe, with a 6 \kms\ velocity offset between the CO and Fe cross-correlation features \citep{lesjak2023}, and Fe emission was also reported from Gemini/GHOST observations \citep{deibert2024}. Using Gemini-S/IGRINS emission observations covering the infrared $H$ and $K$ bands, \citet{sanchez2025} reported detections of neutral Fe, Mg, and Si, as well as H$_2$O, CO, and OH, measuring an enhanced refractory/volatile ratio relative to solar. In the infrared $J$-band, \citep{vansluijs2025} reported detection of Fe emission and tentative detections of Mg and Si features.

In transmission, optical studies have reported detections of Fe, Fe$^+$, and Ti \citep{stangret2022}, as well as the first three lines of the hydrogen Balmer series, which indicate significant atmospheric mass loss \citep{yan2021}. \citet{gandhi2023} measured abundances for 8 transition metals in \wob\ from transmission observations, reporting a slightly sub-solar Fe abundance. \citet{prinoth2023} reported detections of a number of species in \wob, including TiO, and evidence for differences in the spatial distribution of the detected species. \citet{valulato2025} reported detection of Fe in optical transmission in HARPS data, but transmission observations from NIRPS covering $1.0-1.8\rm \ \mu m$ failed to detect \wob, likely as a result of the dramatic increase in $H^{-}$ opacity near $1.4\rm\ \mu m$. 

\citet{sanchez2022} is the most similar analysis to what we present here, and our findings are consistent with what is reported in that work. Compared to the $K$-band observations presented here, the full $H+K$ coverage of IGRINS encompasses many more opacity features from H$_2$O, OH, Fe, and other transmission metals, explaining our tentative/non-detections of these species. 
\subsubsection{\mob}

While \mob\ is relatively bright and accessible from both northern and southern hemispheres, high-resolution transmission studies have been limited by an unfortunately close alignment between the planet cross-correlation track during transit and the cross-correlation feature produced by the Rossiter-McLaughlin effect \citep{casayasbarris2022}. Nevertheless, \citet{scandariato2023} reported detections of neutral Fe, Cr, and Ti in the atmosphere of \mob\ from LBT/PEPSI optical spectra, while
infrared emission observations of \mob\ have consistently detected molecular features from CO and H$_2$O in emission \citep{holmberg2022, ramkumar2023, ramkumar2025}. \citet{ramkumar2023} also reported a detection of Fe in emission, and obtained C/O and metallicity values consistent with solar. \citet{ramkumar2025} subsequently established consistent planetary properties on a timescale of at least two years. In optical emission, \citet{guo2024} reported detections of Fe and Ti. 

The strong CO and weak H$_2$O detections from our analysis are consistent with expectations. Both the Keck/KPIC observations presented here and the VLT/CRIRES+ observation used in \citet{ramkumar2023} cover the $K$ band, with significant gaps between orders. Both sets of observations were similarly obtained prior to secondary eclipse, limiting the potential impact of phase dependent changes when comparing analyses across datasets. The different spectral resolution of CRIRES+ and KPIC is the most significant difference between the two observations. While KPIC has $R = \lambda/\Delta\lambda\sim35,000$, for CRIRES+ $R = \lambda/\Delta\lambda\sim100,000$. This $\sim3\times$ increase in spectral resolution increases the number of resolution elements crossed by the planetary atmosphere over the observations, reducing the impact of self-division of the planetary spectrum during detrending. This may improve the sensitivity to weak-but-ubiquitous H$_2$O lines and significantly improve H$_2$O abundance constraints from CRIRES+ compared to the weak constraints obtained with KPIC data. In the future, using the same pipeline to perform retrievals on both the KPIC and CRIRES+ observations of \mob\ may be useful for clarifying the impact of spectral resolution on the results of HRCCS analysis.

\subsubsection{\tob}

\tob\ was discovered in 2021 by \citet{cabot2021}, and is the least studied of the six objects presented here, with no previously reported detections in the infrared or in emission. In the initial discovery paper, \citet{cabot2021} reported a detection of neutral iron in optical transmission observations, while \citet{basinger2025} subsequently reported detections of Fe and Fe$^+$ in optical transmission observations from LBT/PEPSI, along with tentative detections of neutral Cr and Ni. \citet{simonnin2024} reported detections of 14 species from optical transmission observations with MAROON-X, with significant velocity offsets between species biasing attempts to retrieve abundances. 

While these results from optical transmission observations of \tob\ do not directly inform analysis or interpretation of the infrared emission spectrum of \tob, both the infrared analysis presented here and the previous optical results are consistent with \tob\ as a typical UHJ in the $2000\ \rm K < T_{eq} < 3000\ K$ regime. A strong thermal inversion is present, the infrared spectrum is dominated by emission features from CO, with weak features from H$_2$O strongly impacted by dissociation, and the optical spectrum is dominated by emission features from neutral and singly ionized transition metals. This description applies to all of the objects observed in this work, with the possible exception of the substantially hotter \knb. Obtaining precise abundance constraints is complicated by the effects of dissociation on the vertical mixing profiles and circulation-driven kinematic offsets between species, which appear to be ubiquitous in UHJs generally.

\subsection{Planet continuum level}

The maximum-likelihood and median spectra from the free retrievals plotted in Figure \ref{fig:specs} differ from the expected $F_p/F_s$ assuming both planet and star are blackbodies at their respective equilibrium and effective temperatures. For \ktb, \wtb, and \tob\, the continuum level is less than the blackbody by a factor $\sim$2--4, while for \wob\ and \mob\ the retrieved continuum level is greater than the blackbody value by a similar factor, and for \knb\ the median and maximum-likelihood models straddle the expected continuum. The equilibrium retrievals generally return similar continuum levels to the free retrievals, with the exception of the maximum-likelihood equilibrium models for \ktb\ and \tob, both of which show significantly higher continuua Both the retrieved and blackbody continuua are nearly flat over the entire observed bandpass, changing by $\lesssim$20$\%$. This poses a particular challenge for HRCCS due to the loss of continuum information from the median division step in data processing. Over wide bandpasses, e.g. combining $H$+$K$ or $K$+$L$ data for retrieval, the change in the continuum $F_p/F_s$ over the observed wavelength range is sufficient to recover the continuum level during retrieval \citep[e.g.][]{line2021, finnerty2025hd209}. For $K$-band data alone, the nearly flat $F_p/F_s$ continuum results in the retrieval being effectively insensitive to a multiplicative rescaling of the template spectrum. 

The relative insensitivity to continuum level has important implications for absolute temperatures and molecular abundances. HRCCS is sensitive to the strength of planetary lines relative to the stellar continuum. Planetary line strengths are in turn set by both the molecular abundance of the relevant species and by the thermal contrast of the $P-T$ profile, while the absolute temperature of the $P-T$ profile sets the planet continuum level. This impact of this on our ability to constrain the $P-T$ profile can be seen in the $P-T$ profiles drawn from the posterior plotted in Figure \ref{fig:pts}. While the absolute temperature of the different drawn profiles show a scatter of up to 1000 K for a single target, the change in temperature between the top and bottom of the atmosphere appears to be somewhat better constrained.

The impact of the continuum insensitivity is not isolated to the $P-T$ profile. Without a well defined continuum level, a degeneracy can arise between $P-T$ profile contrast and molecular abundances. Large $P-T$ contrasts with low molecular abundances can produce line features relative to the total star+planet continuum level of similar strength as lower $P-T$ contrasts with high abundances. Further degeneracies may be introduced by dissociation, as the molecular abundance gradient can also lead to the formation of spectral features even for an isothermal $P-T$ profile \citep{finnerty2025k20}. These challenges are reflected in the broad \cpo\ posteriors shown in Figure \ref{fig:abunds}, which we discuss in detail below. 

For all of the targets reported here, there are discrepancies in the stellar temperature and radius on the order of $5-10\%$ between different published references. For example, for MASCARA-1~A\citet{hooton2022} reports $\rm T_{eff} = 7490\pm150$ K and $\rm R_s = 2.08\pm02\ R_\odot$, in good agreement with the values reported in \citet{talens2018m1}, while the \citet{exofop3} temperature is $7800\pm100$ K and radius is $\rm R_s = 2.02\pm0.06\ R_\odot$. For WASP-33~A, the reported effective temperatures are in good agreement, but the reported stellar radii range from $1.444\pm0.034\rm\ R_\odot$ \citep{collier2010} to $1.602\pm0.06\rm\ R_\odot$ \citep{exofop3}. TOI-1518~A is an extreme case, with a 2500 K difference between effective temperatures reported in the \citet{exofop3} and \citet{cabot2021}. This uncertainty in the stellar parameters propagates to the planet radius, which is measured relative to the host star radius, and to the planetary equilibrium temperature, which depends on the host star temperature, radius, and mass via the the semi-major axis. For WASP-33~b, the reported radii range from $1.497\pm0.05\rm\ R_J$ \citep{collier2010} to $1.63\pm0.06\rm\ R_J$, corresponding to a $\sim20\%$ change in the projected surface area of the planet. The accumulation of these potential sources of error could be responsible for a factor $\sim1-2$ discrepancy in the expected $F_p/F_s$. Furthermore, we did not consider the effects of limb or gravity darkening in detail, which could lead to discrepancies between the model stellar flux and the actual stellar flux incident on the targets.  

As discussed in \S \ref{sec:res}, we have not included H$^-$ opacity in the atmospheric retrievals. \citet{finnerty2023} showed that H$^-$ results in a uniform drop in the strength of planet lines relative to the continuum of $\sim10\%$ for K-band observations \wtb. H$^-$ likely has a stronger impact on the hotter UHJs, particularly \knb~\citep{jacobs2022}, potentially forcing the retrieval to lower abundances and/or thermal contrasts in order to match the reduction in line strength caused by H$^-$ opacity. This may also impact the continuum as a result of the previously discussed coupling between line strengths, abundances, and the $P-T$ profile. Better constraints on H$^-$ opacity in HRCCS analysis may be possible by including $J$ and $H$ band spectra, as the H$^-$ opacity shows a sharp change at 1.4 \micron\ corresponding to the change from bound-free to free-free absorption \citep{lenzuni1991, parmentier2018}. 

\subsection{Atmospheric Circulation}

As \citet{finnerty2025k20} discusses, obtaining wind speeds from emission observations over a limited phase range is challenging due ephemeris and radial velocity uncertainties, which can lead to Doppler shifts comparable to expected wind speeds in UHJs. In principle, a global day-to-night wind would lead to a net redshift for observations near secondary eclipse. Additionally accounting for planetary rotation (or a super-rotating jet) will complicate this pattern, assuming that the dayside/afternoon longitudes generally dominate the emission spectrum. Just before and after eclipse, rotation/jets would lead to a slight redshift compared to the Keplerian ephemeris, while closer to quadrature the effect would be a slight blueshift prior to eclipse and a slight redshift post eclipse. This pattern of apparent velocity offsets has been recently observed in \knb\ \citep{zhang2026}. 

Our retrievals reveal no clear kinematic trends indicative of a common global circulation pattern. Using the retrieved median and maximum-likelihood parameters from the free retrievals, we calculated the difference in velocity between the assumed ephemeris and the posterior. The retrieved \kp\ and \vsys\ from the chemical equilibrium retrievals are consistent with the values obtained from the free retrieval for all targets except \knb, and using the equilibrium values would not significantly alter this discussion. For \ktb, the free retrieval median parameters give a shift of $-2.1$ \kms\ at the start of the time series, decreasing to $0.0$ \kms\ by the end, while the maximum-likelihood shift changes from $-1.7$ \kms\ to $-1.0$ \kms. For \wtb, the median parameters give a shift from $-8.3$ \kms\ to $-0.1$ \kms\ over the time series, while maximum-likelihood parameters give a shift of $-7.9$ \kms\ to $-0.4$ \kms. For \knb, the median shift starts at $+0.6$ \kms\ and ends at $-7.1$ \kms, while the maximum-likelihood shift goes from $+0.7$ \kms\ to $-11.9$ \kms. The median parameters from the \wob\ retrieval give a shift from $-5.0$ \kms\ to $+2.0$ \kms, while the maximum-likelihood values go from $6.2$ \kms\ to $+3.2$ \kms. The \mob\ median parameters give a shift from $-9.1$ \kms\ and $-9.6$ \kms, while the maximum-likelihood values give $-9.1$ \kms\ throughout. The \tob\ median shift is $+4.2$ \kms\ to $+6.5$ \kms, while the maximum-likelihood shift is $+0.5$ \kms\ to $+7.1$ \kms. Both the absolute value of these offsets and the change in the offset over the time series are generally similar to or smaller than than the 9 \kms\ instrument resolution of KPIC/NIRSPEC, and within the plausible range for ephemeris uncertainties. Furthermore, there is no clear correlation between the velocity offset and the observed orbital phase, which we would expect to see if the retrieved offsets were primarily due to a common UHJ circulation pattern. 

While we fit for a rotational broadening kernel, the LSF fitting procedure involves systematic uncertainties at the $\sim10\%$ level, discussed in \citet{Finnerty2022, finnerty2025hd143, finnerty2025k20}. While the retrievals consistently provide bounded constraints on the rotational broadening, the physical significance of these values is questionable. The LSF is only measured to $\sim10\%$ precision, which corresponds to a systematic uncertainty in the broadening of approximately 4.5 \kms. This value is similar to the expected wind speeds and rotation rates in UHJs. As a result, apparently plausible constraints in the broadening parameters may not actually be measuring broadening of planetary features, but rather is accounting for errors in measurement of the instrumental LSF. These results demonstrate HRCCS retrievals are sensitive to subtle changes in the shapes of observed spectral lines at scales comparable to the instrument resolution, but obtaining accurate measurements of rotational broadening will require more precise LSF calibration than was available with KPIC. Such calibration could be provided in the future by laser frequency combs. 

Our retrievals use a 1D radiative transfer model, which assumes there is no kinematic offset between different species. This assumption has been shown to be false in optical observations of UHJs \citep[e.g.][]{kesseli2022}. \citet{wardenier2024} found a $\sim3$ \kms\ velocity offset between CO and H$_2$O features in transmission observations of the UHJ WASP-121~b, and that the cross-correlation peaks for H$_2$O and CO vary differently with orbital phase. These findings suggest that the distinct spatial distributions of H$_2$O, OH, and CO resulting from molecular dissociation is leading to differences in the radial velocity of the corresponding emission features at the $\sim3$ \kms\ level. While the spectral resolution of the KPIC observations is 9 \kms, this may still lead to subtle model mismatch, biasing inferred abundances particularly for H$_2$O and OH due to their substantially weaker feature strength compared with CO. In the case of \ktb, the peaks in the \kpvsys\ maps for the H$_2$O and CO models shown in Figure \ref{fig:kpvsys_contours} are not significantly offset, but the weak OH contour is slightly offset. \wtb\ and \wob\ show minor offsets in the contour centers for H$_2$O and CO, while \mob and \tob\ show more significant offsets between species, suggesting this phenomenon is ubiquitous in UHJs. Simulation work is needed to assess the effect such kinematic differences will have on atmospheric retrievals, and development of radiative transfer tools which can account for these offsets will likely be necessary for accurate analysis of data from instruments with large bandpasses and $R\sim100,000$, such as Keck/HISPEC \citep{hispec}. 

\subsection{Atmospheric composition}\label{ssec:coz}

\begin{figure*}
    \centering
    \includegraphics[width=0.95\linewidth]{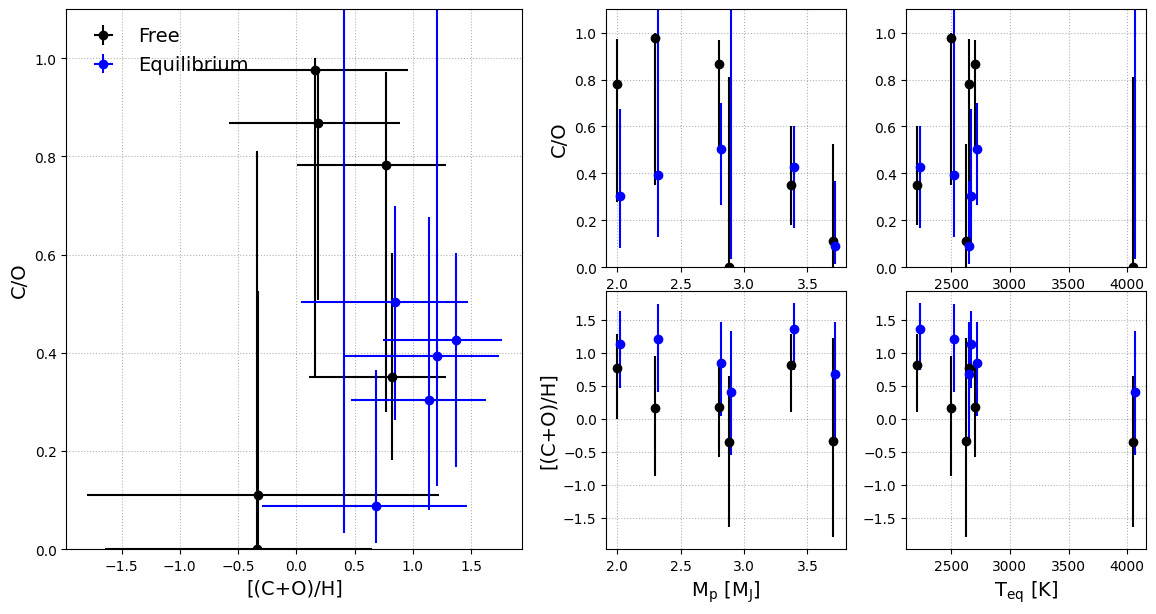}
    \caption{C/O versus volatile abundance (left), C/O versus planet mass and equilibrium temperature (upper row), and volatile abundance versus planet mass and equilibrium temperature (lower row). The free retrieval values are shown in black and the equilibrium retrievals in blue. The equilibrium values have been slightly offset in mass and temperature for clarity. None of these parameters appear to be significantly correlated. Compared to the free retrievals, the equilibrium retrievals tend to prefer slightly lower C/O and higher volatile abundances, leading to weak clustering in the low-C/O, high-volatile region of the left plot which is not seen in the free retrieval case. The uncertainties in these parameters are too large to definitively establish or reject any correlations between these parameters. }
    \label{fig:bulk}
\end{figure*}

The constructed C/O and metallicity posteriors in Figure \ref{fig:abunds} show that we cannot meaningfully constrain either parameter from the presented $K$ band data, which is readily apparent in the multi-object distribution plotted in Figure \ref{fig:bulk}. Even when our retrievals provide simultaneous constraints on the abundances of CO and H$_2$O, the dominant carbon and oxygen carriers in UHJ atmospheres, degeneracies between species abundances and $P-T$ parameters controlling the vertical mixing profile lead to poor constraints on C/O. While an individual retrieval can sometimes provide an upper or lower limit on the C/O ratio, comparing the equilibrium and free retrievals for the same target often reveals a significant discrepancy in the resulting posteriors. These differences occur despite Figure \ref{fig:vmrs} demonstrating that our free retrieval approach can closely replicate the vertical mixing profiles of equilibrium chemistry models.

While the retrieved C/O ratios and metallicities show a wide scatter, the best-fit spectra plotted in Figure \ref{fig:specs} are consistent, with the exception of \knb. This suggests that the spectra of UHJs in the $\rm 2200\ K < T_{eq} < 3000\ K$ range are generally dominated by emission from CO, with weak H$_2$O features significantly impacted by molecular dissociation. Similar spectra have been reported for other UHJs in this temperature range, in particular WASP-18~b \cite{brogi2023} and WASP-121~b \citep{smith2024}. 

Our results demonstrate that assuming chemical equilibrium may not be sufficient to accurately constrain bulk composition in UHJs. Previously, \citet{brogi2023} demonstrated that fixed-with-altitude retrievals will substantially over-estimate C/O ratios in UHJs compared to assuming equilibrium chemistry due to dissociation. In practice, using mixing profiles from a chemical equilibrium code is assuming both that the atmosphere is in chemical equilibrium and that the chemistry code being used to calculate vertical abundance profiles is accurate. Our results demonstrate that very different C/O ratios can nevertheless result in very similar final spectra due to changes in the $P-T$ profile and other parameters. This raises the possibility that minor differences in how the chemical equilibrium assumption is implemented could significantly impact the retrieved C/O ratio.

Figure \ref{fig:bulk} shows that compared to the free retrieval, the chemical equilibrium retrievals tend to prefer slightly higher volatile abundances and lower C/O ratios. Neither the C/O ratio nor the volatile abundance appears to correlate with planet mass or equilibrium temperature for either mixing profile treatment. Assuming equilibrium appears to reduce the scatter in the C/O vs volatile abundance space compared with the free retrieval results, with the equilibrium models tending to prefer solar-to-subsolar C/O and solar-to-supersolar volatile abundances. More precise constraints on both parameters are necessary to conclude these properties are correlated in the UHJ population.

While in cooler hot Jupiters, $K$ band observations are sufficient for measuring the C/O ratio \citep{finnerty2024}, these results demonstrate that this is not the case for UHJs. \citet{brogi2023} retrieved C/O and metallicity for the UHJ WASP-18~b using Gemini/IGRINS observations over the $H$ and $K$ bands with similar spectral resolution to our KPIC observations, which suggests that the additional H$_2$O and OH opacity features in the $H$ band are critical to obtaining such constraints for UHJs. The $L$ band also has extensive H$_2$O and OH features, potentially providing another avenue for constraining UHJ composition in conjunction with CO measurements from the $K$ band.  

While our observations are insufficient to constraint C/O, combining the KPIC data with optical observations may offer some insight into the refractory/volatile ratio for these targets through constraints on transition metal abundances. The wide wavelength coverage of such a joint analysis may also improve constraints on the $P-T$ profile, breaking the degeneracy between the $P-T$ profile and species abundances. Comparing CO to a refractory species with a similarly uniform global distribution may enable refractory/volatile constraints and formation inferences while sidestepping issues posed by non-homogeneous spatial distributions of other species.

\subsection{CO isotopologues}

We include opacity from both $^{12}\rm CO$ and $^{13}\rm CO$ in the free retrievals. In contrast with \citet{line2021, finnerty2023, finnerty2024, finnerty2025k20}, we fit the $^{13}\rm CO$ abdundance in the same way as the other molecular abundances, rather than fitting for the $^{12}\rm CO/^{13} CO$ isotopologue ratio. For \ktb, this does not appear to impact the results, with both this analysis and \citet{finnerty2025k20} preferring upper limits on $^{13}\rm CO$. In contrast, the retrievals presented here prefer an upper limit on $^{13}\rm CO$ for \wtb, while \citet{finnerty2023} weakly preferred a near-solar isotopologue ratio, though with a long tail to lower values.

For all targets except \knb, our retrievals prefer an upper limit on the $^{13}\rm CO$ abundance, which results in upper limits on the isotopologue ratio consistent with or lower than the solar system value. The 95\% upper limits from our retrievals are $\log (^{12}\rm CO/^{13}CO) > 1.8$ for \ktb, $\log (^{12}\rm CO/^{13}CO) > 1.9$ for \wtb, $\log (^{12}\rm CO/^{13}CO) > 2.9$ for \wob, $\log (^{12}\rm CO/^{13}CO) > 0.8$ for \mob, and $\log (^{12}\rm CO/^{13}CO) > 1.4$ for \tob. For comparison, the solar system value is $\log (\rm ^{12}C/^{13}C) = 1.9$ \citep{milam2005}, and the local ISM value is approximately $\log (\rm ^{12}C/^{13}C) = 1.8$ \citep{woods2009}, both of which are similar to the obtained upper limits.

\knb\ is the only target for which we do not obtain a constraint on the $^{12}\rm CO$ abundance from the free retrieval. While the retrieval weakly prefers a large $\rm ^{13}CO$ abundance, the marginalized posterior for $\rm ^{12}CO$ spans the full range of the prior, indicating its abundance is not well constrained. As a result, we do not obtain a constraint on the isotopologue ratio for \knb, with 5\% and 95\% limits on the log of the isotope ratio of $-6.2$ and $+5.5$, respectively. 

It is unclear how a spectrum could arise with detectable $^{13}\rm CO$ features but not show similar features from $^{12}\rm CO$. While fractionation processes can lead to increased relative abundances of $^{13}\rm C$ isotopologues in certain regions of protoplanetary disks, these processes do not result in $\rm ^{12}C/^{13}C < 1$ \citep{woods2009}. The higher mass of $^{13}\rm CO$ leads to a slightly stronger bond than in $^{12}\rm CO$, which could push dissociation slightly higher into the atmosphere, but this slight difference in dissociation pressure leading to $^{13}\rm CO$ features dominating over spectral features from much more abundant $\rm ^{12}CO$ is implausible. The unexpected features of the \knb\ retrieval are more plausibly explained as artifacts of a generally weak detection due to overall weaker molecular features than those seen in the cooler UHJs.

\section{Summary and Conclusions}\label{sec:conc}

We present a uniform analysis of 6 ultra-hot Jupiters observed with Keck/KPIC in $K$-band thermal emission, four of which have not been published previously. We perform a set of retrievals to fit for the $P-T$ profiles, abundances, and kinematic offsets, taking a parameterized approach to the H$_2$O and OH vertical mixing profiles to account for dissociation, and another set of retrievals assuming chemical equilibrium. All six targets are detected using cross-correlation. Five of the six objects show $K$-band spectra dominated by CO emission features beyond 2.3 \micron, with comparatively weak H$_2$O features evident at shorter wavelengths. Only \ktb clearly prefers the presence of OH features, but the OH abundance constraints are limited by the low opacity of this species in the $K$ band. The five CO-dominated objects all have equilibrium temperatures in the $\rm\ 2200\ K < T_{eq} <  3000\ K$ range, suggesting that planets in this regime generally have $K$-band spectra dominated by CO emission, with weak H$_2$O and/or OH features strongly impacted by thermal dissociation of these species. The only observed target not to fit this picture is \knb, which has $\rm T_{eq} \sim 4000\ K$. While the free retrieval for \knb\ suggests significant contamination of the spectrum, the equilibrium retrieval prefers an atmosphere dominated by Fe emission features, consistent with previous results in the literature and suggesting \knb\ is hot enough to significantly dissociate even CO.

Our efforts to constrain C/O ratios were limited by the impact of thermal dissociation and the sensitivity of the C/O ratio inference to the exact vertical mixing treatment. The weak $K$-band H$_2$O features can be fit either by a low H$_2$O abundance and high-altitude dissociation, or by high H$_2$O abundances and dissociation lower in the atmosphere. While the latter scenario is expected under chemical equilibrium, our free-retrieval approach does not consistently prefer this solution, most notably for \ktb\ and \tob. This is likely due to a combination of signal-to-noise and the relatively narrow bandpass of the $K$-band-only observations introducing degeneracies between the $P-T$ profile and molecular abundances.

Our results provide several avenues for future improvements for retrieval pipelines to better analyze observations of UHJs:

\begin{itemize}
    \item Equilibrium retrievals or semi-analytical approaches to vertical mixing profiles offer more consistent constraints on C/O with current data, but distinguishing between dissociation pressure and abundance remains challenging, and seemingly minor differences in mixing profile treatments may significantly impact results. While high-SNR observations can enable free retrieval of dissociation pressures in UHJs, success requires very high quality data, and introduces significant degeneracies between parameters. Combining optical and infrared observations may help to break degeneracies in the $P-T$ profile and improve constraints.
    \item Spectral grasp makes a significant difference to the ability to constrain H$_2$O and OH abundances in UHJs. Observations covering the full $H$ and $K$ bands at similar spectral resolution to the $K$-band data from KPIC provide significantly better constraints, particularly for OH.  
    \item Abundance retrievals for UHJs need to account for kinematic offsets between species. Our single-molecule \kpvsys\ maps routinely show features at $\rm SNR\sim3$  slightly offset from the peak of the maximum-likelihood model. Retrievals which use a single reference frame for all opacity sources may return biased abundances if e.g. H$_2$O is in a slightly different reference frame from CO. The size of opacity tables in computer memory will make this challenging to efficiently implement.
    \item Uncertainties in stellar radial velocities and planetary ephemerides limits the use of short-baseline emission observations to constrain wind speeds and circulation patterns. While the retrievals often prefer offsets from the assumed planet velocities, the origin of these offsets cannot be clearly established. Observations with a wider phase coverage may offer a better ability to separate Keplerian motion from circulation effects.
\end{itemize}

Upcoming instruments and facilities will dramatically improve the sensitivity of ground-based high-resolution spectroscopy in the next decade. For UHJs, detection of individual emission lines will be within reach for the first time. Understanding and addressing the limitations of current analysis tools in these extreme targets will lay the groundwork to take full advantage of this unprecedented sensitivity. 

\begin{acknowledgments}
We thank the anonymous referee whose detailed and insightful comments improved the quality of this paper. 
L. F. was a member of UAW local 4811 when most of this work was performed, and is now a member of UM-PRO. L.F. acknowledges the support of the W.M. Keck Foundation, which also supports development of the KPIC facility data reduction pipeline. The contributed Hoffman2 computing node used for this work was supported by the Heising-Simons Foundation grant \#2020-1821. Funding for KPIC has been provided by the California Institute of Technology, the Jet Propulsion Laboratory, the Heising-Simons Foundation (grants \#2015-129, \#2017-318, \#2019-1312, \#2023-4597, \#2023-4598), the Simons Foundation (through the Caltech Center for Comparative Planetary Evolution), and the NSF under grant AST-1611623. D.E. acknowledges support from the NASA Future Investigators in NASA Earth and Space Science and Technology (FINESST) fellowship under award \#80NSSC19K1423, as well as support from the Keck Visiting Scholars Program (KVSP) to install the Phase II upgrades for KPIC. J.W.X is thankful for support from the Heising-Simons Foundation 51 Pegasi b Fellowship (grant \#2025-5887). 

This work used computational and storage services associated with the Hoffman2 Shared Cluster provided by UCLA Institute for Digital Research and Education’s Research Technology Group. L.F. thanks Briley Lewis for her helpful guide to using Hoffman2, and Paul Molli\`ere for his assistance in adding additional opacities to petitRADTRANS. 

The data presented herein were obtained at the W. M. Keck Observatory, which is operated as a scientific partnership among the California Institute of Technology, the University of California and the National Aeronautics and Space Administration. The Observatory was made possible by the generous financial support of the W. M. Keck Foundation. W. M. Keck Observatory access was supported by Northwestern University and the Center for Interdisciplinary Exploration and Research in Astrophysics (CIERA). The authors wish to recognize and acknowledge the very significant cultural role and reverence that the summit of Mauna Kea has always had within the indigenous Hawaiian community.  We are most fortunate to have the opportunity to conduct observations from this mountain. 

This research has made use of the NASA Exoplanet Archiv and Exoplanet Follow-up Observation Program (ExoFOP; DOI: 10.26134/ExoFOP5) website, which is operated by the California Institute of Technology, under contract with the National Aeronautics and Space Administration under the Exoplanet Exploration Program.

\end{acknowledgments}

\vspace{5mm}
\facilities{Keck:II(NIRSPEC/KPIC)}

\software{astropy \citep{astropy:2013, astropy:2018},  
          \texttt{corner} \citep{corner},
          \petit\ \citep{prt:2019, prt:2020}}

\appendix
\section{Corner Plots}\label{app:corner}

\begin{figure}
    \centering
    \includegraphics[width=1.0\linewidth]{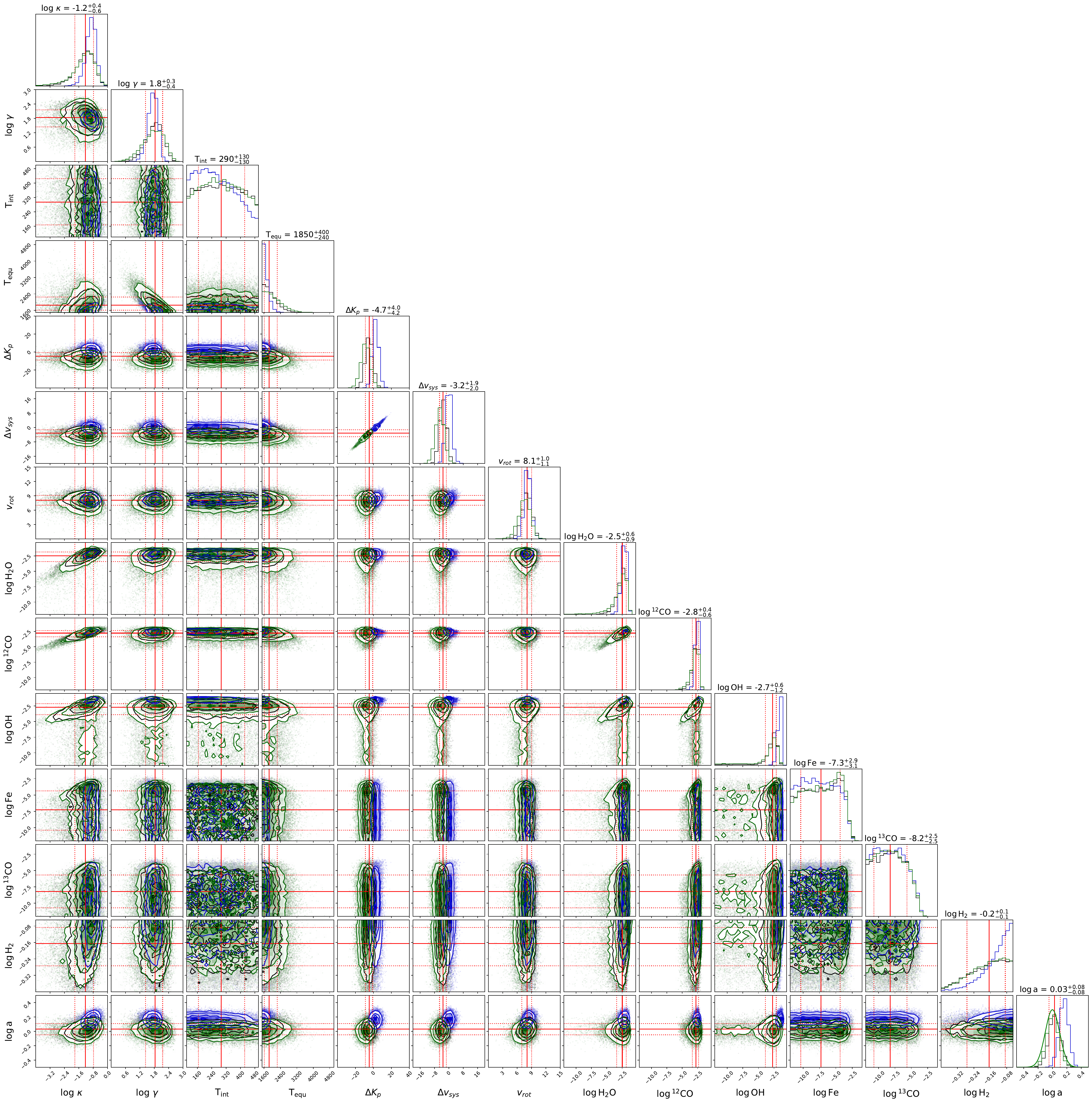}
    \caption{Full corner plot for \ktb. Red solid lines indicate the medians, while red dashed lines indicate the bounds of the marginalized 68\% confidence interval. We discuss these results in Section \ref{sec:res}.}
    \label{fig:cornerk20}
\end{figure}

\begin{figure}
    \centering
    \includegraphics[width=1.0\linewidth]{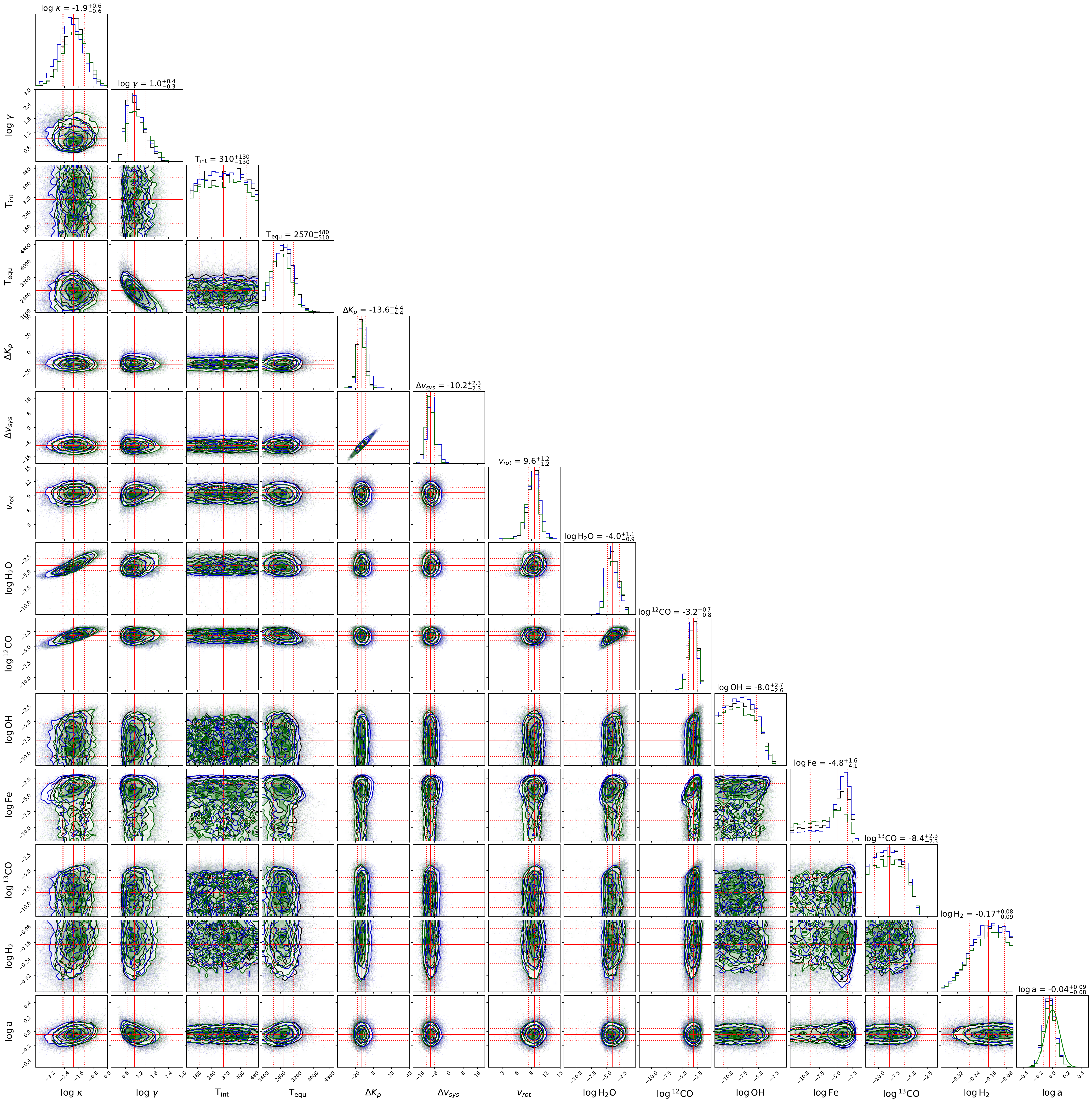}
    \caption{Full corner plot for \wtb. Red solid lines indicate the medians, while red dashed lines indicate the bounds of the marginalized 68\% confidence interval. We discuss these results in Section \ref{sec:res}.}
    \label{fig:cornerw33}
\end{figure}

\begin{figure}
    \centering
    \includegraphics[width=1.0\linewidth]{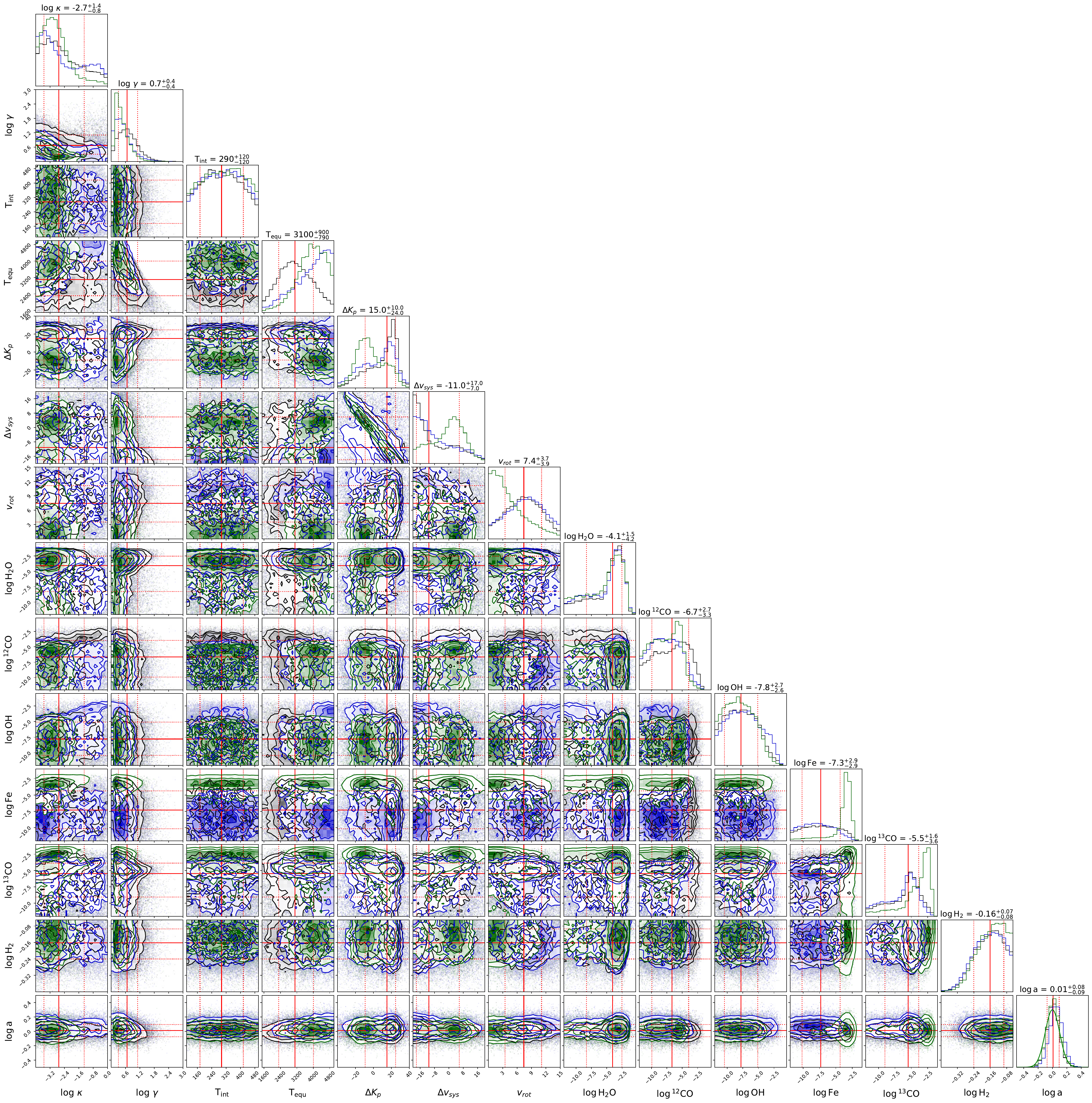}
    \caption{Full corner plot for \knb. Red solid lines indicate the medians, while red dashed lines indicate the bounds of the marginalized 68\% confidence interval. We discuss these results in Section \ref{sec:res}.}
    \label{fig:cornerk9}
\end{figure}

\begin{figure}
    \centering
    \includegraphics[width=1.0\linewidth]{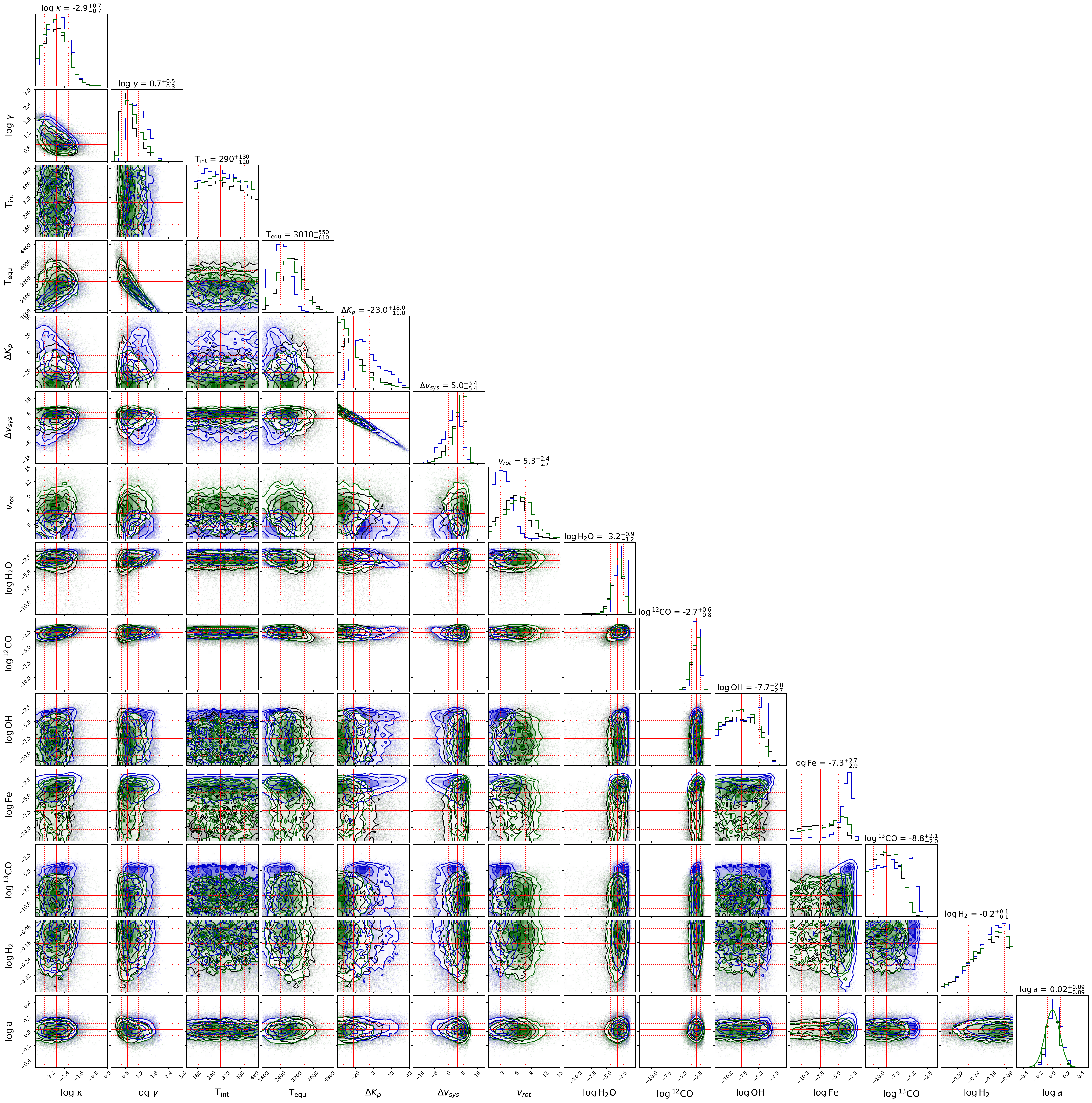}
    \caption{Full corner plot for \wob. Red solid lines indicate the medians, while red dashed lines indicate the bounds of the marginalized 68\% confidence interval. We discuss these results in Section \ref{sec:res}.}
    \label{fig:cornerw18}
\end{figure}

\begin{figure}
    \centering
    \includegraphics[width=1.0\linewidth]{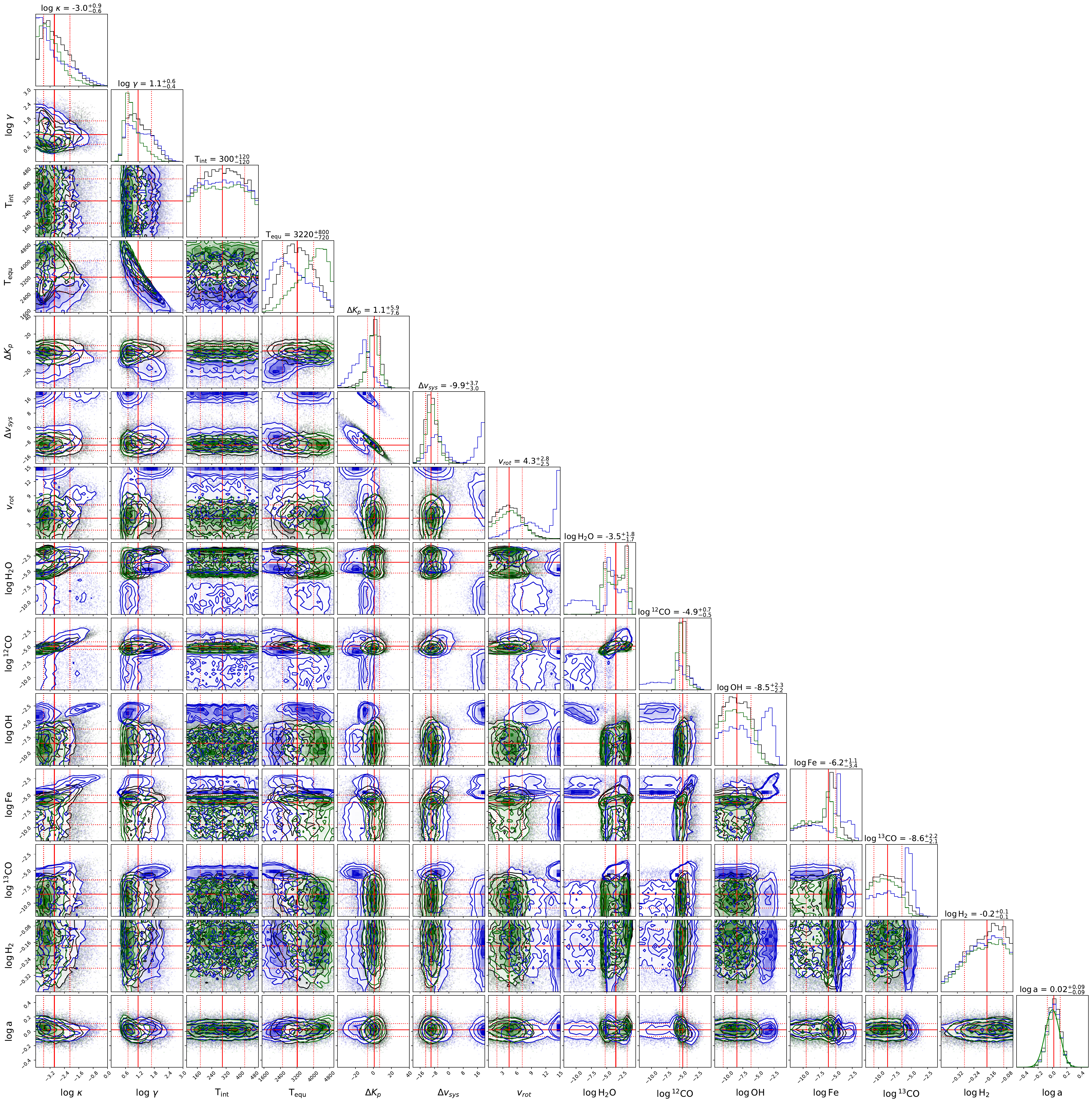}
    \caption{Full corner plot for \mob. Red solid lines indicate the medians, while red dashed lines indicate the bounds of the marginalized 68\% confidence interval. We discuss these results in Section \ref{sec:res}.}
    \label{fig:cornerm1}
\end{figure}

\begin{figure}
    \centering
    \includegraphics[width=1.0\linewidth]{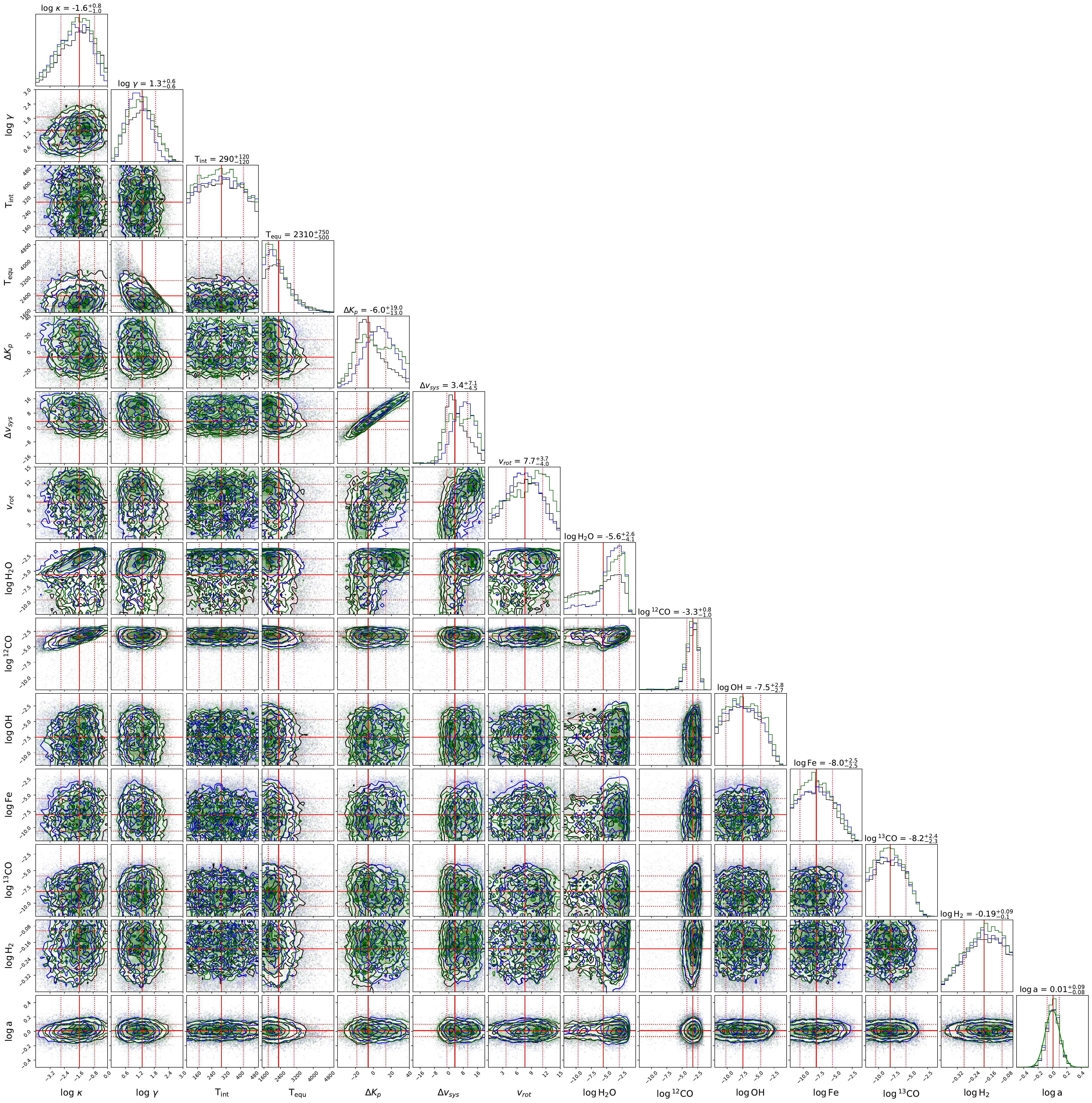}
    \caption{Full corner plot for \tob. Red solid lines indicate the medians, while red dashed lines indicate the bounds of the marginalized 68\% confidence interval. We discuss these results in Section \ref{sec:res}.}
    \label{fig:cornert15}
\end{figure}

Figure \ref{fig:cornerk20} presents the corner plot for the \ktb\ retrievals. Figure \ref{fig:cornerw33} presents the corner for the \wtb\ retrievals, Figure \ref{fig:cornerk9} presents the corner for \knb, Figure \ref{fig:cornerw18} the corner for \wob, Figure \ref{fig:cornerm1} the corner for \mob, and finally Figure \ref{fig:cornert15} shows the corner for \tob. For all targets, we show the retrieved posteriors omitting 4, 6, and 8 principal components in blue, black, and green, respectively. The printed values and uncertainties are the medians and $\pm34\%$ confidence intervals from the six component retrievals. The posteriors are generally consistent as the number of omitted components is changed, indicating the PCA is not strongly impacting the retrieval analysis.

\begin{figure}
    \centering
    \includegraphics[width=0.3\linewidth]{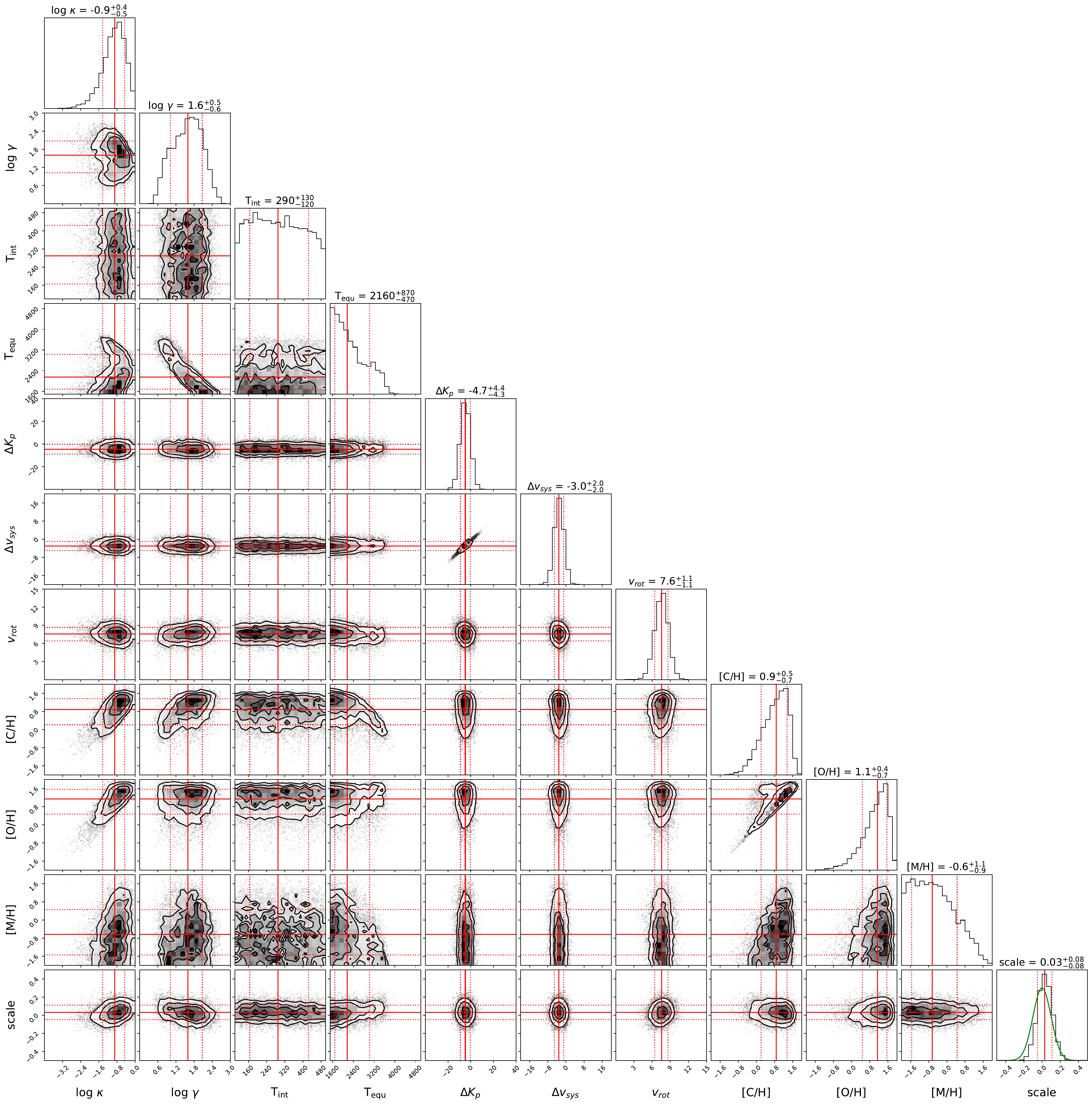}
    \includegraphics[width=0.3\linewidth]{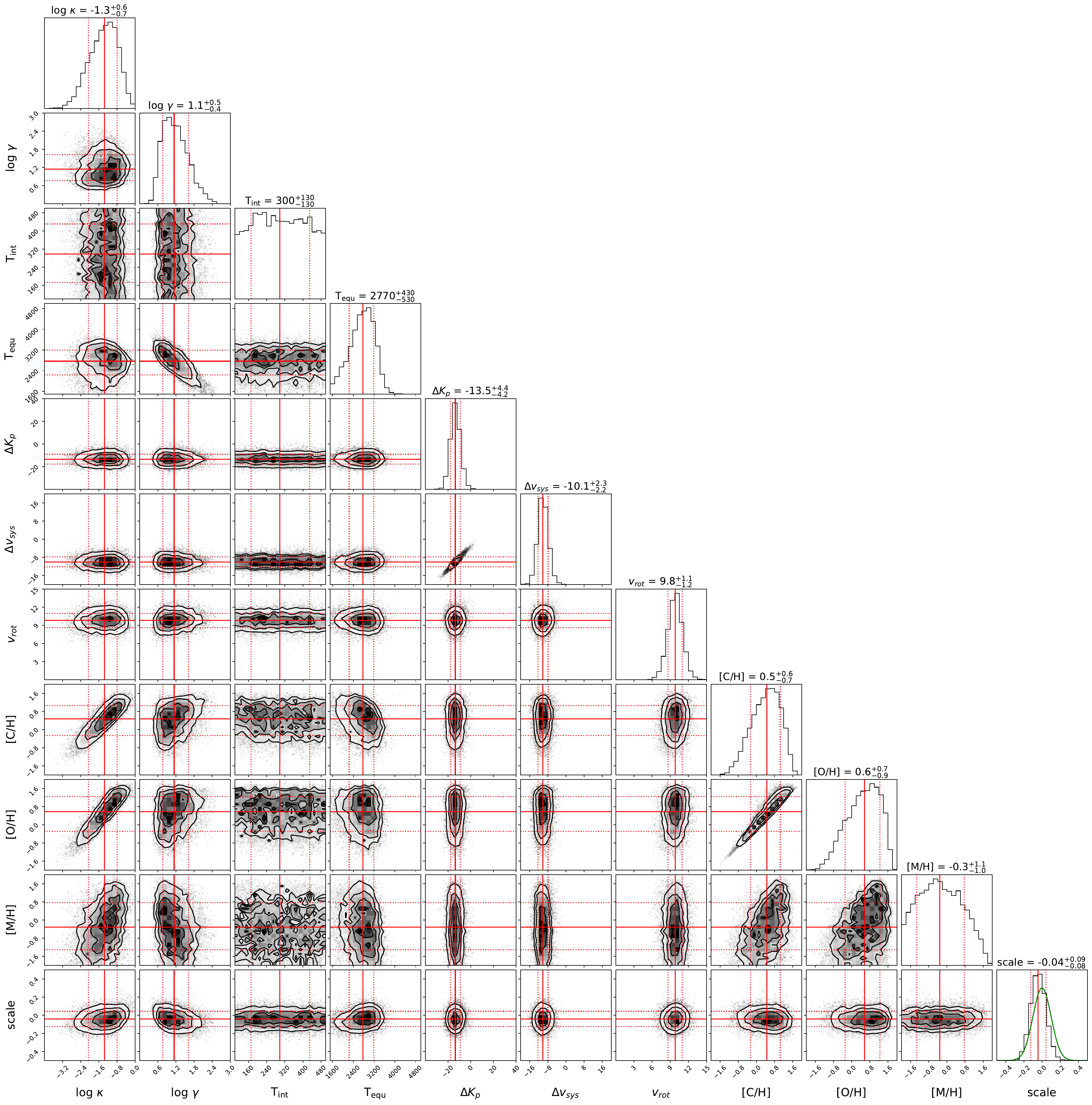}
    \includegraphics[width=0.3\linewidth]{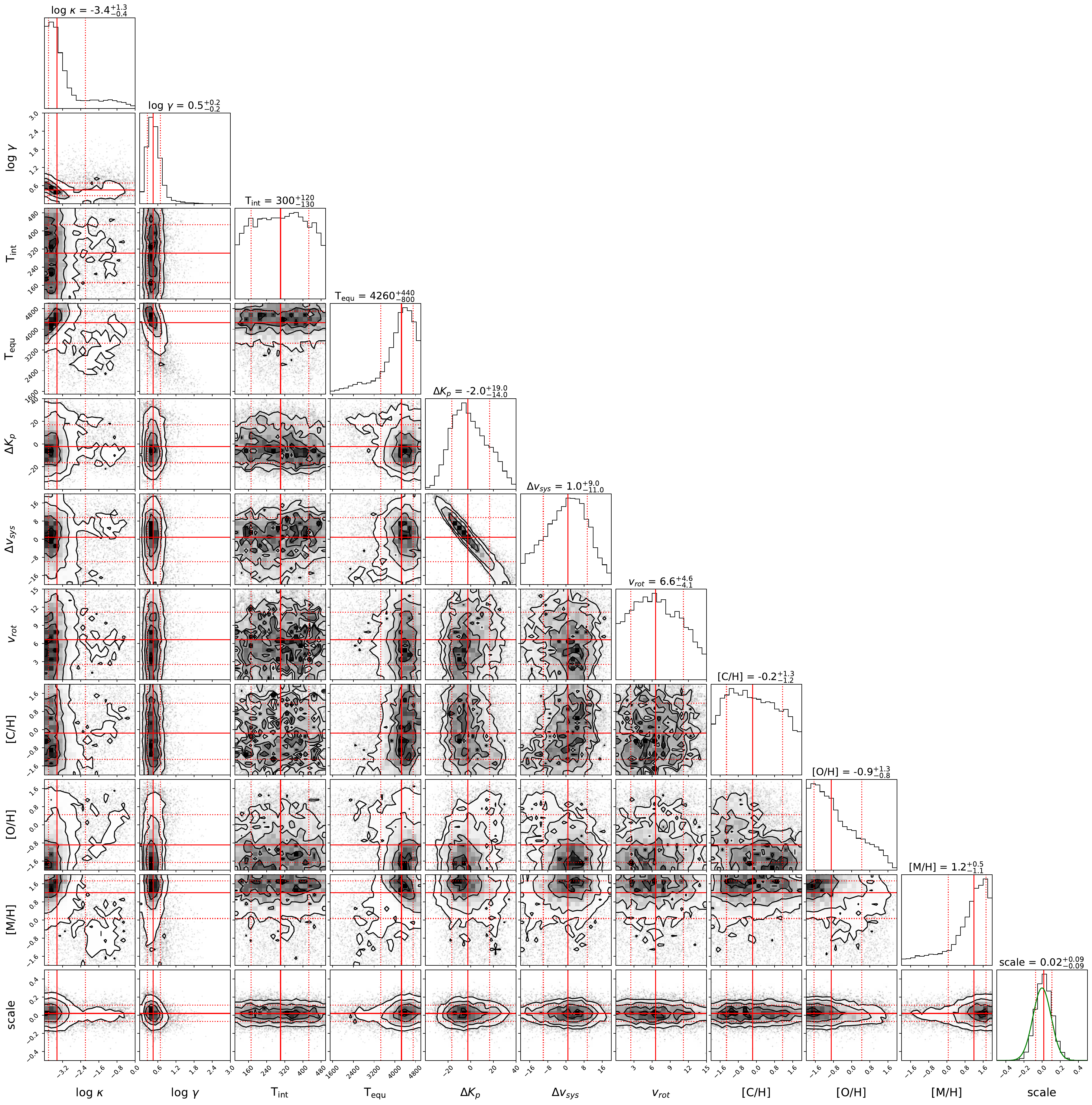}
    \includegraphics[width=0.3\linewidth]{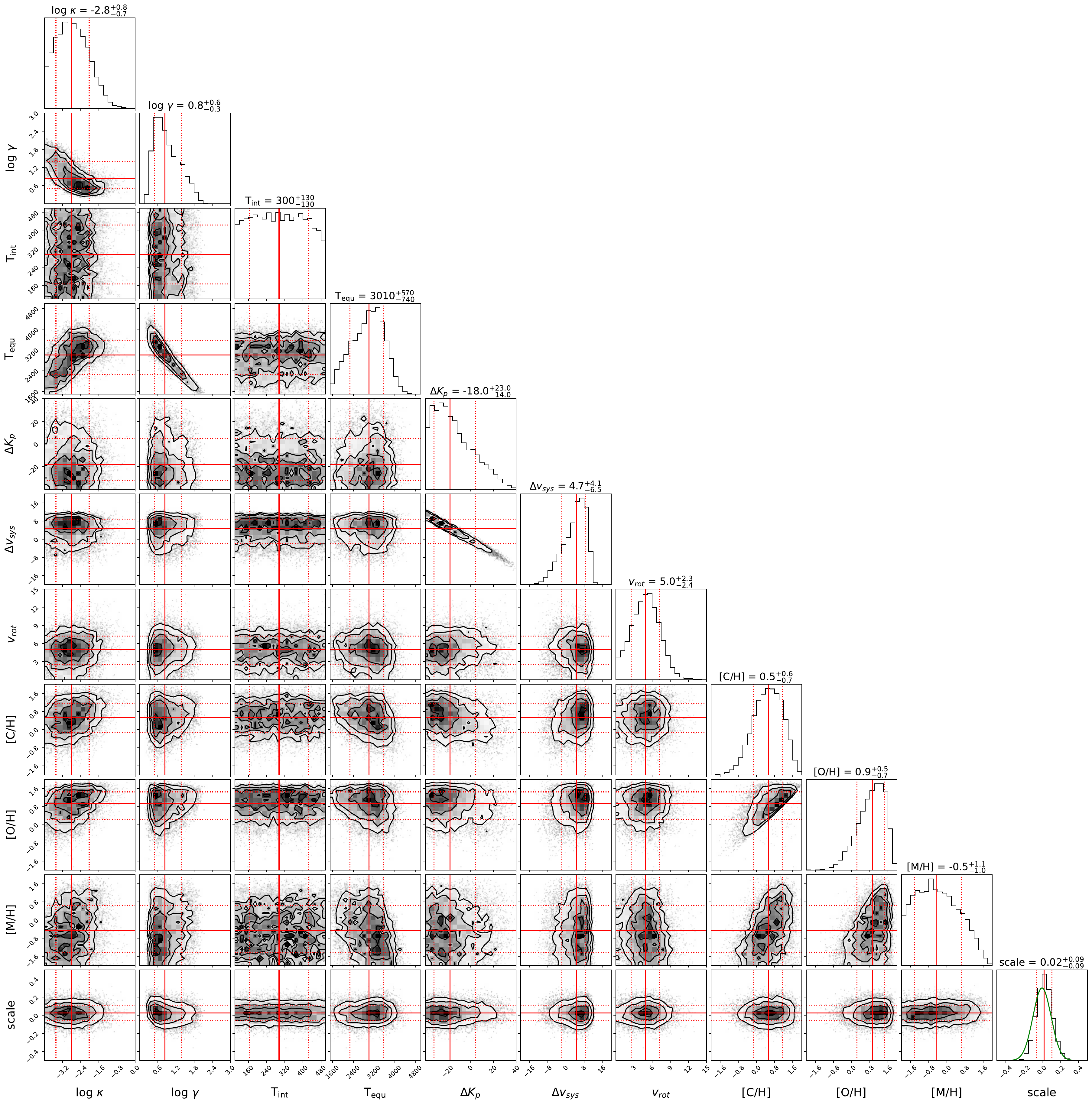}
    \includegraphics[width=0.3\linewidth]{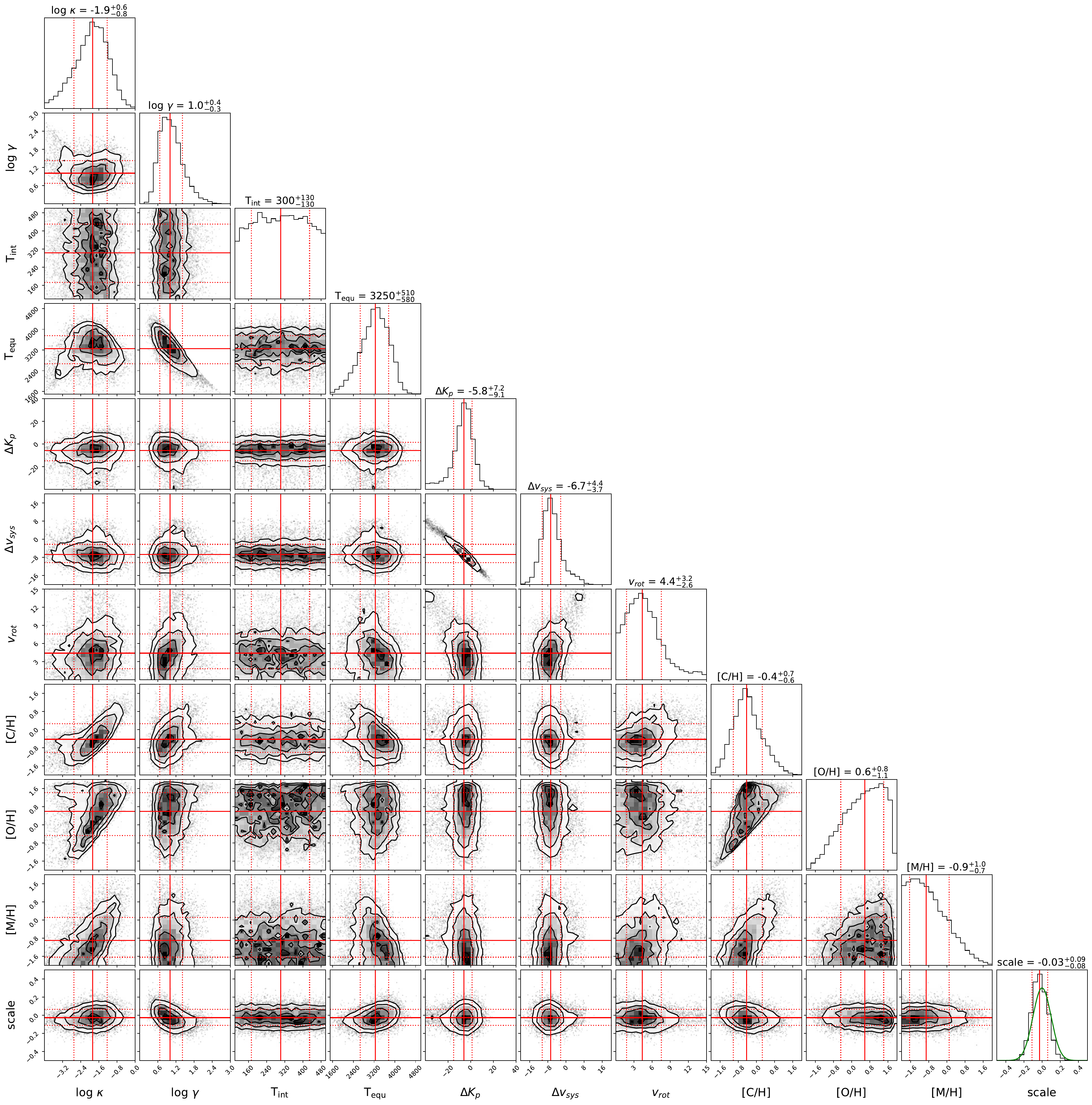}
    \includegraphics[width=0.3\linewidth]{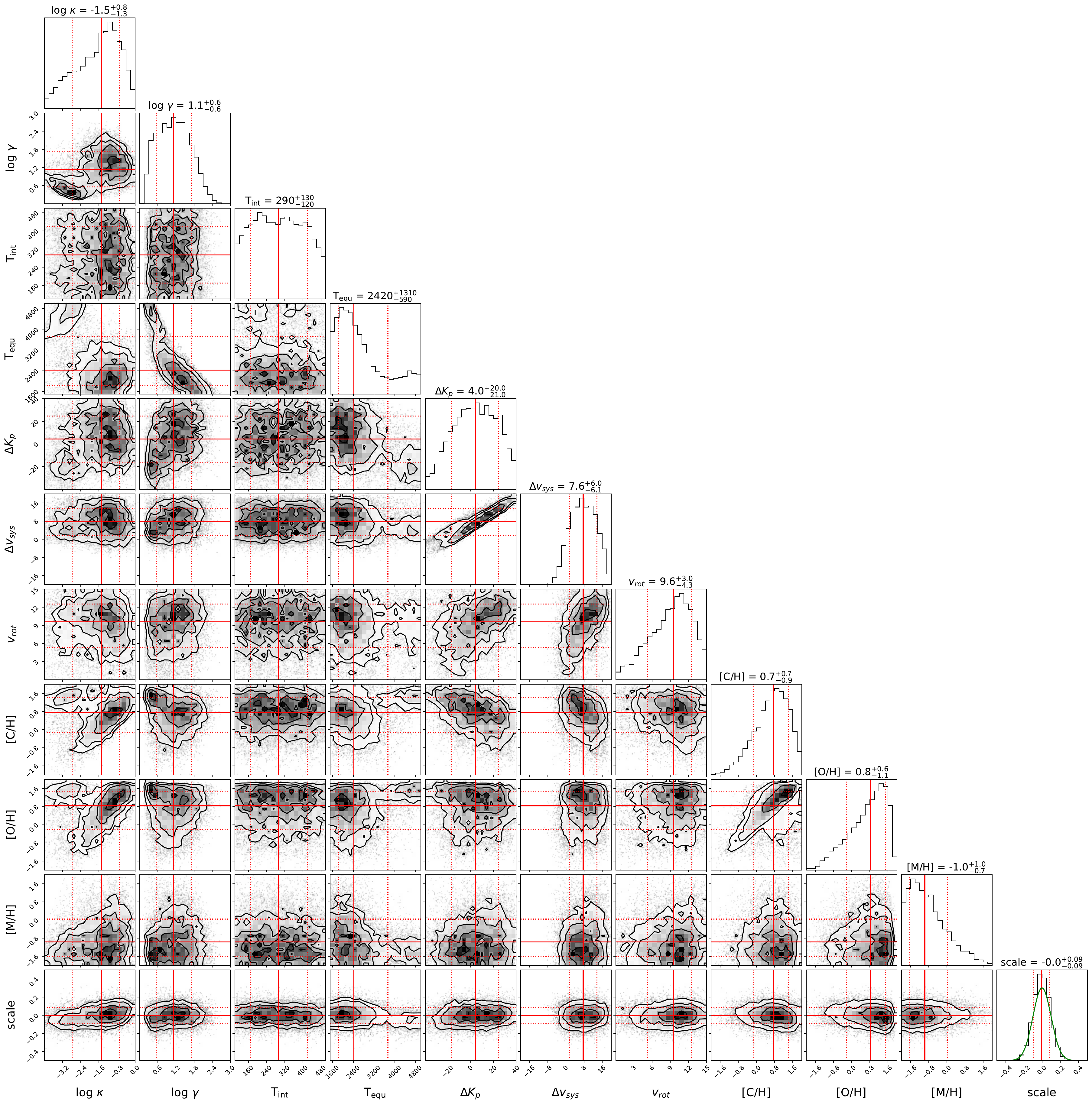}
    \caption{Full corner plot for the equilibrium retrievals. Red solid lines indicate the medians, while red dashed lines indicate the bounds of the marginalized 68\% confidence interval. We discuss these results in Section \ref{sec:res}. Plots correspond, from left to right, top to bottom, \ktb, \wtb, \knb, \wob, \mob\, and \tob, respectively. }
    \label{fig:eqcorners}
\end{figure}

Figure \ref{fig:eqcorners} shows the corner plots from the equilibrium retrievals. Equilibrium retrievals were performed omitting six principle components. 

\clearpage
\bibliography{exoplanetbib}{}
\bibliographystyle{aasjournalv7}

\end{CJK*}
\end{document}